%% file: BA_revision_final.tex
\documentclass[12pt,a4paper]{article}
\usepackage[margin=3cm]{geometry}
\usepackage{graphicx}
\usepackage{color}
\usepackage{natbib}
\usepackage{amsmath}
\usepackage{eurosym}
\usepackage{url}
\usepackage{fancyhdr}
\usepackage{amsthm}
\usepackage{amsmath}
\usepackage{natbib}
\usepackage[colorlinks,citecolor=blue,urlcolor=blue,filecolor=blue,backref=page]{hyperref}
\usepackage{graphicx}

\usepackage{amsmath}
\usepackage{amssymb}
\usepackage{algorithm}
\usepackage{booktabs}
\usepackage{subcaption}
\usepackage{color}

\usepackage{multibib}
\newcites{App}{References}

\numberwithin{equation}{section}
\theoremstyle{plain}
\newtheorem{thm}{Theorem}[section]
\newcommand{\cluster}{cluster}

\newcommand{\comment}[1]{{#1}}
\include{newcommands}
\newcommand{\commentSF}[1]{#1} % very last changes in the final version by Sylvia
\newcommand{\commentBG}[1]{#1} % OLD changes in the final version by Gertraud and Bettina
\newcommand{\commentF}[1]{#1}  % OLD changes in the final version by Sylvia
\newcommand{\commentdel}[1]{#1} % OLD changes, still in the paper, don't use anymore
\begin{document}

	\title{{\centering Generalized mixtures of finite mixtures and
			telescoping sampling}}
	
	\author{Sylvia Fr\"uhwirth-Schnatter\thanks{WU Vienna University of
			Business and Economics}, Gertraud Malsiner-Walli\thanks{WU
			Vienna University of Business and Economics}~{}and\\ Bettina
		Gr\"un\thanks{WU Vienna University of Business and
			Economics}\\} %\vspace{0.5cm}}
	%	\normalsize{WU Vienna University of Business and Economics}\\
	
	\date{} %\today
	
	\maketitle

%\title{Generalized mixtures of finite mixtures and telescoping sampling}
%\runtitle{Generalized MFMs and telescoping sampling}
%\thankstext{T1}{The authors gratefully acknowledge support from the \textit{Austrian Science Fund (FWF)}, grant P28740,  \commentBG{and through \textit{WU Projects}, grant IA-27001574.}}

%\begin{aug}
%\author{\fnms{Sylvia} \snm{Fr\"uhwirth-Schnatter}\thanksref{addr1}\ead[label=e1]{sylvia.fruehwirth-schnatter@wu.ac.at}},
%\author{\fnms{Gertraud} \snm{Malsiner-Walli}\thanksref{addr1}\ead[label=e2]{gertraud.malsiner-walli@wu.ac.at}}
%\and
%\author{\fnms{Bettina} \snm{Gr\"un}\thanksref{addr1}%
%\ead[label=e3]{bettina.gruen@wu.ac.at}}
%
%\runauthor{S. Fr\"uhwirth-Schnatter et al.}
%
%\address[addr1]{Institute for Statistics and Mathematics,
%Vienna University of Economics and Business,
%Welthandelsplatz 1, 1020 Wien, Austria
%\printead{e1}
%\printead*{e2}
%\printead*{e3}
%}
%\end{aug}

\begin{abstract}
Within a Bayesian framework, a comprehensive investigation of
 mixtures of finite mixtures (MFMs), \comment{i.e., finite mixtures with a prior on the number of components,} is performed. This model class has
applications in model-based clustering as well as for semi-parametric
density \comment{estimation} and requires suitable prior specifications and
inference methods to exploit its full potential.
We \commentF{contribute % to Bayesian mixture analysis
by} considering a
generalized class of MFMs % containing static and dynamic MFMs
where the hyperparameter $\edK$ of a symmetric Dirichlet prior on the %component
\comment{weight distribution   %either is fixed or
depends on the number of components. We show that this model class
may be regarded as a Bayesian non-parametric mixture outside the class of Gibbs-type \commentBG{priors}.}
 We emphasize the distinction between the number of components $K$
of a mixture and the number of clusters $\Kn$, i.e., the number of
filled components given the data. In the MFM model, $\Kn$ is a random
variable and its prior depends on the prior on %the number of components
$K$ and \comment{on the hyperparameter $\edK$.  We \commentSF{employ} %propose
a flexible prior distribution
for the number
of components $K$ and derive the  corresponding prior on
the number of clusters $\Kn$ for generalized MFMs.}
% derive computationally feasible formulas to calculate this implicit prior.
 % and link MFMs to Bayesian non-parametric mixtures.
%
 For posterior \commentF{inference % of a generalized MFM,
 we} propose the novel
telescoping sampler which allows Bayesian inference for mixtures with
arbitrary component distributions \commentF{without resorting} to
\comment{reversible jump Markov chain Monte Carlo (MCMC)}
%RJMCMC
methods.  The telescoping sampler explicitly samples the number of
components, but otherwise requires only the usual \commentF{MCMC steps
% for estimating
of} a finite mixture model.  The ease of its application using
different component distributions is demonstrated on several data
sets.

\end{abstract}
%
%\begin{keyword}[class=MSC]
%\kwd[Primary ]{\commentBG{62H30}}
%\kwd[; secondary ]{\commentBG{65C40}}
%\end{keyword}
%
%\comment{
%\begin{keyword}
%\kwd{Bayesian mixture\commentBG{s}}
%%\kwd{Dirichlet prior}
%\kwd{Dirichlet process mixtures}
%\kwd{sparse finite mixtures}
%\kwd{Pitman-Yor process mixtures}
%\kwd{reversible jump MCMC}
%%\kwd{prior distribution}
%\kwd{Gibbs-type priors}
%\end{keyword}}

% \input{part_introduction}
% \newpage
% \tableofcontents
\section{Introduction}

The present paper contributes \commentF{to %the methodology on
Bayesian  mixture
analysis} where the number of components $K$ is unknown and a prior on
$K$ is specified.  This % model
\commentF{class} of mixtures of finite mixtures
(MFMs) has a long tradition in Bayesian mixture \commentF{modeling %, see %, e.g.,
\citep{ric-gre:bay,nob:pos, mcc-yan:how} and has
%recently received new attention, see
gained recent attention by}
 \commentBG{\citet{mil-har:mix,gen-etal:pro,xie-xu:bay},
among others.}
%\comment{and also \citet{arg-etal:isinf} from a Bayesian non-parametric

Previously considered MFMs differ with respect to \commentF{prior specifications}
 on $K$ and the component weights. We combine the
different approaches to a \emph{generalized MFM} model
specification.
%This generalized MFM also encompasses mixture models where the prior on $K$ is degenerate putting all mass on a fixed or infinite value of $K$.
 We base our considerations on the crucial
distinction between the \textit{number of components} $K$ in the
mixture \commentSF{distribution} and the \textit{number of clusters} $\Kn$ in the data
which is defined as the number of \commentF{\lq\lq filled\rq\rq\ mixture components used to generate the observed data}.
 %which is defined as the number of components used to generate the observed data, i.e., the number of ``filled'' mixture components.
 This fundamental distinction between $K$ and $\Kn$ \comment{has always been prevalent in
Bayesian non-parametric (BNP) mixture analysis, see, e.g., the recent work by \citet{arg-etal:isinf}.}
In \comment{applied} finite mixture analysis, \comment{however,} it is still common
 to assume that $K$ and $\Kn$ are the same \comment{entity},
despite  \commentSF{earlier} work by \citet{nob:pos}, %who investigates the effect of empty components on the marginal likelihood of a MFM,
\citet{mcc-yan:how}, %who interpret $K$ as the potential number of species in the population and $\Kn$ as the number of observed species,
and, more recently, \citet{mil-har:mix}.
 % In the present paper, we focus on the implicit prior on $\Kn$ induced by specific choices for the prior on $K$ and the component weights in order to characterize the different MFM specifications and their suitability for clustering and semi-parametric density approximation.

Dirichlet process mixtures (DPMs) are the most popular BNP \comment{mixture} approach.
%both for density estimation and for clustering.
Their focus \comment{naturally} lies on inference on the number of
clusters, with \comment{$K$ being fixed at $+\infty$}.
For DPMs,  the number of
clusters grows as $\Kn \sim \alphaDP \log (N)$ as the number of observations $N$
increases.  Doubt about the usefulness of DPMs for clustering has been
voiced for many years and, indeed, \citet{mil-har:sim} \comment{proved}
inconsistency of DPMs for the number \commentBG{of} clusters for the simple case
of univariate Gaussian mixtures with unit variances.
As a two-parameter alternative to DPMs, \comment{Pitman-Yor process mixtures were introduced
in the BNP literature by \citet{pit-yor:two}.}
\commentBG{\citet{mal-etal:mod,mal-etal:ide}} introduced sparse finite mixtures
(SFMs) in the context of \comment{applied} finite mixture analysis.
 As shown by \citet{deb-etal:are}, both model classes are closely connected.
% As opposed to MFMs,
SFMs choose a \comment{fixed, clearly overfitting value of $K$
\commentSF{in the spirit of \citet{rou-men:asy}}
and
a symmetric Dirichlet prior on the weight distribution
%, $ \etav_K|K=\Kfix \sim \Dirinv{K}{\eFM_{\Kfix}}$
with a very small hyperparameter $\eFM_K$.} \comment{Whereas $K$ is fixed,
this choice allows} the number of clusters $\Kn$ to be a random variable taking values smaller than $K$.
% with high probability, both a priori and a posteriori.
% Empirical investigations in \citet{mal-etal:mod} and subsequent papers show that
\comment{However, the larger
  $K$, the smaller  $\eFM_{K}$ has to be},  motivating the \lq\lq dynamic\rq\rq\
   SFM introduced in \citet{fru-mal:fro}, where
$\eFM_{K} = \alpha /K$ \comment{was chosen with $\alpha$ being a hyperparameter} independent of
$K$.

The class of generalized MFMs  we introduce in this paper
is a finite mixture model with a prior on $K$, where
 the hyperparameter \commentF{$\eFM_{K}$ % of the symmetric Dirichlet prior
 may change as a} function of $K$.
We \commentF{consider % in particular
two special} cases of this specification. The \emph{static}  MFM uses a fixed value $\eFM_{K} \equiv \eFM$. % $\eFM$ for the hyperparameter $\eFM_{K}$.
The \emph{dynamic} MFM \commentSF{uses $\eFM_{K} = \alpha /K$ and can be regarded
as a dynamic SFM} % varies with $K$,
with a prior on $K$. This MFM specification, %was already
\commentF{considered previously  in
\citet{mcc-yan:how}, is} less common in applied \comment{finite} mixture analysis
than the static MFM.
%The dynamic MFM encompasses standard mixture models for small number of components, SFMs for increasing number of components and finally DPMs as the number of components grows toward infinity. Mixing over these different mixture models increases the flexibility of the model class to capture differences in the number of components and clusters as well as partitions.
  %
 \citet{mcc-yan:how} conjecture that the static and dynamic versions of
the MFM \commentSF{are quite} different. We \commentSF{shed} light on this by investigating
the exchangeable partition
probability function (EPPF), i.e., the prior induced on the random partition of
the data \citep{pit:exc} \commentBG{by}  the generalized MFM and discuss
its specific form  for static and dynamic MFMs.
 As \commentSF{shown} in the seminal work by  \citet{gne-pit:exc},
 the static MFM considered in
\citet{ric-gre:bay} and \citet{mil-har:mix}  is equivalent \commentBG{to}
a BNP mixture with \commentBG{a Gibbs-type prior} on the random partitions.
Based on the EPPF of the generalized MFM,
we show that the static MFM is the only mixture within this class that induces a Gibbs-type prior.
Any specification where the hyperparameter \commentF{$\eFM_{K}$}
% of the symmetric Dirichlet prior
varies with $K$ leads to a  BNP mixture beyond  Gibbs-type priors. We \commentSF{focus %in particular
on the dynamic MFM}
where \commentF{$\eFM_{K} = \alpha /K$} % this hyperparameter
\commentBG{is inversely proportional to} the number of
components and show that it converges to a DPM
\commentF{with concentration parameter $\alpha$}, if the prior $p(K)$ puts all mass on $+\infty$. Hence, while staying within the finite mixture framework, the dynamic  MFM  can be regarded as  a \lq\lq natural generalization\rq\rq\ of the celebrated Dirichlet process prior beyond the class of Gibbs\commentBG{-}type priors.

\commentF{We propose the three-parameter beta-negative-binomial
  distribution as a prior on the
number of components $K$ which \commentSF{unifies priors}
proposed in \citet{ric-gre:bay,nob:pos,cer:new,mil-har:mix,gra-etal:los}.}
\commentSF{Building on \citet{ant:mix,nob:pos,gne-pit:exc}, we derive the implicitly induced prior on the number of clusters $\Kn$ for generalized MFMs.} %specification.
%We develop computational feasible formulas for the determination of this prior conditional on sample size given the MFM specification, i.e., the prior on $K$ and on the component weights.

%We also link the different specifications of MFMs to BNP mixtures, such as DPMs and Pitman-Yor process mixtures, and SFMs.

A tremendous challenge for Bayesian mixtures with an unknown number of
components is practical statistical inference.  To this aim,
\citet{ric-gre:bay} introduced \commentF{reversible jump Markov chain Monte
Carlo % methods
(RJMCMC)} for \commentSF{static} MFMs with univariate Gaussian components.
\commentSF{Exploiting that static MFMs are Gibbs-type priors, \citet{mil-har:mix}  introduced sampling techniques from BNP statistics to finite mixture analysis.
Applying the Chinese restaurant process (CRP) sampler of  \citet{jai-nea:spl_2004,jai-nea:spl_2007}, }
they sample the partitions and, in this
way, the number of \cluster s $\Kn$ and infer the number of components $K$
in a post-processing step by linking the distribution of \commentF{$K$ to the distribution of $\Kn$.}
% the number of components to the distribution of the number of clusters.

\comment{In this paper, we} introduce a novel MCMC algorithm \commentSF{for generalized MFMs} called
\emph{telescoping sampling} that updates simultaneously \commentF{the number of
\cluster s $\Kn$ and the number of components $K$} during sampling
without \commentSF{resorting} to RJMCMC. \commentSF{As opposed to CRP sampler, telescoping sampling also works outside the class of Gibbs-type priors.}
\commentSF{Sampling $K$ %The updating of
% the number of components \comment{within} our sampler
% is totally generic as it
only depends on the current partition of the data and is
independent} of the component parameters.  This makes our sampler a
most generic inference tool for finite mixture models with an unknown
number of components which can be applied to arbitrary mixture
families.  Our sampler is easily implemented, for instance, for
% mixtures of multivariate Gaussian distributions
   multivariate Gaussian mixtures  with an unknown number
of components, and thus provides an attractive alternative to RJMCMC
which is challenging to tune in higher dimensions, see, e.g., \citet{del-pap:mul}.

The paper is structured as follows.  In Section~\ref{sectionmod}, we
present the generalized MFM model and  derive the EPPF.
Section~\ref{section3} proposes the beta-negative-binomial
   \commentF{as a prior on $K$} %$p(K)$
  and  derives the prior on the number of clusters \commentBG{$\Kn$} for a generalized MFM.
    Section~\ref{sectionbridge} \comment{discusses connections between applied finite mixture analysis based on MFMs}       % SFMs, MFMs with DPMs
      and BNP mixtures.
    Our novel MCMC sampler % for mixtures with an unknown number of components $K$
is presented in
Section~\ref{sectiontele} and is benchmarked against RJMCMC
and the CRP sampler  % applied in \citet{mil-har:mix}..
 in Section~\ref{sec:empir-demonstr}.
 \commentF{Additionally,  MFMs  with various uni- and multivariate component densities
  are applied both to artificial and real data}
%are presented and the performance of static and dynamic MFMs in combination with different prior specifications  is investigated in a  simulation study with artificial data}
 of varying dimension and  sample size. Section~\ref{sec:disc-final-remarks} concludes.

\section{\commentF{Generalized mixtures of finite mixture models}} \label{sectionmod}

\subsection{Model formulation}  \label{sectionmoddef}

Consider $N$ observations $\ym=(\ym_1, \ldots, \ym_N)$ of a uni- or
multivariate continuous or discrete-valued variable.
The generalized MFM is defined in a hierarchical way:
\begin{align} \label{eq:MFM}
K &\sim p(K),\\
\eta_1,\ldots,\eta_K |K,\edK &\sim \Dirinv{K}{\edK},   \nonumber \\
\phi &\sim p(\phi), \nonumber \\
\btheta_{k}|\phi &\sim p(\btheta _{k}|\phi),   \text{ independently for } k=1,\ldots,K,  \nonumber  \\
S_i|K ,\eta_1,\ldots,\eta_K &\sim \Mulnom{1;\eta_1,\ldots,\eta_K},  \text{ independently for } i=1, \ldots,N , \nonumber\\
\ym_i|K,S_i=k,\btheta_k &\sim f(\ym_i|\btheta_{k}), \text{ independently for } i=1, \ldots,N,   \nonumber
\end{align}
where % $\boldeta_K=(\eta_1,\ldots,\eta_K)$ are the component weights,
$S_i$ is the latent allocation variable of observation $\ym_i$, and
$f(\ym_i|\btheta_{k})$ is the parametric density of component $k$.
Model~\eqref{eq:MFM} depends on a sequence $\gammav = \{\gamma_K\}$ of
positive numbers which defines for each $K$ the hyperparameter of the
symmetric Dirichlet prior
$\boldeta_K | K,\gamma_K \sim \Dirinv{K}{\edK}$ on the component
weights \comment{$\boldeta_K =(\eta_1,\ldots,\eta_K)$}.  % only
  \commentF{The component parameters $\btheta_{k}$ are independent conditional on the (random) hyperparameters $\phi$. In
  combination with the invariance of \commentBG{the }symmetric Dirichlet prior % with a single parameter $\edK$
 % induces an exchangeable prior on the component weights  with the prior considered }
  the prior specification is therefore invariant to
  label-switching.}

Model~\eqref{eq:MFM} contains the finite mixture model with a prior on
the number of components $K$ studied by \citet{ric-gre:bay} and
\citet{mil-har:mix}, who termed this model a mixture of finite
mixtures (MFM), as that special case where \commentF{$\edK\equiv \eFM$}.
%, with $\eFM$ a known constant.
%
 As noted by \citet{mil-har:mix}, assuming the same $\eFM$ for all $K$
is a ``genuine restriction'' which considerably simplifies the
derivation of the implied partition distribution -- a crucial
ingredient to their inference algorithm.  \citet{mcc-yan:how} extend this ``static''
MFM with constant $\eFM$ by
%considering a sequenc $\gammav = \{\gamma_K\}$ and in particular
specifying  a ``dynamic''
MFM where $\edK=\alpha/K$ \commentBG{is inversely proportional to} $K$ and depends on a
hyperparameter $\alpha$, i.e.,
$ \boldeta_K|K, \alpha \sim \Dirinv{K}{\alpha/K}$.

For a given $K$, $\Kn$ is defined as the number of components that
generated the data, i.e., $\Kn=\sum_{k=1}^K \indic{N_k>0}$, where
$N_k =\Count{i:S_i=k}$ counts the observations generated by component
$k$. In the following we refer to $\Kn$ as the number of \cluster s.
Including a prior $p(K)$ leads to both $\Kn$ and $K$ being random a
priori. As opposed to the common perception that for a finite mixture
$\Kn$ given $K$ is deterministic and equal to $K$,
\commentF{we show in Section~\ref{section3} that
the sequence of hyperparameters $\gammav = \{\edK\}$ has a crucial
impact on the induced prior of the data partitions and the number of
clusters $\Kn$}.
 For a static MFM with $\eFM=1$
\citep{ric-gre:bay,mil-har:mix}, e.g., the prior expected number of
clusters, $\Ew{\Kn|N, \eFM=1}$, is indeed close to $\Ew{K}$ for many
priors $p(K)$ with finite mean, even for small $N$.  However, having
$\edK$ decrease with increasing $K$ induces randomness in the prior
distribution of $\Kn$ given $K$, allowing for a gap between $\Kn$ and
$K$ for a wide range of $\alpha$ and $N$ values.

Under model~\eqref{eq:MFM}, the joint distribution of the data
$\ym=(\ym_1, \ldots, \ym_N)$ has a representation as a countably
infinite MFM with $K$ components:
\begin{align}  \label{mix:distym}
  p(\ym)  &=   \sum_{K=1}^\infty  p(K) \prod_{i=1}^N p(\ym_i|K) , \quad   p(\ym_i|K)= \sum_{k=1}^K  \eta_k f(\ym_i|\thetav_k).
\end{align}
\noindent
The type of mixtures which are summed over in (\ref{mix:distym}) vary
with the prior parameter $\gamma_K$ of the component weights.
% Large values of $\gamma_K$ imply mixtures with balanced component sizes.
Using a symmetric Dirichlet prior, a priori the mean
of the component weights given $K$ is equal to a vector of dimension
$K$ with values $1/K$. However, the variance decreases with increasing
\comment{hyperparameter $\gamma_K$  and thus more prior mass} is assigned to balanced weight
distributions.  \comment{On the other hand, the variance increases
and  the component weights a priori}
 become more unbalanced with decreasing values of $\gamma_K$.
For a static MFM with $\edK\equiv \eFM$, mixtures of a
similar type are combined.  For a dynamic MFM with $\gamma_K=\alpha/K$,
mixtures favoring different component size distributions are combined:
standard mixture models with balanced components, which emerge for
small $K$, are mixed with SFMs for moderate $K$ and finally, as $K$
goes to infinity, with DPMs favoring extremely unbalanced component
sizes. \comment{As will be shown in Section~\ref{sectioneppf}, the}
 dynamic prior on the component weights increases the flexibility of
the prior induced on the partitions and $\Kn$ \comment{and leads outside the
family of Gibbs-type priors}.  Moreover, a hyperprior
on $\alpha$, \comment{to be discussed in Section~\ref{sectionpra}},
 achieves \comment{additional} adaptivity of the induced prior \commentSF{on} the partitions
to the data at hand.

\subsection{The EPPF and the prior distribution of  cluster sizes} \label{sectioneppf}

The MFM model~\eqref{eq:MFM} induces through the latent indicators
$\Siv=(S_1,\ldots,S_N)$ a \comment{random} partition $\cP=\{\cC_1,\ldots,\cC_{\Kn}\}$
of the $N$ data points into $\Kn$ \cluster s where each \cluster\
\comment{$\cC_j$} contains all observations generated by the same mixture
component, i.e., $S_i=S_j$ for all $\ym_i, \ym_j \in \comment{\cC_j}$, \comment{see
\citet{lau-gre:bay}. In the tradition of \citet{pit:exc},} we
derive in Theorem~\ref{lemma1} the prior partition probability
function $p(\parti |N, \gammav)$ of a generalized MFM \comment{for} % with
a given
sequence $\gammav = \{\edK\}$ and discuss static MFMs with
$\edK \equiv \eFM$ and dynamic MFMs with $\edK=\alpha/K$ as special
cases.  In addition, we derive the prior distribution
$p(N_1, \ldots, N_{\Kn}|N, \gammav)$ of the {\em labeled} cluster
sizes $N_j=\card{C_j}$, \comment{where the $\Kn$ clusters in $ \cP$ are arranged in some exchangeable random order and we assign label 1 to the first cluster, label 2 to the second cluster and \commentF{so forth % till label $\Kn$ is assigned to the last cluster, see
\commentBG{\citep{pit:com}}}.\footnote{\comment{One such order is arrangement in order of appearance \citep{pit:som}, where the first observation $\ym_1$ belongs to the first cluster and for each $j =2, \ldots \Kn$, the first observation not assigned to $\cup_{\ell=1}^{j-1} \cC_\ell$ belongs to cluster $\cC_j$. However, any other exchangeable random ordering will do.}}}
%\commentdel{This distribution is helpful in deriving the induced prior on the number of clusters $p(K_+|N, \gammav)$.}
%
\begin{thm}\label{lemma1}
  For a generalized MFM with proper prior $p(K)$ and
  $\boldeta_K|K ,\gammav \sim \Dirinv{K}{\edK}$, the probability mass function
  $p(\parti |N, \gammav)$ of the set partition
  $\parti= \{\parti_1, \ldots, \parti_{\Kn}\}$ and the prior
  distribution $p(N_1, \ldots, N_{\Kn}|N, \gammav)$ of the labeled cluster
  sizes are given by:
\begin{align}
\label{mixparti}
p(\parti |N, \gammav)
  &=  \sum_{K=\Kn}^\infty  p(K)   p(\parti | N, K, \edK  ), \\
  \label{mixpartiK}
 p(\parti |  N, K,  \edK )  &= \frac{ V_{N, \Kn}^{K, \edK}}{ \Gamfun{ \edK } ^{\Kn} }    \prod_{j=1}^{\Kn}  \Gamfun{N_j +\edK }   , \comment{\qquad \mbox{where } N_j=\card{C_j},}\\
  \label{mixpK}
p(N_1, \ldots, N_{\Kn} | N, \gammav)
&=   \frac{  N!}{  \Kn !}   \sum_{K=\Kn}^\infty    p(K) \frac{ V_{N, \Kn}^{K, \edK}}{ \Gamfun{ \edK } ^{\Kn} }
  \prod_{j=1}^{\Kn}   \frac{ \Gamfun{N_j +\edK }}{ \Gamfun{N_j + 1}} ,\\
\label{mixpKV}
\displaystyle  V_{N,\Kn }^{K, \edK}  & =  \frac{ \Gamfun{\edK K} K ! }{ \Gamfun{\edK K+N} (K- \Kn )!} .
\end{align}
\end{thm}
Being a symmetric function of the \cluster\ sizes \comment{$(N_1, \ldots, N_{\Kn})$}, $ p(\parti|N, \gammav)$ is
an EPPF \citep{pit:exc}  and defines an exchangeable random partition of the $N$ data points \comment{for the class of generalized MFMs}.
\comment{The EPPF is instrumental for understanding the mathematical properties of the implied partitions and is a main object of interest in BNP mixtures, see, e.g., \citet{lij-pru:mod}.}

\comment{An important class of BNP mixture models are mixtures relying on Gibbs-type random probability measures, or {\em Gibbs-type priors}, introduced in the seminal work by \citet{gne-pit:exc}. They are considered the most natural  generalization of \commentF{DPMs}
% Dirichlet process mixtures,
as they allow \commentF{better} control of the clustering behavior, see the excellent work of \citet{deb-etal:are}. Under a Gibbs-type prior, the EPPF takes a specific product form which allows to study the EPPF of a generalized MFM in this regard.}
 Relying on \citet{gne-pit:exc}, \cite{gne:spe} and \citet{deb-etal:asy}, among others, \citet{mil-har:mix} show that a static MFM  induces a Gibbs-type prior on the partitions. Indeed, for $\edK \equiv \eFM$ the EPPF in (\ref{mixparti}) takes \commentBG{the} following
product form:
\begin{align}   \label{parMFM}
p(\parti   | N,  \eFM )  &=    {V} ^\eFM _{N, {\Kn}}  \prod_{j=1}^{\Kn}  \frac{ \Gamfun{N_j+\eFM} }{ \Gamfun{\eFM} } ,
\end{align}
where  ${V}^\eFM _{N, {k}} =  \sum_{K=\comment{k}}^\infty p(K)  \frac{K ! \Gamfun{\eFM K}}{(K- k )! \Gamfun{\eFM K + N}} $ satisfies \commentBG{the} following recursion for $k=1, \ldots, N-1$ % \citep{mil-har:mix}
\commentSF{(see Appendix~\ref{proof} in the supplementary material for a proof):}\commentdel{\footnote{\comment{Note that the normalization
$ \tilde{V} ^{\commentBG{\eFM}} _{N,k} = \eFM ^k {V} ^\eFM _{N,k} $  is needed to represent (\ref{parMFM}) as the common EPPF of \commentBG{a} Gibbs-type prior: $p(\parti   | N,  \Kn=k)= \tilde{V} ^{\commentBG{\eFM} }_{N,k} \prod_{j=1}^k W_{N_j}$, where $ W_\ell=  (1+ \eFM)_{(\ell-1)!} = \frac{\Gamfun{\ell +  \eFM}}{ \Gamfun{1+\eFM} } $ are the rising factorials.}}}
\begin{align}   \label{recMFM}
\comment{ {V} ^\eFM _{N,k} =  (N + \eFM k )  {V} ^\eFM _{N+1,k}  + {V}^\eFM_{N+1,k+1}.}
\end{align}
\comment{However, for a generalized MFM with $\edK$ depending on $K$, %such as a dynamic MFM with $\edK=\alpha/K$,
we obtain a mixture model with a partition structure beyond Gibbs-type priors.}
For a dynamic MFM, we establish in
Theorem~\ref{theorem1} that the EPPF $p(\parti | N, \alpha) $ can be expressed
explicitly in relation to a DPM with precision parameter $ \alpha$,
for which the EPPF is given by the Ewens distribution:
\begin{align} \label{Ewens}
  \fEwens (\parti | N, \alpha ) & = \frac{ \alpha ^{\Kn} \Gamfun{\alpha} }{
  \Gamfun{\alpha + N}} \prod_{j=1}^{\Kn} \Gamfun{N_j} .
\end{align}

\begin{thm}\label{theorem1}
For a dynamic MFM with  $\edK=\alpha/K$,  the EPPF  $p(\parti  | N, \alpha) $ can be expressed  as:
\begin{align}  \label{mixpNKK}
  p(\parti | N, \alpha) &
=   \fEwens (\parti |N, \alpha ) \times  \sum_{K= K_+ }^ {\infty}  p(K )  R_{\bN,\Kn}^{K, \alpha} ,\\
   R_{\bN,\Kn}^{K,\alpha} &= \prod_{j=1}^{\Kn}   \frac{\Gamfun{N_j +  \frac{\alpha}{K} } (K -j+1)  }
   { \Gamma(1+\frac{\alpha}{K})\Gamfun{N_j}K}, \nonumber
\end{align}
where $\fEwens (\parti | N, \alpha )$ is the probability mass function
(pmf) of the Ewens distribution and $\bN$ is the vector of induced
cluster sizes $(N_1,\ldots,N_{K_+})$.
\end{thm}

\comment{It follows from Theorem~\ref{theorem1} that dynamic MFMs can
  be regarded as a \lq\lq natural generalization\rq\rq\ of the
  celebrated Dirichlet process prior beyond the class of Gibbs\commentBG{-}type
  priors.} Theorems~\ref{lemma1} and \ref{theorem1} (which are proven
in Appendix~\ref{proof}) are exploited \comment{further} in
Section~\ref{section3} to derive the induced prior on the number of
clusters $p(K_+ | N, \gammav)$ and in Section~\ref{sectionbridge} to
\comment{investigate connections between applied finite mixture
  analysis based on} MFMs and commonly used BNP mixtures \comment{in
  more depth}.

\section{The prior distributions of $K$ and $\Kn$}\label{section3}

% Important insights into how to specify the priors of $K$ and the
% sequence $\gammav$ for the prior on the component weights are gained
% by investigating how these prior specifications influence the implicit
% prior on the number of clusters $\Kn$ and other characteristics of the
% partition function.

This section proposes a suitable choice for
$p(K)$ and derives % computationally feasible formulas for determining
the implicit prior of $\Kn$ in dependence of \commentF{$p(K)$, the hyperparameters
$\gammav$ and $N$}
% the priors on $K$ and the component weights
\comment{for a generalized MFM}.

\subsection{Choosing the prior on the number of components
  $K$} \label{sectionpK}

\begin{figure}[t!]
\centering
\includegraphics[width=0.85\textwidth, trim = 0 5 0 5, clip]{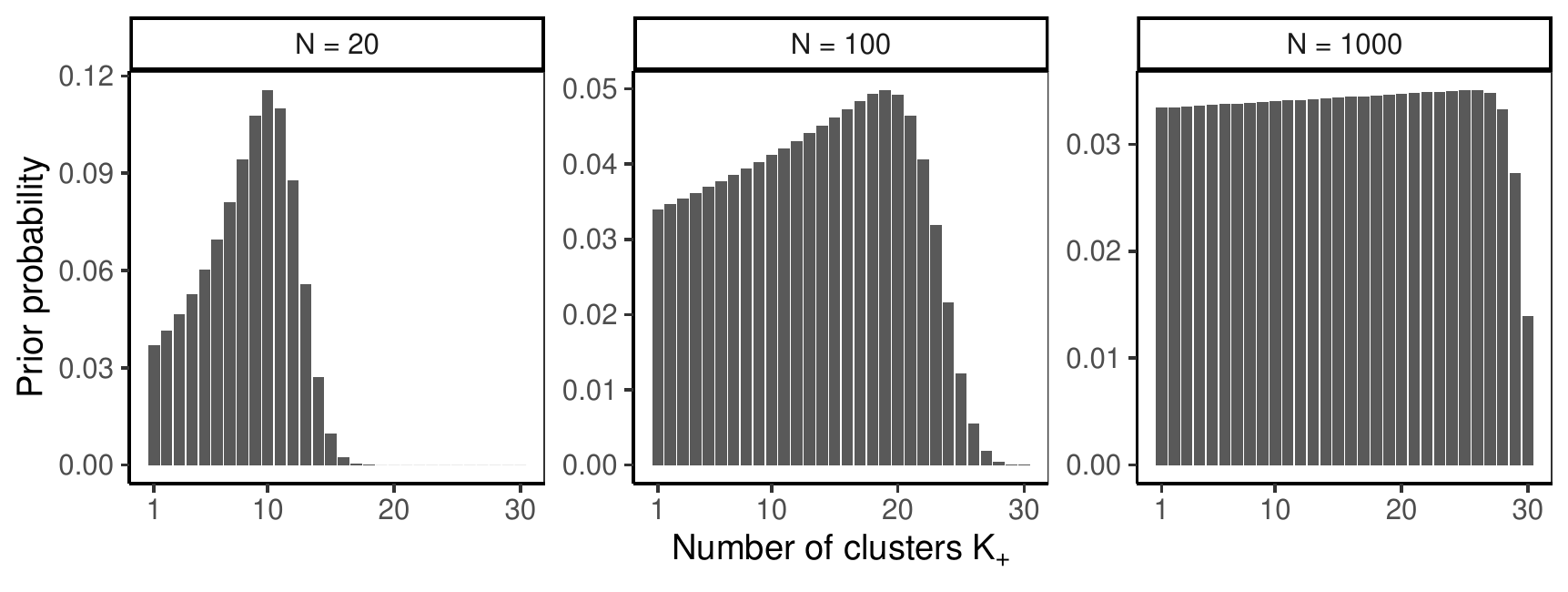}
\caption{The implicit prior $p(\Kn|N, \eFM=1)$ on the number of
  clusters $\Kn$ for the static MFM under the uniform prior
  $K \sim \cU\{1,30\}$ for various data sizes,
  $N=20, 100, 1000$.}\label{fig:UNIFORM}
\end{figure}

%\commentF{In this section, we discuss the choice of the prior $p(K)$ on the number of components $K$.}
\commentF{In their seminal paper,}
\citet{ric-gre:bay} suggest a uniform prior $K \sim \cU\{1,K_{\max}\}$
\comment{for a static MFM} % in combination
with $\gamma_K\equiv 1$. \commentF{However, depending on $N$, the  prior on $\Kn$
\commentSF{might}  be surprisingly informative and far from a uniform distribution.
Figure~\ref{fig:UNIFORM} shows the implied prior
$p(\Kn|N, \eFM=1)$ for a static MFM under the  prior
$K \sim \cU\{1,30\}$ for various data sizes $(N=20, 100, 1000)$. Evidently,
 the prior mode depends on $N$  and only for larger $N$ approximately a uniform prior results.}

\citet{nob:pos} shows that, as an alternative to the uniform prior,
any proper prior $p(K)$ which satisfies $p(K)>0$ for all $K \in \NNN $ %\ge 1$
can be adopted.  While most discrete probability distributions include
zero, in a mixture context the prior $p(K)$ has to exclude zero. This
is often achieved by truncating the pmf at one, e.g.,
\citet{nob-fea:bay} use the Poisson distribution $K \sim \Poisson{1}$
restricted to $\{1,2, \ldots, K_{\max}\}$.  However, it is more
convenient to work with the translated prior $K-1 \sim p_t$, where the
pmf \commentF{$p(K)=p_t(K-1)$ is obtained by evaluating the translated pmf at $K-1$},
as for translated priors hierarchical priors can be \comment{more} easily introduced.
 We propose \commentSF{a translated prior, where $K-1 \sim \BNB{\alphaNB, \pia,\pib}$}
 follows the
beta-negative-binomial (BNB) distribution which is a hierarchical
generalization of the Poisson, the geometric and the negative-binomial
distribution. The \commentSF{corresponding pmf
%of the translated BNB distribution $K-1 $
is} given by:
\begin{align}
  p(K) &=  p_t(K-1)=
  \frac{\Gamfun{\alphaNB + K -1}\Betafun{\alphaNB+\pia, K -1 + \pib}}{ \Gamfun{\alphaNB} \Gamfun{K} \Betafun{\pia,\pib}}  .
\label{BNB}
\end{align}
Appendix~\ref{app:beta-negat-binom} provides the
hierarchical derivation of the prior \comment{and illustrates the
shapes for various parameter values.
For %a $\BNB{\alphaNB, \pia,\pib}$-prior with
$\pia>1$, the \commentSF{expectation $ \Ew{K}= 1+ \alphaNB\pib /(\pia-1)$ is finite}.
 Prior (\ref{BNB}) generalizes the % generalized Waring
prior derived by \cite{cer:new} for the Gnedin-Fisher model} and the prior derived
by \citet{gra-etal:los} from loss-based considerations which can be
regarded as a \commentF{$\BNB{1, \pib, \pia}$ prior.}
% BNB prior, $K - 1 \sim \BNB{1, \pib, \pia}$.
In their
applications, \citet{gra-etal:los} apply \commentF{the $\BNB{1,1,1}$ prior
% $ K-1 \sim \BNB{1,1,1}$
with % density $ p(K) = 1/(K(K+1))$ and
no finite moments.}

The three-parameters \commentF{$\alphaNB$, $\pia$ and
$\pib$ of the} $\BNB{\alphaNB, \pia,\pib}$ prior allow simultaneous control over the
\commentF{expectation and the}
tails of $p(K)$ and the implied prior $p(\Kn| N,\gammav)$ and its
expectation $\Ew{\Kn|N,\gammav}$. Priors $p(K)$ with finite expectation
imply \commentF{that % the prior expectation
$\Ew{\Kn|N,\gammav}$} is finite,
even for increasing $N$.
\commentF{In a clustering context, we propose to use the prior
$K-1 \sim \BNB{1, 4, 3}$ with $\Ew{K}=2$.}
The induced prior on $p(\Kn | N, \gammav)$
is investigated in more detail in
  Section~\ref{sectionprclust}  \commentF{and
differs considerably} from previous choices such as the geometric or the uniform
distribution. \commentF{The $\BNB{1, 4, 3}$ prior leads to a weakly
informative prior on $\Kn$ which is concentrated on moderate number
of clusters and exhibits fat tails to ensure that also a high number of
clusters may be estimated.}

% On the other hand, \citet{mil-har:mix} recommend that $p(K)$ should not have too heavy tails.
% In order to a priori encourage a sparse estimation of the number of clusters, we propose to use the prior $K-1 \sim \BNB{1, 4, 3}$ with $\Ew{K}=2$.

\subsection{The induced prior on the number of clusters $\Kn$}  \label{sectionprclust}

%So far, the explicit form of the prior on the number of clusters $p(\Kn| N ,\gammav)$ was unknown for MFMs, with a few exceptions.
\comment{In applied mixture analysis, we often aim at partitions of the data with a finite, but a priori random number of clusters $\Kn$. Since the number $K$ of components is random a priori for a MFM, this induces $\Kn$ to be random as well, but the induced prior $p(\Kn| N)$ on $\Kn$ does not necessarily coincide with the prior $p(K)$ for a finite number of observations $N$.}
 The  \commentSF{induced prior} $p(\Kn| N)$ %on the number of clusters
has been derived earlier for various mixture models.
 For DPMs,
\citet{ant:mix} provides the prior %distribution
of $\Kn$ as
$\fEwens (\Kn| N,\alpha) = \frac{\Gamfun{\alphaDP} }{ \Gamfun{N+
    \alphaDP}} s_{N,\Kn}$, where
$s_{N,\Kn} = \sum_{\parti} \prod_{j=1}^{\Kn} \Gamfun{N_j}$ is the Stirling
number of \commentBG{the} first kind.
\citet[Proposition~4.2]{nob:pos} gives the prior on $\Kn$ for a standard finite mixture,
while \citet{gne-pit:exc} derive $p(\Kn| N)$ for Gibbs-type priors.
 %
%\comment{For example, recently   report the prior on $K_+$ for static MFM,} while \citet{mil-har:mix}  derive a recursion under the Poisson prior $K-1 \sim \Poiss{\lambda}$ for a static MFM by building on \citet{gne-pit:exc} and \citet{gne:spe}.
%\citet{gne:spe} show that ADD simplifies considerably under the Poisson prior $K-1 \sim \Poiss{\lambda}$
% The same results is provided by \citet{arg-etal:isinf} for the special case of ADD?
%
%\comment{For example, recently   report the prior on $K_+$ for static MFM,} while \citet{mil-har:mix}  derive a recursion under the Poisson prior .

 \comment{Building on this literature,
we derive the prior $p(\Kn| N,\gammav)$ %of the number of clusters $\Kn$
for generalized MFMs
under arbitrary priors $p(K)$.
%\comment{Moreover, we indicate an efficient way to compute this prior.}
 One way to \commentF{obtain this prior is summing the EPPF
\eqref{mixparti}} over all partitions $\parti$:
\begin{align} \label{Knsterl}
 \Prob{\Kn=k| N, \alpha } &=  \sum_{K=k}^\infty  p(K )
  V_{N, k}^{K, \edK} (\edK )^k  S_{\commentBG{N},k} ^{-1, \edK} ,
\end{align}
where the  $ S_{\commentBG{N},k} ^{-1, x}= \sum_{\parti} \prod_{j=1}^k \Gamfun{N_j +  x}/\Gamfun{1+x}$ are the generalized Stirling numbers of the second kind. Alternatively,}
Theorem~\ref{Theorem4} derives $p(\Kn| N,\gammav)$ from the prior of the {\em labeled}
cluster sizes $p(N_1, \ldots, N_{\Kn} |N,\gammav)$ given in
(\ref{mixpK}).

%For a dynamic MFM with $\edK=\alpha/K$, this yields a generalization of the
%prior of \cite{ant:mix}:
%\begin{align}
% \Prob{\Kn=k| N, \alpha } &=    \frac{ \alpha  ^{k}   \Gamfun{\alpha} }{ \Gamfun{\alpha + N}} \sum_{K=k}^\infty  p(K )
% S_{n,k} ^{-1, \alpha/K} \prod_{j=1}^k \frac{ (K -j+1)  }  { K},  \nonumber
%\end{align}
%where the  $ S_{n,k} ^{-1, x}= \sum_{\parti} \prod_{j=1}^k \Gamfun{N_j +  x}/\Gamfun{1+x}$ are the generalized Stirling numbers of the second kind.

\begin{thm}\label{Theorem4}
  For a generalized MFM with priors $p(K)$ and
  $\boldeta_K|K, \gammav \sim \Dirinv{K}{\edK}$, the prior of the
  number of \cluster s $\Kn$ conditional on the sample size $N$ is
  given for $k=1,2,\ldots, N$ by:
  \begin{align} \label{Ppos}
    \Prob{\Kn = k |N, \gammav} &= \frac{N!}{ k!}
    \sum_{K=k}^\infty p(K) \frac{ {V}_{N, k}^{K, \edK}}{ \Gamfun{ \edK } ^{k}
    } C^{K, \edK}_{N,k},
\end{align}
where,  for each $K$, ${V}_{N, k} ^{K, \edK}$ has been defined in (\ref{mixpKV}) and
 $ C^{K, \edK}_{N,k} $  is given by summation over the labeled cluster sizes $(N_1,\ldots, N_k)$:
   \begin{align} \label{PposCk}
    C^{K, \edK}_{N,k} &=  \sum_{\substack{N_1,\ldots, N_k > 0\\N_1+\ldots+N_k=N}}  \prod_{j=1}^k      \frac{ \Gamfun{N_{j} +\edK }} {\Gamfun{N_j + 1} }.
   \end{align}
  \end{thm}

 \noindent \comment{By matching (\ref{Knsterl}) and  (\ref{Ppos}), we find that $C_{N,k} ^{K, \edK}$ is related to the generalized Stirling numbers $S_{\commentBG{N},k} ^{-1, \edK}$ through
  \begin{align}  \label{CSter}
 \frac{N!}{\Gamfun{1+ \edK}^k  k! } C_{N,k} ^{K, \edK} =   S_{\commentBG{N},k} ^{-1, \edK}.
\end{align}
We found it convenient to compute $C_{N,k} ^{K, \edK}$ recursively through Algorithm~\ref{KNMFM}.
The recursion is straightforward to implement and scales well for large $N$, see \citet{gre-etal:spy,gre:fip} and Appendix~\ref{proof} for mathematical derivations.}

\comment{\begin{algorithm}[b!]
  \caption{Computing $C_{N,k} ^{K, \edK}$   for a general\commentBG{ized} MFM.}  \label{KNMFM}
  \footnotesize
  \begin{enumerate}
  \item Define the vector $\cv_{K,1} \in \mathbb{R}^{N}$ and the
    $(N \times N)$ upper triangular Toeplitz matrix $ \Wm_1 $,
    where $ w_n = \frac{ \Gamma(n +\gamma_K )}{\Gamma(n +1)} $,
    $n=1, \ldots,N$,
\begin{align*}
\Wm_1  &= \left(
\begin{array}{cccc}
w_1 &    \ddots   &w_{N-1} & w_{N} \\
    &  w_1        &\ddots                 & w_{N - 1} \\
    &                 & \ddots                & \ddots  \\
    &              &                              &   w_1 \\
\end{array}
\right), & \cv_{K,1}    &=   \left(   \begin{array}{l}  w_N  \\ w_{N-1} \\
\vdots    \\    w_1 \\ \end{array}   \right) .
\end{align*}
\item For all $k \ge 2$,  define  the  vector    $\cv_{K,k} \in
\mathbb{R}^{N-k+1}$  as
\begin{align}  \label{recck}
\cv_{K,k} &= \left( \begin{array}{cc} \bfz_{N-k+1 } &  \Wm_k    \\
\end{array} \right) \cv_{K,k-1},
\end{align}
where $ \Wm_k $ is a $(N-k+1)\times (N-k+1)$ upper triangular
Toeplitz matrix obtained from $ \Wm_{k-1} $ by deleting the first row
and the first column.
\item Then, for all $k \ge 1$,
$C_{N,k} ^{K, \gamma_K} $ is equal to the first element of the vector $\cv
_{K,k}$.
  \end{enumerate}
\end{algorithm}}

For a dynamic MFM with $\edK=\alpha/K$, $C_{N,k}^{K, \edK}$ can be
written as $C^{K, \alpha}_{N,k} $ depending on $K$ and
$\alpha$:
\begin{align} \label{PposD}
\Prob{\Kn = k |N, \alpha } & =
 \frac{N!}{ k!}  \frac{ \alpha  ^{k}   \Gamfun{\alpha} }{ \Gamfun{\alpha + N}}   \sum_{K=k}^\infty   p(K)
C^{K, \alpha }_{N,k}  \prod_{j=1}^k \frac{ (K -j+1)  }  {K \Gamma(1+\frac{\alpha}{K})}.
\end{align}
Putting all prior mass on $K=+\infty$, \commentBG{the} following way to compute
$\fEwens (\Kn| N, \alpha)$ for a DPM emerges from \eqref{PposD},
\begin{align}
  \displaystyle
  \fEwens (\Kn| N, \alpha) &=\frac{N!}{ \Kn!}  \frac{ \alpha  ^{\Kn}   \Gamfun{\alpha} }{ \Gamfun{\alpha + N}}  C^{\infty}_{N,\Kn},
  \label{priorKDPal}
\end{align}
where $ C^\infty_{N,\Kn}$ is independent of $\alpha$ and obtained
through recursion \eqref{recck} with $w_n=1/n$.
 \comment{For a static MFM, Theorem~\ref{Theorem4} simplifies to} the following expression:% for $\Prob{\Kn = k |N, \eFM }$:
\begin{align} \label{PposStaSTa}
\Prob{\Kn = k |N,  \eFM } &=   \frac{N!}{ k!}  \frac{ {V}^\eFM_{N, k}}{ \Gamfun{\eFM } ^{k} }  C ^\eFM  _{N,k},
\end{align}
\comment{where ${V}^\eFM_{N,k}$ is determined recursively from (\ref{recMFM}) and
$C_{N,k}^{K, \edK}$ is written as $C ^\eFM _{N,k}$
independent of $K$ and can be obtained in a single recursion from
(\ref{recck}). Using, again, the normalization
$ \tilde{V} ^{\commentBG{\eFM}} _{N,k} = \eFM ^k {V} ^\eFM _{N,k} $, prior (\ref{PposStaSTa})
is a special case of the prior given in \citet{gne-pit:exc} for Gibbs-type priors:
\begin{align} \label{PposPitGne}
\Prob{\Kn = k |N,  \eFM } &=  \tilde{V} ^{\commentBG{\eFM}} _{N,k}  B_{\commentBG{N},k} (W_{\bullet}),
\end{align}
where $ B_{\commentBG{N},k} (W_{\bullet})$  is the Bell polynomial in the W-structure $W_{\bullet}=\{W_\ell\}$ defined in Footnote~2.\footnote{\comment{This follows from   (\ref{CSter}) and $ S_{\commentBG{N},k} ^{-1, \eFM}= B_{\commentBG{N},k} (W_{\bullet}) $ \citep[Eq.~(1.20)]{pit:com}.}}}
 Finally, putting all prior mass on  $K=\Kfix$,
(\ref{PposStaSTa}) gives the result of
\citet[Proposition~4.2]{nob:pos}  for a standard finite mixture:
\begin{align} \label{PSFM}
\Prob{\Kn = k |N,  K=\Kfix, \eFM } &=   \frac{N!}{ k!}   \frac{{\Kfix} !}{({\Kfix} - k )!} \frac{\Gamfun{\eFM {\Kfix}} }
{  \Gamfun{ \eFM {\Kfix} + N} \Gamfun{\eFM}^k} C^{\eFM }_{N,k}.
\end{align}

\begin{figure}[t!]
\includegraphics[width=0.85\textwidth, trim = 0 5 0 5, clip]{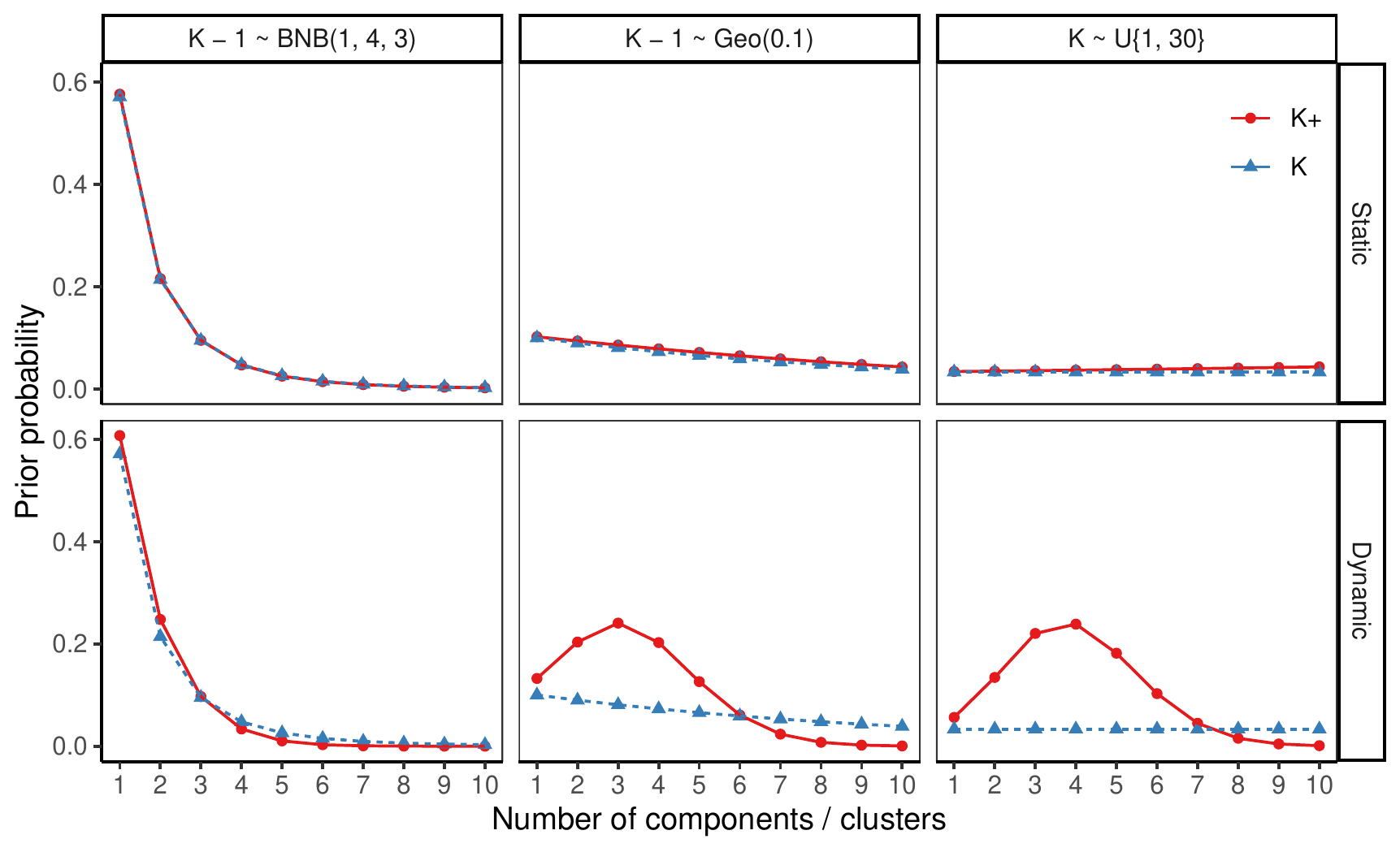}
\caption{Priors of $K$ (dashed blue lines, triangles) and $\Kn$ (solid
  red lines, circles) under the priors $K-1 \sim \BNB{1, 4, 3}$,
  $K-1\sim \Geo{0.1}$ and $K \sim \cU\{1,30\}$ for a static MFM with
  $\gamma = 1$ (top) and dynamic MFM with $\alpha = 1$ (bottom), with
  $N=82$. \label{plot:staticMFM}}
\end{figure}

%
% The induced prior $p(K_+)$ is plotted for various specifications of
% $p(K)$ and $\gammav$ in Appendix~\ref{app:induced-prior-number}. In
% Figure~\ref{plot:staticMFM} $p(K_+)$ is plotted for the negative
% binomial distribution presented in the next section,
% $K -1 \sim \BNB{\mu, 4, 3}$ with $\mu=1,9$, under a static and
% dynamic MFM. Under the static MFM, $p(K_+)$ basically coincides with
% $p(K)$ for both priors $p(K)$, whereas under the dynamic MFM
% $p(K_+)$ considerably differs from $p(K)$ by concentrating the mass
% on smaller values.

%For a MFM the induced prior on the number of clusters $\Kn$ depends on
%the explicit choices for the prior on $K$ and the prior on the
%hyperparameter $\gammav$.

For illustration, Figure~\ref{plot:staticMFM} shows the impact of
 \commentF{various priors $p(K)$ %on the number of components
on the
induced prior $p(\Kn|N,\gammav)$ % on the number of clusters
for static MFMs with $\gamma = 1$ (top row)
and dynamic MFMs with $\alpha = 1$ (bottom row).  The priors $p(K$) in the three columns are
the translated beta-negative-binomial prior $K-1 \sim \text{BNB}(1, 4, 3)$ with $\Ew{K} = 2$ suggested in Section~\ref{sectionpK}, the \commentSF{prior  $K-1 \sim \text{Geo}(0.1)$}
with $\Ew{K} = 10$ suggested by \citet{mil-har:mix} and the uniform \commentSF{prior $K \sim \mathcal{U}\{1, 30\}$} with $\Ew{K} = 15.5$
used by \citet{ric-gre:bay}.}

% (1)
%$K-1 \sim \text{BNB}(1, 4, 3)$, (2) $K-1 \sim \text{Geo}(0.1)$ and (3)
%$K \sim \mathcal{U}\{1, 30\}$.
%
% The first prior $p(K)$ follows a translated beta-negative-binomial
% distribution with mean $\Ew{K} = 2$, i.e., has a small mean
%  and rather fat tails. % We will suggest this prior as a
% generalization of priors previously considered for $K$ and explain the
% parameter choice, inducing a small mean and rather fat tails, in
% Section~\ref{sectionpK}.
% The second prior is the $\text{Geo}(0.1)$ prior with $\Ew{K} = 10$ suggested by \citet{mil-har:mix} and the third \comment{one is} the uniform $\mathcal{U}\{1, 30\}$ prior with $\Ew{K} = 15.5$ used by \citet{ric-gre:bay}.
%
 \commentF{For static MFMs, %(top row)
% only a small gap between $p(\Kn)$ and $p(K)$
% is present for $\Kn$ and $K$ between one and ten and
$p(\Kn) \approx p(K)$ for all three priors for  values
for $\Kn$ and $K$ between one and ten}.  In contrast, for a dynamic MFM % (bottom row),
$p(\Kn)$ and $p(K)$ are only close for the BNB prior which has a small
mean value. For the priors $p(K)$ with larger mean values, $p(\Kn)$
considerably differs from $p(K)$ with mass being pulled towards
smaller values of $\Kn$.  The \comment{corresponding} posteriors of $K$ and $\Kn$
obtained \comment{under these priors for the famous Galaxy data} %an empirical data set
 are shown in
Figure~\ref{plot_galaxy} in Section~\ref{revGalax}.

\section{\comment{Bridging  finite mixtures analysis and BNP mixtures}}
\label{sectionbridge}

\subsection{Connecting SFMs, MFMs and BNP
  mixtures} \label{sectsubbridge}

\commentF{Generalized MFMs} % introduced in Section~\ref{sectionmoddef}
extend both
Dirichlet process mixtures (DPMs) and sparse finite mixtures
(SFMs). By allowing the number of components $K$ to be finite and
random, MFMs provide notably more flexibility in the prior
distribution on the partition space than DPMs and SFMs, similar to popular BNP mixtures
\citep{deb-etal:are}.

SFMs result as that special case of MFMs, where $p(K)=\indic{K=\Kfix}$
puts all prior mass on \commentBG{a fixed number of components} $\Kfix$. \comment{It follows from  Theorem~\ref{theorem1}
and earlier work by \citet{ish-zar:mar}}
that the prior distribution imposed on the partition space by a SFM
lacks flexibility with increasing $\Kfix $ and approaches the \commentSF{Ewens
distribution (\ref{Ewens})} as $\eFM_{\Kfix}=\alpha/\Kfix$ approaches 0:
\begin{align}  \nonumber % \label{SFMpNKK}
 \lim _{\Kfix \rightarrow \infty} \frac{p(\parti |  \eFM_{\Kfix}=\alpha/\Kfix, \Kfix )}{ \fEwens (\parti |  \alpha=\eFM_{\Kfix} \Kfix ) } &=
 \commentSF{ \lim _{\Kfix \rightarrow \infty}    R_{\bN,\Kn}^{\Kfix,\alpha}}
  %\prod_{j=1}^{\Kn}   \frac{\Gamfun{N_j +  \eFM_{\Kfix} } (\Kfix -j+1)  }   { \Gamfun{N_j}\Gamma(1+\eFM_{\Kfix}) \Kfix }
  = 1.
\end{align}
This implies that SFMs do not easily deal with situations with many,
well-balanced clusters, a behavior that is also observed for DPMs.
By considering $K$ as an additional second parameter following a prior
$p(K)$, \comment{the dynamic MFM emerges \commentF{as a more flexible family than a SFM with $K=\Kfix$ fixed. Dynamic} MFMs  can also be regarded as a more flexible  extension %converges to
of a DPM.  Since $R_{\bN,\Kn}^{K, \alpha}$ in Theorem~\ref{theorem1} converges to 1 as $K$ increases, putting all prior mass on $K=+\infty$ yields the Ewens distribution as limiting case. Thus, DPMs
\comment{result} %may be regarded
as the limiting case of a dynamic MFM where the prior $p(K)$
increasingly concentrates all prior mass at $K=+\infty$. }

\comment{Several close connections between MFMs and Pitman-Yor
process mixtures (PYM) deserve to be mentioned.}  In Bayesian non-parametrics, mixtures based
on the Pitman-Yor prior $\PY{\betaPY,\alphaPY}$ with
$\betaPY \in [0,1), \alphaPY > -\betaPY $ \citep{pit-yor:two} are a
commonly used two-parameter alternative to DPMs which are the special
case where $\betaPY=0$ and $\alphaPY=\alphaDP$.  There exists a second family of PYMs, where
$\betaPY < 0$ and $\alphaPY= K |\betaPY|$ with $K \in \NNN $ being a
natural number, see \citet{gne:spe} and \citet{deb-etal:are}.  In the corresponding
stick-breaking representation, stick $\stick_{K}=1$ a.s. Hence, this
prior yields a mixture with infinitely many components, of which only
$K $ have non-zero weights, with the symmetric Dirichlet distribution
$\Dirinv{K}{ | \betaPY | }$ acting as prior. Furthermore, \commentBG{at most}
$K$ components can be populated.  The EPPF of a PYM (with $K$ known) reads:
\begin{align}  \nonumber %\label{pripiPY}
  p(\parti|N, \betaPY,\alphaPY) &=
  \frac{\Gamfun{\alphaPY} }{ \Gamfun{N+ \alphaPY}}  \prod_{j=1} ^{\Kn} \left(  \alphaPY +  \betaPY(j-1) \right) \frac{\Gamfun{N_j-\betaPY} }{ \Gamfun{1- \betaPY}}.
\end{align}
\comment{By matching EPPFs (and using \commentF{$\Gamfun{1-\betaPY} % = -\betaPY \Gamfun{-\betaPY}
= |\betaPY| \Gamfun{|\betaPY|}$}), it is evident that a finite mixture with
\commentF{$K$ known  and  % hyperparameter
    $\edK>0$} is equivalent to a mixture with a $\PY{-\edK, K \edK}$ prior,
    as proven in \citet{gne-pit:exc}. This equivalence of SFMs and
PYMs} provides a theoretical explanation of the
empirical finding that SFMs can lead to more sensible cluster
solutions than DPMs, see, e.g., \citet{fru-mal:fro}.

\comment{Even more interesting connections to BNP mixtures arise for MFMs, where $K$ is random.
As pointed out by \citet{mil-har:inc} and proven much earlier by \citet{gne-pit:exc}, for a static MFM, the dual BNP mixture is a Gibbs-type prior which arises from mixing a
 $\PY{-\eFM, K \eFM}$ prior over the concentration parameter
$ \alphaPY_K=K \eFM$, while the reinforcement parameter $\betaPY = -\eFM$ is fixed. The Fisher-Gnedin model studied in \cite{gne:spe} is equivalent
to a static MFM with $\eFM=1$ and
$K-1 \sim \Poiss{\lambda}$. The static MFM is also a special case of the class of mixtures based on normalized independent finite point processes recently introduced by \citet{arg-etal:isinf}.}

On the other hand,
for a dynamic MFM, the prior partition distribution of the dual BNP mixture lies outside of the family of Gibbs-type priors, as it arises from mixing
a $\PY{-\alpha/K, \alpha}$-prior over the reinforcement
\commentSF{parameter $\betaPY_K= - \alpha/K$}, while the concentration parameter
$ \alphaPY = \alpha$ is fixed,  see also the discussion in \citet{deb-etal:are}.
 As shown in \citet{pit:som},
  a system of predictive distributions  emerges from the EPPF, quantifying the probability that a new observation $\ym_{N+1}$ belongs to any of the $\Kn=k$ existing clusters in $\parti = \{ \parti_1, \ldots, \parti_{k}\}$ or creates a
  partitions $\parti \new = \{ \parti_1, \ldots, \parti_{k}, \parti_{k+1} \}$
 with a  new cluster $\parti_{k+1}$ of size $N_{k+1}=1$.
 % Based on Theorem~\ref{theorem1},
 \commentSF{For a dynamic MFM the prior probability to introduce a new cluster for $\ym_{N+1}$ is given by (see Appendix~\ref{proof} for a proof):}
  \begin{align}  \label{dMFMpre}
  \Prob{ \ym_{N+1} \in \parti_{k+1}|\bN, \Kn=k, \alpha  } &=&
   \frac{\alpha }{\alpha + N }
  \commentSF{  \left( 1 - k \cdot \frac{ \sum_{K= k }^ {\infty}  p(K )/K  R_{\bN ,k}^{K, \alpha}}
  { \sum_{K= k }^ {\infty}  p(K )  R_{\bN ,k}^{K, \alpha}} \right). }
% \times  \frac{ \sum_{\Kna= 1 }^ {\infty}  p(k + \Kna ) \Kna  R_{\bN ,k}^{K+\Kna, \alpha}} { \sum_{K= k }^ {\infty}  p(K )  R_{\bN ,k}^{K, \alpha}},
\end{align}
This probability \commentSF{(bounded by the predictive probability $\alpha/(\alpha + N )$ of a DPM)} not only depends on $N$ and \commentSF{the current number of clusters $\Kn$}, which characterizes Gibbs-type priors \citep{deb-etal:asy}, but also on the occupation numbers $N_1, \ldots, N_{\Kn}$.
This
confirms  once more that dynamic MFMs, while staying within the finite mixture framework,
are an example of a general random partition prior \citep{deb-etal:are}.

%while staying within the finite mixture framework, generalized  MFMs are a flexible class with close connections to BNP mixture models.

\subsection{Comparing static and  dynamic MFMs and DPMs} \label{sectionpa}

% Static and dynamic MFMs and DPMs have in common that they strongly
% favor homogeneity for very small values of, respectively, $\gamma$ and
% $\alpha$. As proven in Appendix~\ref{priorhomo}, the prior probability
% $\Prob{\Kn = 1 |N,\gammav }$ converges to 1 as $\alpha \rightarrow 0$
% or $\eFM \rightarrow 0$.  The more $\gamma$ and $\alpha$ increase, the
% smaller is $\Prob{\Kn = 1 |N,\gammav}$ for a fixed $N$ and
% heterogeneity with more than one cluster is allowed a priori, see
% Figure~\ref{fig:Knfunhomo} in Appendix~\ref{priorhomo}.

In the following we compare the induced priors on the number of
clusters  and the partitions for static and dynamic MFMs and
DPMs in more detail \commentF{and  investigate the influence of the prior on $K$ and,
respectively, the
hyperparameters $\gamma$ % for static
and $\alpha$.}
%\commentS{Results indicate the additional flexibility of dynamic MFMs  compared to static MFMs and DPMs.}

\begin{figure}[b!]
  \centering
  \includegraphics[width=0.85\textwidth, trim = 0 5 0 5, clip]{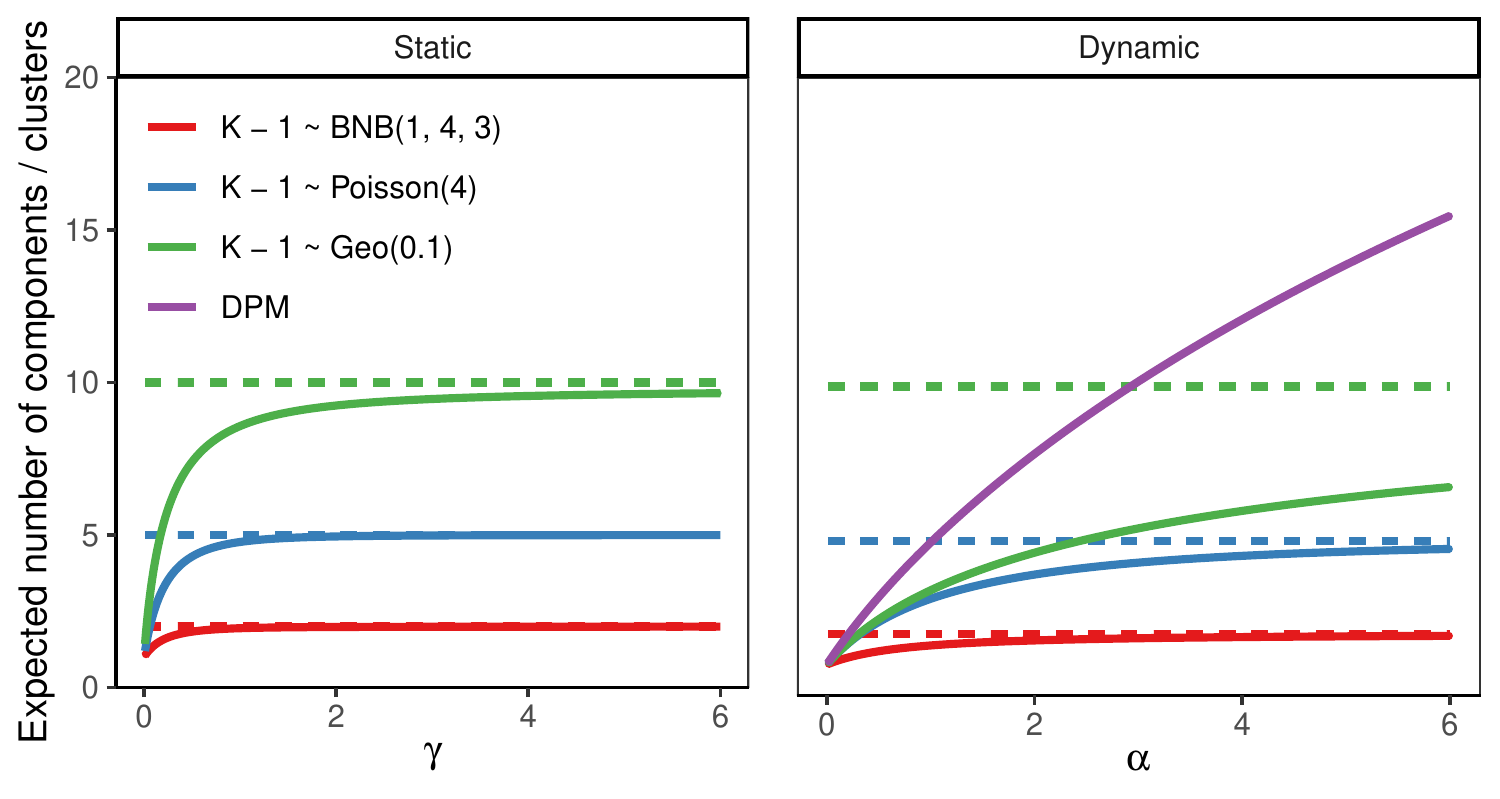}
  \caption{Prior expectations $\Ew{\Kn|\gamma , N}$ for static MFMs
    (left) and $\Ew{\Kn| \alpha, N}$ for dynamic MFMs (right) as
    functions of $\gamma$ and $\alpha$ for $N=100$ under the priors
    $K -1 \sim \BNB{1, 4, 3}$, $K-1 \sim \Poi{4}$, and
    $K-1 \sim \Geo{0.1}$ in comparison to a DPM. For each prior
    $p(K)$, the prior expectation $\Ew{K}$ is plotted as a horizontal
    dashed line.}\label{fig:KnfunhypDyn}
\end{figure}

Regarding the prior on the number of clusters $\Kn$, a fundamental
question is whether a MFM allows $\Kn$ to be different from $K$
a priori, as for DPMs (where $K=\infty$).  To gain further
\commentF{understanding, % in this regard,
we} plot in Figure~\ref{fig:KnfunhypDyn}
the expectation of the induced prior \commentF{$p(\Kn|N,\gammav)$ % of the number of clusters
as a function of % the hyperparameters
$\gamma$} (for static
MFMs) and $\alpha$ (for DPMs and dynamic MFMs) for $N=100$ under
various priors $p(K)$.
For both classes of MFMs, the gap between the expected number of
clusters, $\Ew{\Kn|N,\gammav},$ and the expected number of components,
$\Ew{K}$, decreases for increasing $\gamma$ or $\alpha$.
However, for dynamic MFMs the decrease is much slower and, even as
$\alpha$ increases, a considerable gap remains between
$\Ew{\Kn|N,\gammav}$ and $\Ew{K}$.  This is the effect of linking
$\gammav$ to $K$ through $\gamma_K=\alpha/K$, thus avoiding that $K_+$
increases too quickly as $K$ increases. This implies that the
influence of the prior on $K$ on the induced prior on $\Kn$ is
attenuated for an extended range of $\alpha$ values.

As emphasized by \citet{gre-ric:mod}, beyond the induced prior on
$\Kn$, the {\em conditional} EPPF, induced for a given number of
clusters $\Kn=k$,
\begin{align} \label{PNjonit}
p(N_1, \ldots, N_k |N, \Kn = k,\gammav )&= \frac{\Prob{N_1, \ldots, N_k|N,\gammav} }{ \Prob{\Kn = k |N,\gammav}},
\end{align}
is important for \commentF{comparing mixture models}.  This
prior allows a deeper understanding of the impact of choosing
$\gammav$ for MFMs on the partition distribution.

For a DPM, the conditional EPPF can be expressed using
Theorem~\ref{Theorem4} as:
\begin{align}\label{partmainDPM}
 p_{\DP} (N_1, \ldots, N_k |N, \Kn = k) & = \frac{1}{ C ^\infty _{N,k}}  \prod_{j=1}^{k}   \frac{1}{ N_j} ,
\end{align}
and is known to be highly unbalanced \citep{ant:mix}, \commentF{favoring
 partitions with some small values $N_j$}  due to the
factors $1 / N_j$, $j=1,\ldots,k$ \citep{mil-har:mix}.  However, being
independent of $\alpha$, the conditional EPPF cannot be made more
flexible for a DPM.
 \commentF{In contrast,} for a static MFM, the conditional EPPF depends on $\eFM$,\footnote{\commentF{Note that
\cite{mil-har:mix} report an approximate formula for the conditional
EPPF of a static MFM, while our result is exact.}}
\begin{align} \label{partmainSMFM}
  p(N_1, \ldots, N_k |N, \Kn = k, \gamma) &= \frac{1}{ C ^\eFM _{N,k}}  \prod_{j=1}^{k}   \frac{ \Gamfun{N_j +\gamma}}{ \Gamfun{N_j + 1}}.
\end{align}
For $\gamma=1$, the uniform distribution over all partitions of $N$
data points into $\Kn = k$ clusters results.    Varying the
hyperparameter $\gamma$ introduces flexibility in the conditional EPPF
for a static MFM: decreasing $\gamma$ favors more unequal allocations,
increasing $\gamma$ favors partitions with more equal allocations.
 \commentF{The conditional EPPF of a dynamic MFM} is obtained by dividing
\eqref{mixpK} by \eqref{Ppos}:
%, with barely any simplifications being possible:\comment{
%
\begin{align} \label{eq:cEPPFgen}
p(N_1, \ldots, N_k |N, \Kn = k,\alpha)&=  \frac{\displaystyle
\sum_{K=k}^\infty    p(K) \frac{ V_{N, k}^{K, \alpha}}{ \Gamfun{ \frac{\alpha}{K} } ^{k} }
\prod_{j=1}^{k}   \frac{ \Gamfun{N_j + \frac{\alpha}{K} }}
{\Gamfun{N_j + 1}}}{ \displaystyle \sum_{K=k}^\infty p(K) \frac{ {V}_{N, k}^{K, \alpha}}{ \Gamfun{\frac{\alpha}{K}} ^{k}} C^{K, \alpha}_{N,k}}.
\end{align}
\commentF{This conditional EPPF % of a dynamic MFM
depends} both on $\alpha$ and
$p(K)$, whereas the conditional EPPF of a static MFM is independent of
$p(K)$. Thus, having a second parameter $K$, dynamic MFMs \commentBG{are} more
flexible than static MFMs regarding the conditional EPPF.
%Also, more evenly distributed clusters are encouraged a priori by increasing $\alpha$, e.g., compare the conditional EPPF for a DPM which corresponds to the limiting case $\alpha = 0$ and the conditional EPPF for $\alpha = 1$ where the uniform distribution emerges.
\commentF{Overall,} in comparison to DPMs, static and dynamic MFMs induce more
flexible prior structures both on the prior of the number of clusters
and on the partition distribution, \commentF{see \citet{gre-etal:spy} for a detailed further investigation}.

 \comment{Additional flexibility} is achieved by adjusting the
hyperparameters $\gamma$ and $\alpha$ to suit the data.  In
Section~\ref{sectionpra}, a hyperprior on $\alpha$ is suggested, to
achieve adaptivity of the induced prior on the partition to the data
at hand.  Also a static MFM can be combined with a prior on $\eFM$,
rather than choosing a fixed value such as $\eFM=1$.

\subsection{Choosing the prior on $\alpha$ for dynamic MFMs} \label{sectionpra}

For a dynamic MFM the parameter $\alpha$ plays a crucial role for the
prior distribution induced on the number of clusters and the
partitions. %A prior on $\alpha$ also helps preserve flexibility in the induced prior on the partitions.
 \comment{On the one hand, the}
 prior should have positive mass
close to zero to allow a priori for a single cluster solution \comment{which corresponds
to homogeneity.} \comment{At the same time,
fat tails should allow a priori larger values of $\Kn$ and partitions
with balanced cluster sizes.}

The DPM literature would suggest a Gamma
distribution $\alpha \sim \cG(a,b)$ \citep[e.g.,][]{esc-wes:bay,
  jar-etal:dir}.  If $a=b\ll 1$, the expectation of $\alpha$ is \commentF{one, while the} variance is large, leading to a vague prior on $\alpha$.
\commentF{For} DPMs this induces a very informative prior on the number
of clusters % $p(K_+|N,a,b)$
which is concentrated on $1$ and $+\infty$
\citep[see][]{dor:sel,mur-swe:sel}.  For dynamic MFMs, such a prior
would -- given its mode at zero -- strongly favor homogeneity, and
fail for data with balanced cluster sizes.
\comment{Instead, we} propose to use the $F$-distribution $\alpha \sim
\cF(\nu_l,\nu_r)$. The two parameters allow to \comment{control %model
 the behavior of the prior}
close to zero and in the tail independently.  Choosing $\nu_r$ small
gives fat tails.  For a finite mean value, given by $\nu_r/(\nu_r-2)$,
but no higher moments, we specify \commentSF{$2<\nu_r\leq 3$}. Choosing a % different,
small value for $\nu_l$ allows independent control over the prior
probability of homogeneity. Since the mode is given by
\comment{$(\nu_l-2)\nu_r/(\nu_l (\nu_r+2))$,} choosing $\nu_l>2$ avoids a
spike at 0.
% Choosing a mode close to one induces maximal average standard deviation of the relative entropy for a wide range of priors $p(K)$ (see Figure~\ref{fig:compmarg}).
In our empirical analysis, we
use $\alpha \sim \cF(6,3)$.

%\newpage

\section{Inference algorithm: Telescoping sampling} \label{sectiontele}

A novel sampling method called \emph{telescoping sampling} is
introduced for a Bayesian analysis of finite mixtures with an
unknown number of components which is related to,
% and, at the same time,
\commentF{but also} fundamentally different from RJMCMC \citep{ric-gre:bay} and the
CRP sampler \citep{jai-nea:spl_2004,jai-nea:spl_2007} applied in \citet{mil-har:mix}.

Similar to \citet{jai-nea:spl_2004,jai-nea:spl_2007}, the telescoping sampler is a
trans-dimensional Gibbs sampler which exploits the EPPF of a MFM given
in (\ref{mixparti}).  However, we do not work with the marginal EPPF
$p(\parti |N, \gammav)$, as \citet{mil-har:mix} do, but use a second
level of data augmentation where we introduce the unknown number of
components $K$, in addition to the partition $\parti$, as a latent
variable. \commentSF{This allows to apply the telescoping sampler outside
the framework of Gibbs-type priors. We
explicitly include} $K$ in the sampling scheme as in
\citet{ric-gre:bay}. However, rather than using RJMCMC, $K$ is sampled
conditional on $\parti$ from the conditional posterior
$p(K| \parti, \edK) \propto p(\parti | N, K, \edK ) p(K) $ which is
obtained by combining the {\it conditional} EPPF
$p(\parti | N, K, \edK )$ provided in (\ref{mixpartiK}) with the prior
$p(K)$:
\begin{eqnarray} \label{postKK}
 p(K| \parti,  \edK) \propto  p(K) \frac{K !}{(K- K_+ )!} \frac{\Gamfun{\edK K} }{ \Gamfun{N+\edK K}  \Gamfun{\edK } ^{K_+}}
\prod_{j: N_j>0}    \Gamfun{N_j + \edK},
\end{eqnarray}
for $K=K_+, K_+ +1,\ldots$, where $\Kn$ is the number of clusters in $\parti$.

While \citet{mil-har:mix} use \eqref{postKK} for static MFMs
\commentSF{to infer $K$ %the number of components
in a post-processing step, the
telescoping (TS) sampler %developed in this paper
integrates} \eqref{postKK} into a trans-dimensional Gibbs sampler \commentSF{for
generalized MFMs and samples} $K$ and the partitions $\parti$
(including $\Kn$) in different blocks.
 \commentSF{Since $K \ge \Kn$ by definition,  the number of
empty components $K-\Kn$
varies over the iterations of the sampler, taking zero or a larger
value. The} difference between $K$ and $\Kn$ behaves similar to a
telescope which can also be stretched or pulled together; hence the
name of the sampler.
 Full details of the TS sampler are provided for dynamic MFMs
 %with $\gamma_K = \alpha/K $
 in Algorithm~\ref{TELE}. The TS sampler can be
applied with minor modifications to \commentF{static MFMs}
%with $\gamma_K \equiv \eFM $
(see Algorithm~\ref{TELEstat} in
Appendix~\ref{sec:telescoping-sampler}).  In both cases, the
hyperparameter $ \wehyp=\alpha$ or, respectively, $\wehyp = \eFM$ is
assumed to be unknown.

\begin{algorithm}[t!]
  \caption{Telescoping sampling for a dynamic MFM.}  \label{TELE}
  \footnotesize
  \begin{enumerate} \itemsep -1mm
  \item Update the partition $\cP$ by   sampling
    from $p(\Siv|\etav_K,\thetav_1,\ldots, \thetav_K,\ym)$:	\\[-8mm]
    \begin{enumerate} \itemsep -1mm
    \item Sample $S_i$, for $i=1,\ldots,N$,  from
      $	\Prob{S_i=k|\etav_K,\thetav_1,\ldots, \thetav_K, \ym_i, K}  \propto \eta_k f(\ym_i|\thetav_k)$,  $k=1, \ldots,K$.
    \item Determine $N_k=\#\{i|S_i=k\}$ for $k=1, \ldots,K$, the
      number $\Kn = \sum_{k=1}^K I\{N_k>0\}$ of non-empty components and
      \commentF{relabel %the components
      such} that the first $K_+$ components are
      non-empty.
    \end{enumerate}
  \item Conditional on $\cP$,
    update  the parameters  of the  (non-empty) components:	\\[-8mm]
    \begin{enumerate} \itemsep -1mm
    \item For the (filled) components $k=1,\ldots,K_+$, sample
      $\btheta_{k}|\Siv,\ym,\phi$ from
      \begin{align*}
        p(\thetav_k|\Siv,\ym,\phi) &\propto   p(\thetav_k|\phi)\prod_{i:S_i=k} f(\ym_i|\thetav_k).
      \end{align*}
    \item Sample the hyperparameter $\phi$ (if any) conditional on
      $K_+$ and $\thetav_1,\ldots,\thetav_{K_+}$ from
      \begin{align} \label{sampphi}
        p(\phi|\thetav_1,\ldots,\thetav_{K_+},K_+) &\propto  p(\phi) \prod_{k=1} ^{K_+}  p(\thetav_k|\phi).
      \end{align}
    \end{enumerate}
  \item Conditional on $\cP$,  draw new values  of $K$ and $\alpha$:	\\[-8mm]
    \begin{enumerate} \itemsep -1mm
    \item Sample $K$ from
      \begin{align}  \label{postKK2}
        p(K|\cP,\alpha)  &\propto  p(K)
	\frac{ \alpha ^{\Kn}   K!}{K ^{\Kn} (K-K_+)!} \prod_{k=1}^{K_+} \frac{\Gamma(N_k+\frac{\alpha}{K})}{\Gamma(1+ \frac{\alpha}{K})}	,
        \quad K=\Kn, \Kn +1,\ldots.
      \end{align}
    \item Use a random walk Metropolis-Hastings step with proposal
      $\log(\alpha^{\new}) \sim \Normal{\log(\alpha),s_\alpha^2}$ to
      sample $\alpha|\cP,K$
      from
      \begin{align*}
        p(\alpha|\cP,K) &\propto p(\alpha)\frac{ \alpha ^{\Kn} \Gamma(\alpha)}{\Gamma(N+\alpha)}
        \prod_{k=1}^{K_+} \frac{\Gamma(N_k+\frac{\alpha}{K})}{\Gamma(1+ \frac{\alpha}{K})}.
      \end{align*}
    \end{enumerate}
  \item \commentF{Conditional on $K,  \Siv, \alpha$ and $\phi$, add $K-K_+$ empty components and update
  $\etav_K$:}  \\[-8mm] %the weight distribution  :	
    \begin{enumerate}	\itemsep -1mm	
    \item If $K> K_+$, then add $K-K_+$ empty components (i.e., $N_k=0$
      for $k=K_+ +1,\ldots,K$) and sample $\thetav_k|\phi$ from the
      prior $p(\thetav_k|\phi)$ for $k=K_+ +1,\ldots,K$.
    \item Sample $\etav_K|K,\alpha, \Siv \sim \Dir{e_1,\ldots,e_K}$,
      where $e_k=\alpha/K + N_k $.
    \end{enumerate}
  \end{enumerate}
\end{algorithm}

Very conveniently, due to the conditional independence between the
parameters $\thetav_k$  in the (non-empty) clusters and the number of
components $K$, given the partition $\cP$, %the number of components
$K$ is sampled from the conditional posterior $p(K|\cP, \comment{\edK})$ %\wehyp)$
\comment{given in (\ref{postKK})} without any
reference to the specific component distribution. Hence, the TS
sampler is straightforward to implement and very generic, since the
conditional posterior $p(K|\cP, \comment{\edK})$ % $p(K|\cP, \wehyp)$
% of $K$
does not depend on the component parameters.  This makes our sampler a most
generic, easily implemented algorithm for finite mixture models with
simultaneous inference on the unknown number of components and the
unknown number of \cluster s for a wide range of component
models. This greatly simplifies the application of MFMs in new
application contexts allowing for arbitrary component distributions
and extensions with hierarchical priors.  In contrast, the
challenge to design good moves for \commentF{RJMCMC is legendary}.
But also for CRP samplers \commentSF{(which are confined to static MFMs)}, the creation of new clusters requires
knowledge of the marginal likelihood which depends on the chosen
mixture family and might be difficult to work out for more complex
mixtures.

More specifically, the TS sampler is a partially marginalized sampler,
moving back and forth between sampling from the mixture posterior
distribution
$p(K,\Siv, \etav_K, \thetav_1, \ldots,$ $\thetav_K, \phi, \wehyp |\ym)$,
which lives in the augmented parameter space of the mixture
distribution, and sampling from the collapsed posterior
$p(K,\cP, \thetav_1, \ldots, \thetav_{\Kn},\phi, \wehyp|\ym)$, which
lives in the set partition space and is marginalized with respect to
the parameters of the empty components, the weight distribution
$\etav_K$ and all allocations $\Siv$ that induce the same set
partition $\cP$.
 The full mixture posterior
$p(K,\Siv, \etav_K, \thetav_1, \ldots, \thetav_K, \phi, \wehyp |\ym)$
is proportional to
\begin{align}  \label{postall}
\prod_{k:N_k>0} p(\ym^{[k]}| \thetav_k)   p(\thetav_k| \phi)  \prod_{k:N_k=0}  p(\thetav_k| \phi)
\prod_{k=1}^K  \eta_k ^{N_k+\eFM_K-1} \frac{\Gamfun{K \eFM_K}}{ \Gamfun{ \eFM_K}^K }  p(\phi) p(K) p(\wehyp),
\end{align}
where $\ym^{[k]}$ are the $N_k>0$ observations in cluster $\cC_k$ of
the partition $\cP=\{\cC_1,\ldots,\cC_{K_+}\}$ implied by $\Siv$
\comment{(after reordering such that the $\Kn$ non-empty clusters appear first)}.
% and $\eFM_K=\alpha/K$ for dynamic and $\eFM_K\equiv \eFM$ for static MFMs.
The posterior (\ref{postall}) lends itself to the conditional sampling
Step~1 of the TS sampler which is a standard step for finite mixtures
with $K$ known.  The TS sampler is related to conditional samplers for
infinite mixtures insofar, as all indicators $\Siv$ are sampled
jointly due to the conditional independence of $\Siv$ given
$\etav_K,\thetav_1,\ldots, \thetav_K,\ym$. As opposed to this, the CRP
sampler applied in \citet{mil-har:mix} is a single-move sampler
updating the allocation of each observation one-at-a-time.

\commentF{Integrating (\ref{postall})} with respect to the weight
distribution $\etav_K$, the parameters $\thetav_k$ of \commentBG{the}
\commentF{empty components} % \commentBG{$k=\Kn+1,\ldots,K$} %($N_k=0$)
and all allocations $\Siv$ that induce the \commentF{same partition}
$\cP$ yields (after suitable relabeling) the collapsed posterior which
lives in the set partition space:
\begin{multline}
p(K,\cP, \thetav_1, \ldots, \thetav_{\Kn},\phi, \wehyp |\ym)
\propto
\prod_{k=1}^{\Kn} \commentSF{p(\ym^{[k]}|\thetav_k)}   p(\thetav_k| \phi)   \frac{\Gamfun{K \eFM_K}}{ \Gamfun{ \eFM_K}^K } p(\phi) p(K) p(\wehyp) \\
\cdot  \int   \prod_{k:N_k=0}  p(\thetav_k| \phi)  d ( \thetav_{\Kn+1}, \ldots,\thetav_K)
\sum_{\Siv: \Siv \in \cP} \int    \prod_{k=1}^K  \eta_k ^{N_k+\eFM_K-1}  d \etav_K \\
=\prod_{k=1}^{\Kn} \commentSF{p(\ym^{[k]}|\thetav_k)}   p(\thetav_k| \phi)
\frac{K!}{(K-\Kn)!}  \frac{\Gamfun{K \eFM_K} \prod_{k=1}^{\Kn}  \Gamfun{ N_k+ \eFM_K}}{\Gamfun{N+K \eFM_K}  \Gamfun{ \eFM_K}^{\Kn} }  p(\phi) p(K) p(\wehyp). \label{postallcol}
\end{multline}
We see in (\ref{postallcol}) that updating of the parameters
$\thetav_1, \ldots, \thetav_{\Kn}$ and $\phi$ (Step~2) can be
performed independently from updating $K$ and the hyperparameter
$\wehyp$ (Step~3).  It should be noted that the conditional posterior
$p(K| \parti, \wehyp) $ of $K$ given $\parti$ that results from
(\ref{postallcol}) is identical with (\ref{postKK}), verifying the
validity of Step~3(a) (or 3(a*)) in our partially marginalized
sampler. In practice, Step~3(a) (or 3(a*)) is implemented by
  considering \comment{an upper bound $\Kmax$ for % a truncated prior on
  $K$ and sampling $K$  from a multinomial distribution over
  $\{ K_+, \ldots, \Kmax\}$, with the success
  probabilities being proportional to the non-normalized posterior probability of $K$.}
   In the following
  empirical analysis we use a maximum value of $\Kmax=100$.

The sampler returns to conditional sampling from the full mixture
posterior in Step~4(b) (or 4(b*)), by sampling the parameters of the
empty components conditional on $\phi$ and sampling the weight
distribution $\etav_K$ \commentF{from the conventional Dirichlet
posterior distribution.  Using the stick breaking
representation of a finite mixture, with the sticks
following $\stick_k|K,\edK \sim \Betadis{\edK,(K-k)\edK}$,} Step 4(b) (or 4(b*)) can be rewritten in terms of sampling
the sticks from a generalized Dirichlet distribution, see, e.g.,
Algorithm~1 of \citet{fru-mal:fro}.

In order to learn the component parameters, a hierarchical prior
structure is introduced in the Bayesian mixture model (\ref{eq:MFM}).
Basically, in Step~2(b) of the TS sampler, any hierarchical prior
$p(\phi)$ on the model parameters can be used.  For other samplers,
such as the allocation sampler \citep{nob-fea:bay}, the prior
$p(\phi)$ has to be conditionally \commentF{conjugate % in order to be able
to} easily integrate out the component parameters $\thetav_k$. A specific
feature of the TS sampler is that the hyperparameters $\phi$ are
learned in Step~2(b) only from the $\Kn$ \emph{filled} components and
that the parameters of the \comment{$K-K_+$} empty components are sampled subsequently
in Step~4(a) from the conditional prior $p(\thetav_k|\phi)$ for
\comment{$k=\Kn+1, \ldots, K$}.  In this way, the parameters of the filled components inform
the parameters of the empty components.  In our opinion, this is an
elegant way to handle hierarchical priors for component parameters in
a dimension changing framework.

The TS sampler allows for a varying, but conditionally finite model
dimension $K$.  Truncation, however, does not result from slice
sampling \citep{kal-etal:sli}, a popular method for DPMs to turn the
infinite mixture into a conditionally finite one.
The TS sampler adds and deletes components \commentF{as follows.} %in the following way.
Step~3(a) is a birth move, where new components are created, if a
value $K > \Kn$ is sampled.  These components are empty, since we
leave the \comment{filled components in} partition $\cP$ unchanged. \comment{Observations} are allocated to these empty components during the subsequent sweep of the sampler in
Step~1(a).
Components can only disappear, if they get emptied in the allocation
Step~1(a).  Hence, for the TS sampler to work well, the tail
probability $ \sum _{K>\Kn} p(K|\cP,\wehyp )$ cannot be too small, as
this probability controls how many empty components are added in
Step~3(a) (or 3(a*)). The more $p(K|\cP,\wehyp )$ is concentrated at
$\Kn$, the more likely mixing \comment{for $\Kn$ and $K$} will be poor
for the TS sampler.  This is true both for static and dynamic MFMs.
% however, there is an interesting difference between
% the two algorithms in updating $K$ given $\cP$.  While
% $ p(K|\cP,\alpha)$ depends on the cluster sizes $\bf{N}$ for dynamic
% MFMs, $p(K|\cP,\eFM)$ is independent of $\bf{N}$ for static MFMs and only
% informed through $\Kn$ and $\eFM$.

Finally, we allow the hyperparameter %$ \wehyp$
of the weight distribution, either $\alpha$ or $\eFM$, to be an
unknown parameter  estimated from the data under a
hyperprior. % $p(\alpha)$ or $p(\eFM)$.
 $\alpha$ (or $\eFM$) are updated in
Step~3(b) (or 3(b*)), which is the only updating step where a random
walk Metropolis-Hastings step is employed.
%\comment{Through this hyperprior,} one can inform the mixture about the desired level of sparsity, e.g., by choosing a prior with a small prior expectation $\Ew{\alpha}$ (or $\Ew{\eFM}$).

%\input{part_gertraud_bettina}

\section{Empirical demonstrations}\label{sec:empir-demonstr}

\subsection{Benchmarking the telescoping sampler} \label{sec:benchm-telesc-sampl}

We compare the performance of the TS sampler to two other samplers
previously proposed to fit \comment{a} static MFM  with univariate
Gaussian components, namely, reversible jump MCMC
\citep[RJ;][]{ric-gre:bay} and the Jain-Neal split-merge algorithm
\citep[JN;][]{jai-nea:spl_2004,jai-nea:spl_2007,mil-har:mix}.  In contrast to the TS sampler, where in
each iteration both $K$ and $K_+$ are updated, the RJ sampler just
samples $K$ while $K_+$ is calculated a posteriori from the sampled
allocations, and the JN sampler just samples the partitions and thus
$K_+$, whereas the posterior of $K$ is reconstructed in a
post-processing step \commentF{\citep[see][Equation~(3.7)]{mil-har:mix}}.

For this comparison we consider the well-known Galaxy data
\citep{roe:den}, which is a small data set of $N=82$ measurements on
velocities of different galaxies from six well-separated sections of
the space, and fit univariate Gaussian mixtures,
$y_i| S_i=k \sim \Normal {\mu_k, \sigma_k^2},$ with $K$ unknown.
Priors are chosen as in \citet{ric-gre:bay}, namely $p(K)$ is a
uniform distribution \commentBG{$\cU\{1, 30\}$},
$\etav_K|K \sim \Dirinv{K}{\edK}$ with $\edK\equiv 1$ is uniform,
whereas $\mu_k\sim \Normal{m,R^2}$,
$\sigma^2_k \sim \Gammainv{2, C_{0} }$, and
$C_0 \sim \Gammad {0.2, 10/R^2}$, where $m$ and $R$ are the midpoint
and the length of the observation interval. These priors are imposed
for sake of comparison with previous results, but not motivated by
modeling considerations nor selected to favor the TS sampler.

Results were obtained \commentBG{for the RJ sampler} using the Nmix
software provided by Peter Green and for the JN sampler \commentBG{as
  implemented in} \citet{mil-har:mix}\footnote{\commentBG{Both are}
  included in the supplementary material to
  \citet{mil-har:mix}.}. Each sampler was run for 1,000,000 iterations
without thinning after discarding the first 10,000 iterations and \commentBG{using} 100
different initializations.  Table~\ref{tab:Kp_short_gal} summarizes
the posterior $p(\Kn|\by)$ over all 100 runs based on the means for
all three samplers (see Appendix~\ref{app:benchm-telesc-sampl} for
more detailed results). The posteriors estimated by \commentBG{all
  three} samplers are very \commentBG{similar} indicating that the TS
sampler provides suitable draws from this posterior distribution.

\begin{table}[t!]
  \centering
  \begin{tabular}{l@{ }c@{\quad}c@{\quad}c@{\quad}c@{\quad}c@{\quad}c@{\quad}c@{\quad}c@{\quad}c@{\quad}c@{\quad}c@{\quad}c}
    \toprule
    Sampler& 1 & 2 & 3 & 4 & 5 & 6 & 7 & 8 & 9 & 10 & 11 & $\ge 12$\\
    \midrule
    TS & .000 & .000 & .070 & .161 & .228 & .228 & .159 & .087 & .040 & .017 & .006 & .003 \\
    RJ & .006 & .000 & .070 & .161 & .227 & .226 & .158 & .086 & .040 & .017 & .006 & .003 \\
    JN & .000 & .000 & .070 & .162 & .228 & .228 & .159 & .087 & .040 & .017 & .006 & .003 \\
    \bottomrule
  \end{tabular}
  \caption{Galaxy data. Mean estimates over 100 MCMC runs of the
    posterior of $K_+$ for the telescoping (TS), the RJMCMC (RJ) and
    the Jain-Neal (JN) sampler.}\label{tab:Kp_short_gal}
\end{table}

\begin{figure}[t!]
  \centering
  \includegraphics[width=0.85\textwidth, trim = 0 5 0 5, clip]{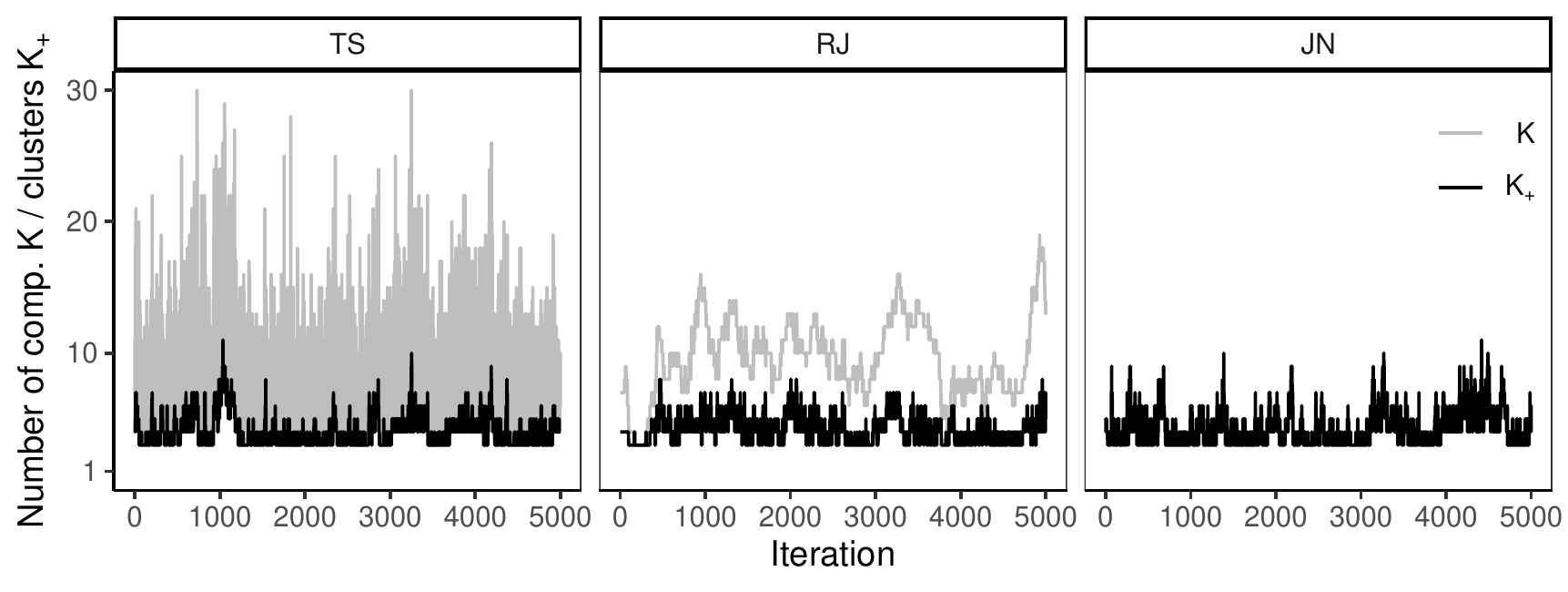}
  \caption{Simulated data, $N=1000$, $\gamma_K \equiv 0.1$,
    \commentBG{all other priors} as in \cite{ric-gre:bay}. Trace plots of
    $K$ (gray) and $K_+$ (black) for the TS, RJ and JN
    sampler.\label{fig:trace}}
\end{figure}

The performance of the three samplers is compared by inspecting the
number of clusters $K_+$ as well as the number of components $K$
obtained for the MCMC iterations, if available.  For this comparison,
we use a simulated data set with a data generating process similar to
the Galaxy data set. \commentBG{We draw $N=1000$ observations from a
three-component univariate Gaussian mixture (see
Figure~\ref{fig:N1000} in Appendix~\ref{app:benchm-telesc-sampl})
%The
%larger sample size is used to provide a more realistic comparison of
%performance for middle-sized data sets where in fact achieving a
%suitable performance is relevant.
 and specify priors} on the component
parameters as used in \cite{ric-gre:bay} for the Galaxy data set and
fit a static MFM with $\gamma = 0.1$.  The smaller value for the
Dirichlet parameter increases the gap between the prior on $K$ and
$\Kn$ and thus improves the mixing of the TS and RJ samplers. Each
sampler is run for 100,000 iterations without thinning. The first 10\%
iterations are omitted as burn-in.

Figure~\ref{fig:trace} \commentBG{shows} a combined trace plot of
$K_+$ and $K$ (if available) for each of the three samplers using the
first 5,000 iterations after \commentBG{omitting} burn-in. In each
trace plot the black line shows how the number of clusters $\Kn$
induced by the sampled partitions varies over the iterations. For the
TS and RJ samplers, in addition, the gray lines show how the number of
components $K$ vary. For the TS sampler, $K$ is sampled given $K_+$,
while for the RJ sampler $K$ changes if components are split or
combined or due to a birth or death of an empty component. This
difference is clearly visible in the trace plots with poorer mixing in
$K$ for the RJ relative to the TN sampler.

\begin{figure}[t!]
  \centering
  \includegraphics[width=0.65\textwidth, trim = 0 5 0 5, clip]{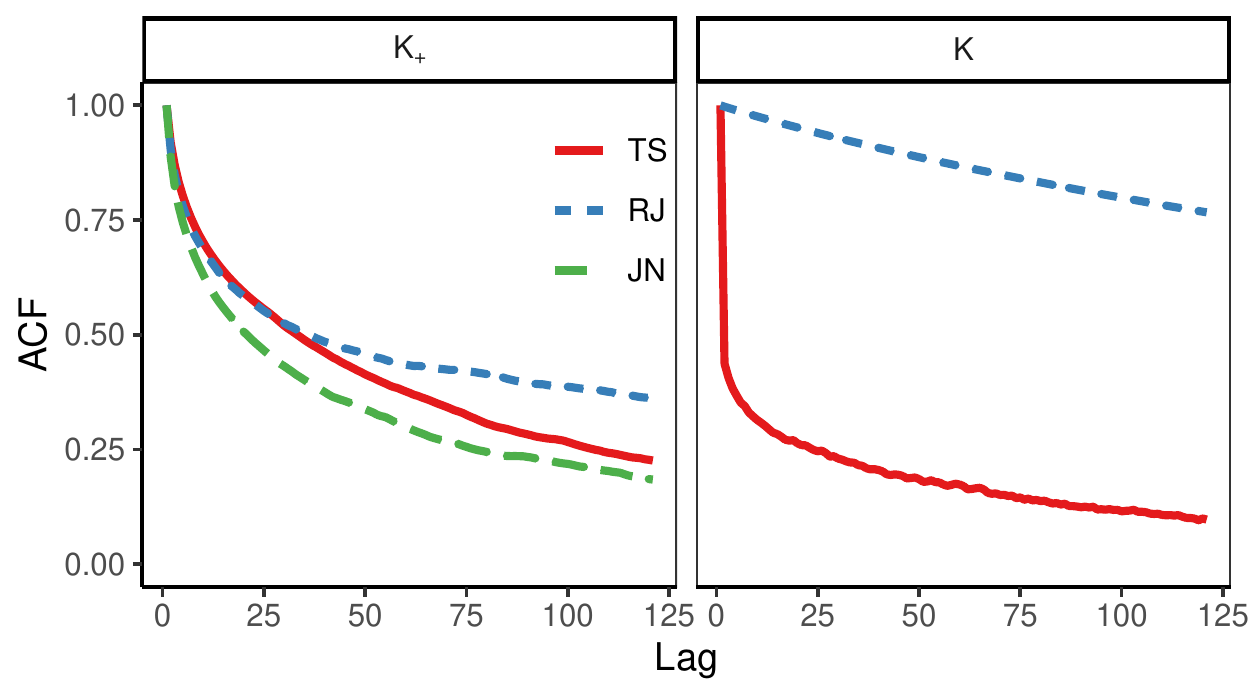}
  \caption{Simulated data, $N=1000$, $\gamma_K \equiv 0.1$,
    \commentBG{all other priors} as in
    \cite{ric-gre:bay}. Auto-correlation function (ACF) for $K_+$
    (left) and $K$ (right) for the TS (solid red line), RJ (dashed
    blue line) and JN (long dashed green line)
    sampler. \label{fig:acf}}
\end{figure}

We assess the efficiency of the three samplers by estimating
auto-correlation functions (ACFs) for the \commentBG{sampled} $\Kn$
and $K$ values (if available) and \commentBG{visualizing them} in
Figure~\ref{fig:acf}. Regarding $\Kn$, the efficiency is rather
comparable over the three samplers, with slight advantages for JN
followed by TS and RJ being \comment{the least efficient}. %last.
\commentBG{Comparing the ACFs for $K$ clearly confirms % indicates
  that TS outperforms RJ.}
%This was
%already visible in the trace plots \comment{in Figure~\ref{fig:trace}}.

The performance comparison indicates that TS is competitive
\commentBG{with the} other samplers, while providing the advantage of
being easily adjusted and immediately applicable for mixtures with
other component distributions or models. Note however, that an
appropriate choice of $\gamma_K$ has an impact on the efficiency of
the sampler as a too large value of $\gamma_K$ prevents that empty
components are created while too small values induce many
(superfluous) additional empty components.
%that the
%performance of the samplers depends on characteristics of the data set
%as well as the prior specifications used and this might impact on a
%situative best choice of sampler.

\subsection{\commentBG{Sensitivity to the prior choice on the number of
  components}} \label{revGalax}

In the following we use the TS sampler to investigate how the
posteriors of $K$ and $\Kn$ vary in dependence of different prior
specifications $p(K, \gamma_K)$ for the Galaxy data set. \commentBG{Although this
data set is very popular in the clustering literature, there is no
consensus on the number of clusters in the sample, see for instance \citet{ait:lik}, \citet{gru-etal:how} and the discussion in
Appendix~\ref{app:revis-galaxy-data}.}
% E.g.,
% \citet{ait:lik}, \comment{for instance,} compares the posterior results for $K$ reported in the
%literature for five different Bayesian analyses of the Galaxy data set
%and reports that the modes of the posterior probabilities for $K$ are
%rather diffuse over the range \comment{4 to 9}.  {Recently, also \citet{gru-etal:how} perform a sensitivity analysis  for the galaxy data to understand the impact of the prior specifications on the clustering result}.

In contrast to these previous
Bayesian analyses, we keep the priors on the component parameters
fixed to those as specified by \cite{ric-gre:bay} for all analyses.
In this way, the impact of the priors on $K$ and the component weights
can be investigated without mixing these effects with those of
different prior specifications on the component parameters. We
consider the static and dynamic MFM with the same priors $p(K)$ and
$\gamma_K$ as specified in Figure~\ref{plot:staticMFM}, i.e.,
  $K-1 \sim \BNB{1, 4, 3}$, $K-1 \sim \Geo{0.1}$, and
  $K \sim \mathcal{U}\{1, 30\}$,  and $\gamma = 1$ for the static MFM
  and $\alpha = 1$ \commentBG{for the dynamic MFM.}
%  The results of our analysis
%indicate how different prior specifications for $p(K, \gamma_K)$ and
%thus implicitly on $p(\Kn|N, \gammav)$ influence the posteriors
%obtained for $K$ and $\Kn$.

\begin{figure}[t!]
  \includegraphics[width=0.85\textwidth, trim = 0 5 0 5, clip]{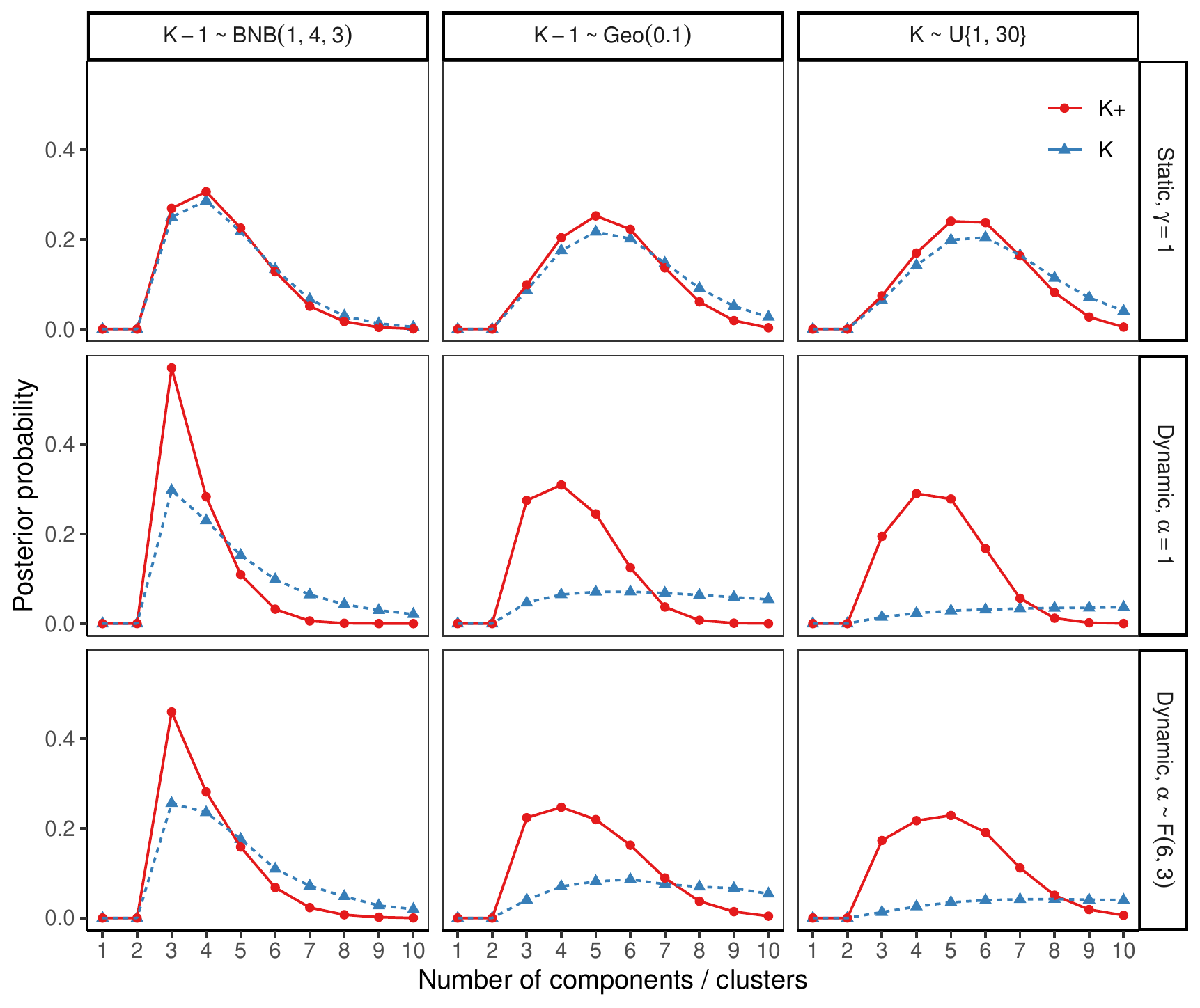}
  \caption{Galaxy data. Posteriors of $K$ (dashed blue lines,
    triangles) and $\Kn$ (solid red lines, circles) under priors
    $K-1 \sim \BNB{1, 4, 3}$ (left), $K-1\sim \Geo{0.1}$ (middle) and
    $K \sim \cU\{1,30\}$ (right) under a static MFM with $\gamma = 1$
    (top) and dynamic MFMs with $\alpha = 1$ (middle) and
    $\alpha\sim \cF(6,3)$ (bottom), for $N=82$. \label{plot_galaxy}}
\end{figure}

In Figure~\ref{plot_galaxy} in the top row, the posteriors of $K$ and
$K_+$ are reported for the static MFM with $\edK\equiv 1$.  \commentBG{The posteriors $p(\Kn|\ym)$ and $p(K|\ym)$ are very similar to
each other regardless of $p(K)$ specified.
%The mode of
%$p(K_+|\ym)$ is close to the mode of $p(K|\ym)$.
%
In contrast,} for the dynamic prior $\gamma_K =1/K$, shown in the
\commentBG{middle} row, the posteriors $p(K|\ym)$ and $p(K_+|\ym)$ differ
considerably.  While the posterior $p(K|\ym)$ becomes flatter compared to fixed
$\gamma=1$, most of the posterior mass of $p(K_+|\ym)$ concentrates on
$K_+$ equal to 3, 4 or 5 which are reasonable values for the number of
clusters in this data set.  Comparing the posteriors of $\Kn$ and $K$
to the corresponding priors in Figure~\ref{plot:staticMFM} indicates
that the posteriors are strongly influenced by the prior
distributions. E.g., the flat prior for $\Kn$ induced by the uniform
distribution and $\gamma_K=1$ (plot in Figure~\ref{plot:staticMFM} on
the top right) results in a posterior of $K_+$ favoring large values
between $4$ and $7$ clusters which clearly overestimates the number of
clusters in this small data set. In contrast, a sparse prior on $\Kn$
in combination with a dynamic MFM favors a sparse estimation of the
number of clusters also a posteriori, see, e.g., the posterior
$p(\Kn|\ym)$ for the $\BNB{1,4,3}$ prior  where three clusters
\commentBG{are estimated}.

\comment{Under the hyperprior} $\alpha \sim \cF(6,3)$, the posterior of $K_+$ looks rather similar
to \comment{assuming that} $\alpha=1$ fixed, see \commentBG{Figure~\ref{plot_galaxy} at the bottom. However, if the
shrinkage prior $\alpha \sim \cG(1,20)$ is specified, the posterior of
$K_+$ becomes completely independent of both the prior and posterior
of $K$, see   Appendix~\ref{app:revis-galaxy-data} where also
results for other specifications on $K$ and the
 weights are reported, in particular a FM, a SFM and
a DPM model.}

Figure~\ref{plot_galaxy} shows that depending on the prior on $K$ and
whether a static or dynamic MFM is specified, the posterior mode of
$p(K_+|y)$ varies. This highlights the impact of the implicitly
specified prior on $\Kn$ on the posterior of $\Kn$. This especially
applies to the Galaxy data set which contains only $N=82$ observations
and has no clear cluster structure. If, \commentBG{in contrast,} there is considerable
information in the data, the posteriors of $\Kn$ for different prior
specifications $p(K)$ coincide, as can be seen in the next section
when analyzing the Thyroid data set.

%Also, interestingly, the heavy-tailed prior $K-1\sim \BNB{1,1,1}$
%with no mean is better able to adapt to this sparse cluster solution
%than the priors with large mean, which seem to be more informative and
%pull the number of clusters towards large values.

%In this case, regardless of $p(K)$, for each prior
%specification three clusters are estimated, while the posteriors of
%$K$ are very flat.
%%
%In Appendix~\ref{app:revis-galaxy-data}, the posteriors of $K$ and
%$K_+$ are also reported for other prior specifications on $K$ and the
%component weights, in particular a FM, a SFM and
%a DPM model.

\subsection{Changing the clustering kernel}\label{sec:chang-clust-kern}

We use the TS sampler to fit dynamic MFMs with different component
distributions, \commentBG{i.e.,} the multivariate Gaussian
distribution and the latent class model for multivariate categorical
data.  This demonstrates how easily the TS sampler can be used
\commentBG{to fit} a MFM regardless of the component distributions.
\commentSF{For $K$} we use the same priors $p(K)$ as in the previous
  section.  It will turn out that a prior specification for $K$ where
$\Ew{K}$ is small and the tails are not too light, in combination with
the dynamic prior $\gamma_K=\alpha/K$ on the component weights and
$\alpha \sim \cF(6,3)$ gives good clustering results.

The final partition is obtained by identifying the models through the
post-processing procedure suggested by \cite{fru:book} and applied in
\cite{mal-etal:mod,mal-etal:ide}. First, the number of clusters
$\hat{K}_+$ is estimated by the mode of the posterior
$p(K_+|\ym)$. Then for all posterior draws where
$K_+^{(m)}=\hat{K}_+$, \commentBG{the component parameters} are
clustered in the point process representation into $\hat{K}_+$
clusters using $k$-means clustering. A unique labeling of the draws is
obtained and used to reorder all draws, including the sampled
allocations. \commentBG{The final partition of the data is then
  determined by the maximum a posteriori (MAP) estimate of the
  relabeled cluster allocations.}

\subsubsection{Multivariate Gaussian mixtures: Thyroid data}\label{sec:mult-gauss-mixt}

The Thyroid data are a benchmark data set for multivariate normal
mixtures included in the $R$ package \textit{mclust}
\citep{scr-etal:mcl}. It consists of five laboratory test variables
and a categorical variable indicating the operation diagnosis (with
three potential values) for 215 patients. A dynamic MFM with
multivariate normal component densities is fitted using a simplified
version of the priors proposed in \cite{mal-etal:mod} \commentBG{for}
the component parameters (for details see
Appendix~\ref{app:mult-gauss-mixt}).
As can be seen in the left-hand column of Table~\ref{tab:thyr}, for
all priors on $K$ the mode of the posteriors for $K_+$ \comment{lies} at three,
even for the uniform prior.  Also the posterior mode of $K$ is three,
indicating that rarely empty components were sampled.
% The posterior expectation
% $\bar \gamma_K$ (see Table~\ref{tab:thyr-gamma} in
% Appendix~\ref{sec:furth-illustr-appl}) is smaller than 1 indicating
% that unequally sized clusters are estimated. In fact,
For the $K-1 \sim \BNB{1,4,3}$ prior, the final partition
obtained through the MAP estimate consists of three clusters with 28,
37 and 150 patients. The ARI of this partition with the known
operation diagnosis is 0.88, which is equal to the ARI of the
\textit{mclust} solution. Overall these results suggest that, if the
data are informative regarding a specific cluster structure, the
clustering result is not susceptible to the prior specification of
$p(K)$.

\begin{table}[t!]
	\centering
	\begin{tabular}{@{}r@{ }|c@{ }c@{ }c@{ }c@{ }|c@{ }c@{ }c@{ }c@{ }}%|c@{ }c@{ }c@{ }c@{}}
		\toprule
		&\multicolumn{4}{|c|}{Thyroid}&\multicolumn{4}{c}{Fear}%&\multicolumn{4}{c}{Eye tracking}
		\\
		$p(K)$  &
		\multicolumn{2}{c}{$p(K_+|\ym)$} &   \multicolumn{2}{c|}{$p(K|\ym)$}&   \multicolumn{2}{c}{$p(K_+|\ym)$} &   \multicolumn{2}{c}{$p(K|\ym)$} %& \multicolumn{2}{c}{$p(K_+|\ym)$} &   \multicolumn{2}{c}{$p(K|\ym)$}
		\\ \midrule
		$\cU\{1, 30\}$ &  3 & [3, 3] & 3 & [4, 19] & 6 & [5, 9] & 30 & [10, 24] \\%&13  &[12, 21]  &119  &[50, 118]\\
		%    $\BNB{1, 1, 1}$  & 3 & [3, 3] & 3 &[3, 5] &2 & [3, 7] & 2 & [3, 14]& 10 & [9, 16] & 11 & [12, 41] \\
		$\Geo{0.1}$  & 3 & [3, 3] & 3 & [3, 7]& 4 & [4, 7] & 5 & [5, 16]\\ %&9 & [9, 15] & 13 & [12, 25]\\
		$\BNB{1, 4, 3}$  & 3 & [3, 3]   & 3 & [3, 4]& 2 & [2, 4] & 2 & [2, 5] \\% &6 & [6, 10] & 7 & [7, 13] \\
		\bottomrule
	\end{tabular}
	\caption{Thyroid  and  Fear data.  Posterior
		inference for $K$ and $K_+$ for a dynamic MFM based on different
		priors $p(K)$ and $\alpha \sim \cF(6,3)$. The posteriors of $K_+$
		and $K$ are summarized by their modes, followed by the 1st and 3rd
		quartiles.}\label{tab:thyr}
\end{table}

\subsubsection{Latent class analysis: Fear data}

\cite{ste-etal:sta} consider data of $93$ children in the context of
infant temperamental research. For each child, three categorical
features are observed, namely motor activity (M) with 4 categories,
fret/cry behavior (C) with 3 categories, and fear of unfamiliar events
(F) with 3 categories, see \cite{fru-mal:fro} for the contingency
table of the data. The scientific hypothesis is that two different
profiles in children are present. To test this, a latent class model
is fitted \commentBG{using} a dynamic MFM \commentBG{with a uniform
  Dirichlet prior on the component parameters.}
Table~\ref{tab:thyr} shows that the prior $K-1 \sim \BNB{1,4,3}$
\commentBG{selects $K_+ = 2$,
% when using the mode to estimate a suitable
%number of clusters
confirming the theoretically expected number of clusters.}
%
% For the priors on $K$ with small expectation the posterior
% mean $\bar \gamma_K$ (see Table~\ref{tab:thyr-gamma} in
% Appendix~\ref{sec:furth-illustr-appl}) is around two or higher
% indicating that almost equally sized clusters are estimated with the
% final partition containing two clusters with 42 and 51 observations.
%
The geometric prior with $\Ew{K}=10$ and the truncated uniform prior,
however, overestimate the number of clusters with the mode of $K_+$ at
4 and 6, respectively.
\commentBG{The results obtained when identifying} the MCMC output from
a dynamic MFM with $K-1 \sim \BNB{1, 4, 3}$ and $\alpha \sim \cF(6,3)$
\commentBG{indicate that the two classes have a rather different
  profile regarding the occurrence probabilities of the categories
  (see Appendix~\ref{app:latent-class-analys}), which coincides with
  the findings in \citet{ste-etal:sta}.}

%\subsubsection{Poisson mixtures: Eye tracking data}
%
%The count data on eye tracking anomalies in 101 schizophrenic patients
%studied by \citet{esc-wes:com} are reconsidered. To capture
%overdispersion and account for excess zeros, \cite{fru:book} analyzed
%the data using a Poisson mixture model. The aim of this analysis is
%not to perform model-based clustering, but to capture the extreme
%unobserved heterogeneity present in this data set (see
%Appendix~\ref{app:poiss-mixt-eye}).
%
%\cite{fru-mal:fro} report that fitting a SFM and a DPM lead to 4 to 5
%clusters. Using the same hierarchical specification for the component
%means, the dynamic MFM provides an approximation of the unobserved
%heterogeneity distribution based on 4 to 6 clusters for priors on $K$
%with small mean values. The other priors on $K$ estimate considerably
%more clusters, see the right-hand column in Table~\ref{tab:thyr}.

% The posterior of $\lambda_k$ (not identified
% draws) for $K-1\sim \BNB{1,4,3}$ and $K_+=6$ is shown in
% Appendix~\ref{app:poiss-mixt-eye}.

\subsection{Investigating the telescoping sampler with artificial data} \label{TSartificial}

\commentBG{We perform a simulation study with artificial data to
  investigate how the TS sampler performs
%in a Bayesian cluster analysis application
  in dependence of sample size $N$, dimension $r$ and number of
  clusters $K_+$.  In addition, we vary the priors for
  $p(K, \gamma_K)$ considering static and dynamic MFMs and in
  particular include the suggested priors $K-1 \sim \BNB{1,4,3}$ and
  $\alpha \sim \cF(6,3)$.}
% to study the performance of the recommended choices.
%	
We sample 100 data sets from a multivariate normal mixture with eight
equally sized components, varying dimension ($r=2,8,12$) and
increasing sample size ($N=400, 4000, 10000$), combining higher values
of the dimension $r$ with larger sample sizes $N$. A detailed
description of the data generating processes of the simulated data as
well as the specified priors $p(K)$ and Dirichlet parameters $\gamma$
and $\alpha$ is given in Appendix~\ref{app:invest-telesc-sampl}.

% We systematically vary the prior specifications for $p(K, \gamma_K)$
% and assess their impact.
% For the

\begin{figure}[b!]
  \centering
  \includegraphics[width=0.85\textwidth, trim = 0 5 0 5, clip]{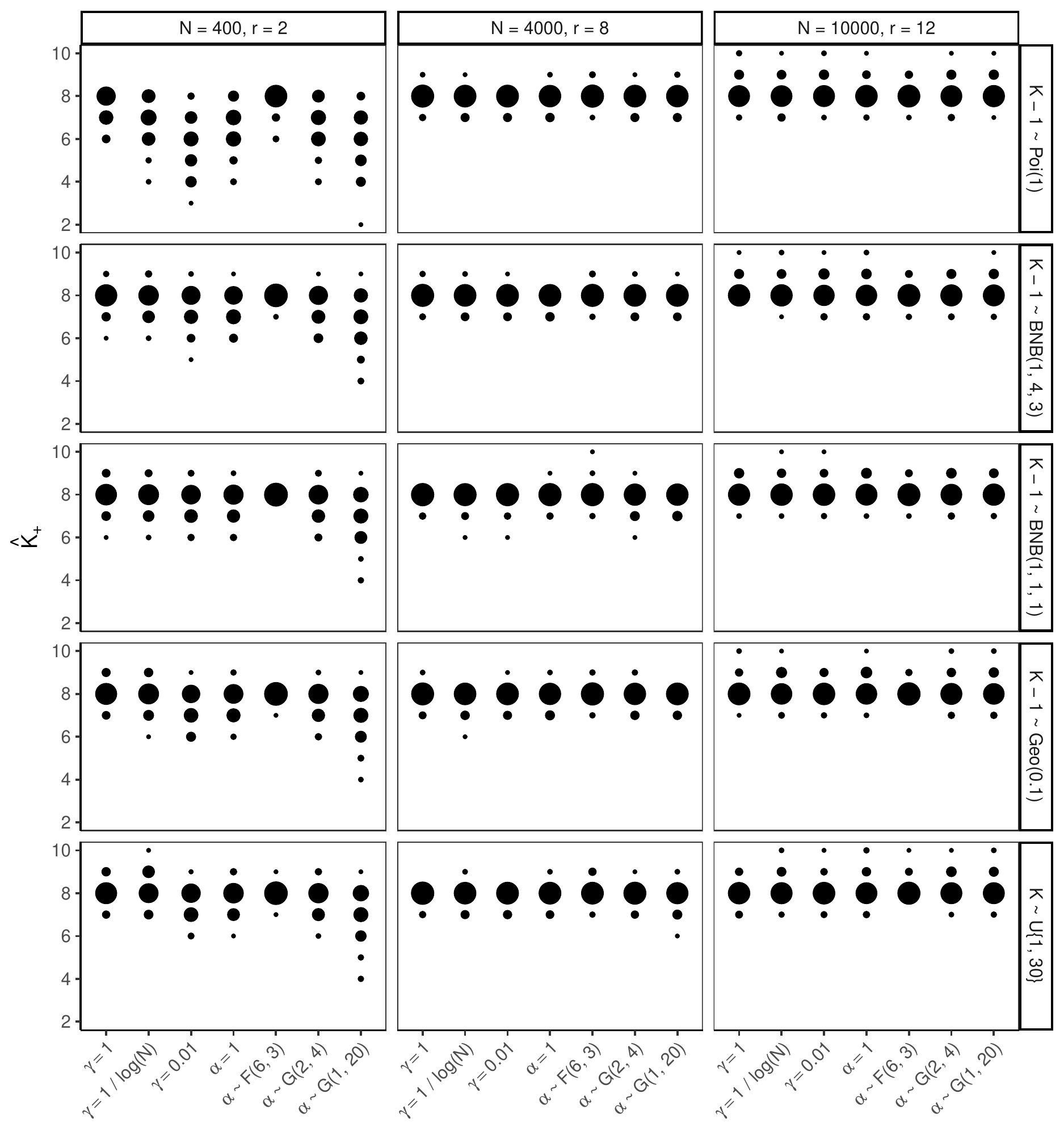}
  \caption{Simulation study. Estimated number of clusters for 100
    artificial data sets drawn from mixtures of multivariate Gaussian
    distributions with eight components. Results based on TS sampling
    for varying \commentSF{sample size $N$ and dimension $r$} (columns), priors on
    $K$ (rows) and Dirichlet parameter \commentBG{values} $\gamma_K$
    ($x$-axis). The size of a bubble point shows the
    \comment{percentage} of artificial data sets \commentBG{with} a
    specific number of clusters estimated.}
  \label{fig:simulation}
\end{figure}

\commentBG{Results are visualized in a bubble plot in
  Figure~\ref{fig:simulation}.
%The different
%data scenarios where sample size $N$ and dimension $r$ are varied are
%given in the columns, while  the different priors on $K$ are displayed in the rows. In
%each panel the results are given for the different priors on the
%Dirichlet parameter $\gamma_K$, indicated on the $x$-axis (the first three specifications refer to the static MFM, the last three to the dynamic MFM specification).
  The area of the bubbles is proportional to the percentage of data
  sets with a specific number of clusters $K_+$ estimated as indicated
  on the $y$-axis.}
The results show how the influence of the prior $p(K, \gamma_K)$
decreases when the information in the sample increases.  If the
information is weak, i.e., for $N = 400$ and $r = 2$, the prior
specifications on $K$ and on $\gamma_K$ have \commentSF{considerable} impact on the
clustering result (first column of Figure~\ref{fig:simulation}).  The
estimated number of clusters $K_+$ tends to be lower for the Poisson
prior regardless of the prior imposed on $\gamma_K$. \commentBG{While}
the Poisson prior with $\lambda = 1$ induces the same prior mean
$\Ew{K} = 2$ \comment{as} the $\BNB{1, 4, 3}$ prior, it has also light
tails. Thus, the fatter tails of the $\BNB{1, 4, 3}$ prior allow to
estimate the number of clusters in the data correctly despite
\commentBG{its} sparsity inducing properties.
Regarding the prior on the Dirichlet parameter $\gamma_K$, the results
of the static MFM clearly indicate that the estimated number of
clusters decreases for decreasing values of $\gamma$. In the dynamic
case, using $\alpha \sim \cF(6, 3)$ gives more \comment{reliable}
results than the other \comment{specifications for $\alpha$}
regardless of the prior on $K$. In contrast, the influence of the
sparsity inducing prior $\alpha \sim \mathcal{G}(1, 20)$ is clearly
visible \comment{across all priors} on $K$, \comment{leading even to
  four estimated} clusters instead of eight. Overall, the results for
the combination $K -1 \sim \BNB{1, 4, 3}$ and $\alpha \sim \cF(6, 3)$
confirm the suitability of this prior specification for determining
the number of clusters in a Bayesian cluster analysis application.
For $N = 4000$ the estimated number of data clusters $\hat{K}_+$ is
equal to eight for nearly all data sets regardless of the prior
specifications. \commentBG{Results are similar for $N = 10000$.}
% with a slight increase of
%spurious additional clusters being detected for the larger sample size
%where also the dimension is higher, i.e., $r = 12$ instead of $r = 8$.

%\input{part_discussion}

\section{Concluding remarks}\label{sec:disc-final-remarks}

Being a finite mixture model where the number of components is
unknown, the MFM model has a long tradition in Bayesian mixture
analysis. Building on this tradition, a key aspect of our work is to
explicitly distinguish between the number of components $K$ in the
mixture distribution and the number of clusters $\Kn$ in the partition
of the data, corresponding to non-empty components given the data.
With this fundamental distinction in mind, we contribute to MFMs both
from a methodological as well as a computational perspective.

Traditionally, the hyperparameter $\gamma$ of \comment{a symmetric} Dirichlet prior on
the component weights is a fixed value, often equal to one.  In this
paper, we investigate in detail a more general MFM specification which
defines the hyperparameter $\gamma_K$ of the symmetric Dirichlet prior
dynamically and dependent on $K$.  We provide theoretical results that
characterize how this specification of a dynamic symmetric Dirichlet prior
 on the component weights influences the induced prior on the number of clusters
and the partition structure. \comment{While a static MFM with fixed  $\gamma$
corresponds to a Bayesian non-parametric mixture within the class of Gibbs-type priors,
our dynamic version where $\gamma_K$ depends on $K$ leads to more a flexible %Bayesian non-parametric
mixture outside the class of Gibbs-type priors}.

Regarding posterior inference, we introduce the novel telescoping (TS)
sampler which is a trans-dimensional Gibbs sampler that simultaneously
infers the posterior on the number of components $K$ and the number of
clusters $\Kn$. As illustrated, for instance, for multivariate
Gaussian mixtures, the TS sampler can be easily implemented for any
kind of component model or distribution.  Based on the TS sampler, in
future work many different kinds of mixture models can be easily
fitted to cluster different types of data which require the use of
specific component distributions and models. \comment{Future work
  should also investigate the potential to improve the computational
  efficiency of the TS sampler, e.g., by reducing the computational
  burden due to the empty components.}\\

\smallskip
\smallskip

\noindent\textit{Acknowledgments}

The authors would like to thank Raffaele Argiento, Pierpaolo De
Blasi, and Annalisa Cerquetti as well as an anonymous reviewer and
the associate editor for valuable suggestions and feedback which
helped to improve this work.

\smallskip
\smallskip

\bibliographystyle{ba}
\bibliography{sylvia_kyoto}

\input{appendix}

\end{document}

%% file: newcommands.tex
\newcommand{\card}[1]{\mbox{\rm card}\left(#1\right)}

\newcommand{\fEwens}{p_{\footnotesize  \DP}}
\newcommand{\Diag}[1]{\mbox{\rm Diag}\left(#1\right)}
\newcommand{\DP}{{\rm DP}}
\newcommand{\PY}[1]{{\cal PY}( #1)}
\newcommand{\Poi}[1]{\mathcal{P}\left(#1\right)}
\newcommand{\Poisson}[1]{\Poi{#1}}
\newcommand{\Poiss}[1]{\mathcal{P}\left( #1 \right)}

\newcommand{\Gammad}[1]{ \mathcal{G}\left(#1\right)}
\newcommand{\NegBin}[1]{\mbox{\rm NegBin}\left(#1\right)}
\newcommand{\alphaNB}{\alpha_\lambda}
\newcommand{\Kmax}{K_{\max} } %sdimension of regression coefficient
\newcommand{\Kfix}{K_{\footnotesize f}} %sdimension of regression coefficient
\newcommand{\wehyp}{\omega}

\newcommand{\NNN}{\mathbb{N}}
\newcommand{\bfz}{{\mathbf{0}}}

\newcommand{\cv}{\boldsymbol{c}}
\newcommand{\Wm}{\boldsymbol{W}}

\newcommand{\Bincoef}[2]{\left( \begin{array}{c}   #1 \\#2  \end{array}\right)}
\newcommand{\Betafun}[1]{B (#1)}
\newcommand{\Geo}[1]{\mbox{\rm Geo}\left(#1\right)}
\newcommand{\pigeo}{\pi}
\newcommand{\pia}{a_\pigeo}
\newcommand{\pib}{b_\pigeo}
\newcommand{\BNB}[1]{\mbox{\rm BNB}\left(#1\right)}

\newcommand{\Normal}[1]{ \mathcal{N}\left(#1\right)}
\newcommand{\new}{^{\text{new}}}

\newcommand{\thetav}{{\mathbf{\boldsymbol{\theta}}}} % notation for component specfic parameter
\newcommand{\EE}{\mbox{\rm E}}
\newcommand{\Ew}[1]{\EE(#1)}   % Expectation of a rv\newcommand{\by}{\mathbf{y}}

\newcommand{\by}{\mathbf{y}}

\newcommand{\bb}{\mathbf{b}}
\newcommand{\bB}{\mathbf{B}}
\newcommand{\bC}{\mathbf{C}}
\newcommand{\bG}{\mathbf{G}}

\newcommand{\bN}{\mathbf{N}}

\newcommand{\bS}{\mathbf{S}}

\newcommand{\btheta}{\boldsymbol{\theta}}

\newcommand{\boldeta}{\boldsymbol{\eta}}
\newcommand{\bmu}{\boldsymbol{\mu}}

\newcommand{\bSigma}{\boldsymbol{\Sigma}}

\newcommand{\cN}{\mathcal{N}}
\newcommand{\cG}{\mathcal{G}}

\newcommand{\cW}{\mathcal{W}}
\newcommand{\cC}{\mathcal{C}}
\newcommand{\cU}{\mathcal{U}}

\newcommand{\cF}{\mathcal{F}}
\newcommand{\cP}{\mathcal{C}}

\newcommand{\stick}{v}

\newcommand{\Betadis}[1]{\mathcal{B}\left(#1\right)}
\newcommand{\ym}{{\mathbf y}} % y multivariate observation

\newcommand{\parti}{{\cal C}}

\newcommand{\Kn}{K _+}
\newcommand{\edK}{\ed{K,0}}

\newcommand{\eFM}{\gamma}
\renewcommand{\edK}{\gamma_K}
\newcommand{\Siv}{{\mathbf{S}}} % indicatoren
\newcommand{\etav}{{\boldsymbol{\eta}}} % vector of weights
\newcommand{\gammav}{{\boldsymbol{\gamma}}} % vector of weights
\newcommand{\Gamfun}[1]{\Gamma (#1)}
\newcommand{\Count}[1]{\#\{#1\}}
\newcommand{\Prob}[1]{\mbox{\rm Pr}\{#1\}}
\newcommand{\Dirinv}[2]{\mathcal{D}_{#1}\left(#2\right)}
\newcommand{\Gammainv}[1]{\mathcal{G}^{-1} \left(#1\right)}
\newcommand{\alphaDP}{\alpha}
\newcommand{\alphaPY}{\theta}

\newcommand{\betaPY}{\sigma}

\newcommand{\Mulnom}[1]{\mbox{\rm MulNom}\left(#1\right)}
\newcommand{\indic}[1]{I{\{#1\}}}
\newcommand{\Dir}[1]{ \mathcal{D}\left(#1\right)}

{\begin{figure}[t!]
		\begin{center}
			\scalebox{#4}{\includegraphics{#3.png}}
			\caption{#1}\label{#2}}{\end{center}\end{figure}}

{\begin{figure}[t!]
		\begin{center}
			\scalebox{#5}{\includegraphics{#3.png}}
			\scalebox{#5}{\includegraphics{#4.png}}
			\caption{#1}\label{#2}}{\end{center}\end{figure}}

{\begin{figure}[t!]
		\begin{center}
			\scalebox{#5}{\includegraphics{#3.png}}\\
			\scalebox{#5}{\includegraphics{#4.png}}
			\caption{#1}\label{#2}}{\end{center}\end{figure}}

{\begin{figure}[t!]
		\begin{center}
			\scalebox{#6}{\includegraphics{#3.png}}\\
			\scalebox{#6}{\includegraphics{#4.png}}\\
            \scalebox{#6}{\includegraphics{#5.png}}
			\caption{#1}\label{#2}}{\end{center}\end{figure}}

{\begin{figure}[t!]
		\begin{center}
			\scalebox{#6}{\includegraphics{#3.png}}
			\scalebox{#6}{\includegraphics{#4.png}}
			\scalebox{#6}{\includegraphics{#5.png}}
			\caption{#1}\label{#2}
		}{\end{center}\end{figure}}

{\begin{figure}[t!]
		\begin{center}
			\scalebox{#7}{\includegraphics{#3.png}}
			\scalebox{#7}{\includegraphics{#4.png}}\\[#8mm]
			\scalebox{#7}{\includegraphics{#5.png}}
            \scalebox{#7}{\includegraphics{#6.png}}
			\caption{#1}\label{#2}
		}{\end{center}\end{figure}}

%% file: appendix.tex
\newpage
\appendix

% Change equation numbering
\setcounter{equation}{0}
\setcounter{figure}{0}
\setcounter{table}{0}
\setcounter{page}{1}

\renewcommand{\thesection}{\Alph{section}}
\renewcommand{\thetable}{\Alph{section}.\arabic{table}}
\renewcommand{\thefigure}{\Alph{section}.\arabic{figure}}
\renewcommand{\theequation}{\Alph{section}.\arabic{equation}}

\begin{center}
\sffamily\LARGE{\bfseries Supplementary material for:\\
  ``Generalized mixtures of finite mixtures and telescoping sampling''}
\end{center}

\section{Mathematical derivations}  \label{proof}

\paragraph*{Proof of Theorem~\ref{lemma1}.}

Let $\Siv=(S_1,\ldots,S_N)$ be the collection of all component
indicators which, for a given $K$, associate each observation $\ym_i$
with the component that generated this data point (see model
\eqref{eq:MFM}).  For any MFM with prior
$\boldeta_K|K, \edK \sim \Dirinv{K}{\edK}$, the marginal prior
$p( \Siv | K, \edK )$ for a fixed $K$ is given by:
\begin{align}
  p( \Siv  |  K, \edK) &=  \int p(\bS|\boldeta_K)p(\boldeta_K|K,\edK  )d\boldeta_K \nonumber\\
  & = \comment{\frac{\Gamfun{ \edK K } }{ \Gamfun{\edK K + N}}
    \prod_{\ell =1}^K   \frac{\Gamfun{N_\ell+\edK}}{\Gamfun{\edK }}}, \label{priorSall}
\end{align}
\citepApp[see, for example][Chapter 3, Equation (3.24)]{fru:book}.
\comment{If we define $\Kn$ as the number of all occupied components with $N_\ell > 0$ and reorder the components such that the non-empty components appear first, with $N_1, \ldots, N_{\Kn}$ being the corresponding occupation numbers, then
$\Siv$ defines a} set partition $\parti= \{\parti_1, \ldots, \parti_{\Kn}\}$ of the data
\commentSF{indices $\{1, \ldots,N\}$} with $N_j=\card{\parti_j}$. There are
$$ \comment{{K \choose \Kn } \Kn ! } = \frac{K !}{(K- \Kn )!}$$
assignment vectors   $\Siv$ that define the same partition $\parti$,
\comment{where the first factor accounts for choosing $\Kn$  among the  $K$ components,
which are labeled $\{1, \ldots, \Kn\}$, while the second factor accounts for all possibilities to relabel these $\Kn$ (non-empty) components.}
Multiplying \eqref{priorSall} by this number yields:
\begin{align*}
p(\parti | N, K, \edK ) &= \frac{ \Gamfun{\edK K} K ! }{ \Gamfun{\edK K+N}   \Gamfun{ \edK } ^{\Kn} (K- \Kn )!}  \prod_{j=1}^{\Kn}  \Gamfun{N_j +\edK }\\
&=  \frac{ V_{N, \Kn}^{K, \edK}}{ \Gamfun{ \edK } ^{\Kn} }    \prod_{j=1}^{\Kn}  \Gamfun{N_j +\edK } ,
\end{align*}
where $V_{N,\commentBG{\Kn}}^{K, \edK}$ is defined as in \eqref{mixpKV}.
Averaging $p(\parti | N, K, \edK ) $ over the prior $p(K)$ yields the
probability mass function (pmf) $p(\parti | N, \gammav)$ given in
\eqref{mixparti}:
\begin{align*}
p(\parti | N, \gammav) &=   \sum_{K=\Kn}^\infty   p(K) p(\parti  | N, K, \edK ).
\end{align*}
For any $\Kn=1,2,\ldots,N$, consider the cluster sizes $(N_1, \ldots, N_{\Kn})$ of
the $\Kn$ non-empty clusters \comment{which are labeled $\{1, \ldots, \Kn \}$.
For this labeling,}
there are
$$ {K \choose \Kn }  = \frac{K !}{ \Kn! (K- \Kn )!}$$
ways to choose $\Kn$ non-empty  among the  $K$ components and
$$  {N \choose {N_1  \cdots N_{\Kn}}} =\frac{N!}{N_1! \cdots  N_{\Kn}! } $$
different ways to assign $N$ observations into clusters of size
$N_1, \ldots N_{\Kn}$.
%Any of these assignments can be represented by an allocation vector $\Siv$ with a prior probability given by \eqref{priorSiv}.
Multiplying \eqref{priorSall} by this number yields:
\begin{align}  \label{mixpKKapp}
p(N_1, \ldots, N_{\Kn}  |  N, K,  \edK ) &=   \frac{  N!}{  \Kn !}  \frac{ V_{N, \Kn}^{K, \edK}}{ \Gamfun{ \edK } ^{\Kn} }
  \prod_{j=1}^{\Kn}   \frac{ \Gamfun{N_j +\edK }}{ \Gamfun{N_j + 1}}.
\end{align}
Averaging over the prior $p(K)$ yields the prior of the labeled
cluster sizes given in \eqref{mixpK}:
\begin{align*}
p(N_1, \ldots, N_{\Kn} | N, \gammav) &=   \sum_{K=\Kn}^\infty  p(K) p(N_1, \ldots, N_{\Kn}  |  N, K,  \edK ).
\end{align*}

\paragraph*{\comment{Derivation of (\ref{recMFM}).}}

\comment{Using $\Gamfun{\eFM K + N} = \Gamfun{\eFM K + N+ 1}/(\eFM K + N)$, we obtain:
\begin{eqnarray*}
  {V}^\eFM _{N, {k}} = \sum_{K=k}^\infty p(K) V_{N,k }^{K, \eFM} =
                         \sum_{K=k}^\infty
p(K)  \frac{K !}{(K- k )!} \frac{ (\eFM K + N)  \Gamfun{\eFM K} }{ \Gamfun{ \eFM K + N + 1}} .
\end{eqnarray*}
Splitting $\eFM K + N =  (\eFM k + N)  + \eFM (K-k)$, we obtain:
\begin{eqnarray*}
{V}^\eFM _{N, {k}} & = &  (\eFM k + N)    \sum_{K=k}^\infty
p(K)  \frac{K !}{(K- k )!} \frac{\Gamfun{\eFM K} }{  \Gamfun{ \eFM K + N + 1}}  \\
&+ & \eFM  \sum_{K=k + 1}^\infty
p(K)  \frac{K !}{(K- k -1 )!} \frac{\Gamfun{\eFM K} }{  \Gamfun{ \eFM K + N + 1}} \\
& = & (\eFM k + N)  {V}^\eFM _{N+1, {k}} + \eFM  {V}^\eFM _{N+1, {k}+1}.
\end{eqnarray*}}

\paragraph*{Proof of Theorem~\ref{theorem1}.}
For a dynamic MFM,
\begin{align*}
  V_{N,\Kn}^{K,\alpha}   &=    \frac{K !}{(K- \Kn )!}  \frac{\Gamfun{\alpha} }{ \Gamfun{\alpha+N}}   ,
\end{align*}
and we obtain the following  EPPF from \eqref{mixparti}:
\begin{align*}
p(\parti  | N, \alpha ) &=   \frac{\Gamfun{\alpha} }{ \Gamfun{\alpha+N}}
\sum_{K=\Kn}^ {\infty}  p(K )  \frac{ K ! }{ \Gamfun{\frac{\alpha}{K} } ^{\Kn} (K- \Kn )! }     \prod_{j=1}^{{\Kn}}   \Gamfun{N_j + \frac{\alpha}{K} } .
\end{align*}
Using $\Gamma(\frac{\alpha}{K})=\frac{K}{\alpha}\Gamma(1+\frac{\alpha}{K})$, we obtain:
\begin{align*}
  \frac{K !}{ (K- \Kn )!}	\prod_{j=1}^{K_+} \frac{\Gamma(N_j+\frac{\alpha}{K})}{\Gamma(\frac{\alpha}{K})}
  &=  \alpha ^{\Kn}   \frac{K !}{K ^{\Kn } (K- \Kn )!}
  \prod_{j=1}^{K_+} \frac{\Gamma(N_j+\frac{\alpha}{K})}{\Gamma(1+ \frac{\alpha}{K})}	.
\end{align*}
Therefore, $p(\parti | N, \alpha)$ can be expressed as in \eqref{mixpNKK}:
\begin{align*}
p(\parti | N, \alpha)
=     \frac{ \alpha  ^{\Kn}   \Gamfun{\alpha} }{ \Gamfun{\alpha + N}}  \prod_{j=1}^{\Kn}  \Gamfun{N_j}
  \sum_{K= \Kn}^ {\infty}  p(K )
    \prod_{j=1}^{\Kn}       \frac{ \Gamfun{N_j + \frac{\alpha}{K} }(K -j+1)  }
   {\Gamma(1+\frac{\alpha}{K})\Gamfun{N_j} K }.
\end{align*}

\paragraph*{Proof of Theorem~\ref{Theorem4} and Algorithm~\ref{KNMFM}.}

The marginal prior $ \Prob{\Kn = k |N, \gammav} $ is obtained by aggregating
the prior pmf $p(N_1, \ldots, N_{k} |N,\gammav)$ of the labeled cluster sizes
$(N_1, \ldots, N_{k} )$ of a partition with $k$ non-empty clusters, given in \eqref{mixpK},
over all cluster sizes $N_1, \ldots, N_{k}$ such that
$N_1+\ldots+N_k=N$.  The resulting prior
$ \Prob{\Kn = k |N, \gammav} $ can be represented as
\begin{align} \label{mixpKmar}
\Prob{\Kn = k |  N, \gammav } &= \sum_{K=k}^ {\infty}  p(K )  \Prob{\Kn = k |N, K, \edK},
\end{align}
where $\Prob{\Kn = k |N, K,\gamma_K} $ is the prior of $\Kn$ for a fixed number of components $K$,
\begin{align*}
  \Prob{\Kn = k |N, K, \edK} &=
   \frac{N!}{ k!}  \times \frac{ V_{N, k}^{K, \edK}}{ \Gamfun{ \edK } ^{k} }  \times C_{N,k}^{K, \edK}, \\
  C_{N,k}^{K, \edK} &= \sum_{\substack{N_1,\ldots, N_k >0\\N_1+\ldots+N_k=N}}  \prod_{j=1}^k \frac{ \Gamfun{N_{j} +\edK }} {\Gamfun{N_j + 1} },
\end{align*}
and the prior uncertainty with respect to $K$ is integrated out. \comment{This
proves Theorem~\ref{Theorem4}.}

The number of terms in $C_{N,k}^{K, \edK}$ is the number of partitions of $N$
into $k$ integer summands with regard to order.  \comment{Algorithm~\ref{KNMFM}
is based on following} recursion %of order $N$ (at each iteration step)
 to compute $C_{N,k}^{K, \edK}$ for $k \in \{1,2,\ldots\}$.  % This makes it feasible to compute $C_{N,k}^{K, \edK}$ also for large values of $N$ both for static and dynamic MFM.
 We write
\begin{align*}
  C_{N,k}^{K, \edK} &= \sum_{\substack{N_1,\ldots, N_k > 0\\N_1+\ldots+N_k=N}}  \prod_{j=1}^k   w_{N_j}, \quad \mbox{\rm where}\quad  w_{n} =\frac{ \Gamfun{ n +\edK }} {\Gamfun{n + 1} }, \,\, n=1,\ldots,N.
\end{align*}
Since the cluster sizes are labeled, this can be written \commentBG{for $k\ge 2$} as:
\begin{align} \label{Pposys}
  C_{N,k}^{K, \edK} &=  \sum_{n=1} ^{N-k+1} w_{n}
  \sum_{\substack{n_1,\ldots, n_{k-1} > 0\\n_1+\ldots+n_{k-1}=N -n}}  \prod_{j=1}^{k-1}   w_{n_j} =
  \sum_{n=1} ^{N-k+1} w_{n}  C_{N-n,k-1}^{K, \edK},
\end{align}
where  $C_{\tilde n , \tilde k}^{K, \edK}$ is defined for  $\tilde k \in \{1, \ldots, K\}$ and  $\tilde n=1,\ldots, N$ as:
\begin{align}  \label{CNtilde}
  C^{K, \edK}_{\tilde n , \tilde  k} &=  \sum_{\substack{\tilde n_1,\ldots, \tilde n_{\tilde k} >0\\\tilde n_1+\ldots+\tilde n_{\tilde k}=\tilde n}}  \prod_{j=1}^ {\tilde k}  \frac{ \Gamfun{\tilde n_{j} +\edK }} {\Gamfun{\tilde n_j + 1} }.
\end{align}
Note that for $\tilde k =1$, $C_{\tilde n , 1}^{K, \edK} = \frac{ \Gamfun{\tilde n  +\edK }} {\Gamfun{\tilde n  + 1} } = w_{\tilde n}$.
Equation~\eqref{Pposys} is equivalent to the following recursive system:
\begin{align} \label{sysCk}
& C_{N,k}^{K, \edK}   =& w_{1 } C_{N-1,k-1}^{K, \edK} +\, & w_{2} C_{N-2,k-1}^{K, \edK}  &            &       \ldots   &  +\,w_{N-k+1} C_{k-1,k-1}^{K, \edK},\\
& C_{N-1,k}^{K, \edK} =&   \nonumber     & w_{1 } C_{N-2,k-1}^{K, \edK}  +\,&  w_{2 } C_{N-3,k-1}^{K, \edK}  &  \ldots  & +\,w_{N-k} C_{k-1,k-1}^{K, \edK},\\
&   & \cdots&          &                        &                 & \nonumber \\
& C_{k+1,k}^{K, \edK} =&   \nonumber    &                     &                      &  w_{1 } C_{k,k-1}^{K, \edK}  & +\, w_{2}C_{k-1,k-1}^{K, \edK}, \\
& C_{k,k}^{K, \edK}   =&   \nonumber     &                    &                        &                 & w_{1 }  C_{k-1,k-1}^{K, \edK}.
\end{align}
Hence, if we define
\begin{align*}
\cv _{K,k}   & =   \left(   \begin{array}{l}   C_{N,k}  ^{K, \edK}   \\  C_{N-1,k}   ^{K, \edK} \\ \vdots    \\   C_{k,k}  ^{K, \edK} \\ \end{array}   \right),
\end{align*}
for all  $k \in \{2,3,\ldots\}$, then we obtain from \eqref{sysCk}:
\begin{align*}
\cv _{K,k} &= \left( \begin{array}{cc} \bfz_{N-k+1 } &  \Wm_k    \\ \end{array} \right)
\cv _{K,k-1}, &
\Wm_k  &= \left(
\begin{array}{ccccc}
w_1 &  w_2     & \ddots   &w_{N-k} & w_{N-k+1} \\
       &  w_1      & \ddots  &\ddots                 & w_{N - k} \\
       &                 & \ddots  &  w_2               & \ddots  \\
      &               &             &  w_1               &w_2  \\
      &              &            &                        &   w_1 \\
\end{array}
\right).
\end{align*}
Obviously, $C_{N,k} ^{K, \edK}$ is equal to the first element of
the vector $\cv _{K,k}$ for all $k \in \{1,2,\ldots\}$.  $\Wm_1$ takes the
form given in \comment{Algorithm~\ref{KNMFM}} and $ \Wm_k $ is obtained from
$ \Wm_{k-1} $ for all $k \in \{2,3,\ldots\}$ by deleting the first row
and the first column.

\paragraph*{\comment{Proof of (\ref{dMFMpre}).}}

Using \citet{pit:som}, we obtain:
\begin{eqnarray*}
  \Prob{ \ym_{N+1} \in \parti_{k+1}|\bN, \Kn=k, \alpha  } &=&
   \frac{p(\parti \new | \commentSF{N+1}, \alpha)}{p(\parti | N, \alpha)}\\
   &=&
   \frac{\fEwens (\parti \new  |\commentSF{N+1}, \alpha )}{\fEwens (\parti |N, \alpha )}
\cdot  \frac{ \sum_{K= k+1 }^ {\infty}  p(K )  R_{\commentSF{\bN \new,k+1}}^{K, \alpha}}
  { \sum_{K= k }^ {\infty}  p(K )  R_{\bN ,k}^{K, \alpha}},
  \end{eqnarray*}
  \commentSF{where $\bN \new =(\bN,N_{k+1})$. Since the $(k+1)$th cluster is of size
  $N_{k+1}=1$, we obtain:}
   \begin{align*}
  R_{\commentSF{\bN \new},k+1}^{K, \alpha}= \prod_{j=1}^{k}   \frac{\Gamfun{N_j +  \frac{\alpha}{K} } (K -j+1)  }   { \Gamma(1+\frac{\alpha}{K})\Gamfun{N_j}K} \frac{K-k}{\commentSF{K}}
  = \frac{K-k}{\commentSF{K}} R_{\bN,k }^{K, \alpha},
   \end{align*}
   \commentSF{and
   \begin{align*}
   \frac{\fEwens (\parti \new  |\commentSF{N+1}, \alpha )}{\fEwens (\parti |N, \alpha )} = \frac{\alpha}{N +\alpha}.
    \end{align*}
    Therefore,}
   \begin{eqnarray*}
  \Prob{ \ym_{N+1} \in \parti_{k+1}|\bN, \Kn=k, \alpha  }
 & = &  \frac{\alpha }{\alpha + N }
 \cdot  \frac{ \sum_{K= k+1 }^ {\infty}  p(K ) \commentSF{(K-k)/K}  R_{\bN ,k}^{K, \alpha}}
  { \sum_{K= k }^ {\infty}  p(K )  R_{\bN ,k}^{K, \alpha}} \\
   & = &  \commentSF{\frac{\alpha }{\alpha + N }
 \cdot  \frac{ \sum_{K= k }^ {\infty}  p(K ) (K-k)/K  R_{\bN ,k}^{K, \alpha}}
  { \sum_{K= k }^ {\infty}  p(K )  R_{\bN ,k}^{K, \alpha}} }.
   \end{eqnarray*}
  \commentSF{Evidently, since $(K-k)/K \leq 1$, $\Prob{ \ym_{N+1} \in \parti_{k+1}|\bN, \Kn=k, \alpha  }$ is bounded by the predictive probability $\alpha/(\alpha + N )$ of a DPM:
    \begin{eqnarray*}
\Prob{ \ym_{N+1} \in \parti_{k+1}|\bN, \Kn=k, \alpha  } =  \frac{\alpha }{\alpha + N }
\frac{ \sum_{K= k }^ {\infty}  p(K ) (K-k)/K  R_{\bN ,k}^{K, \alpha}}
  { \sum_{K= k }^ {\infty}  p(K )  R_{\bN ,k}^{K, \alpha}}  \leq
   \frac{\alpha }{\alpha + N }.
  \end{eqnarray*}
  To make this relation more evident, the predictive probability is expressed as in (\ref{dMFMpre}):
  \begin{eqnarray*}
\Prob{ \ym_{N+1} \in \parti_{k+1}|\bN, \Kn=k, \alpha  }
 & = &  \frac{\alpha }{\alpha + N }
 \cdot \left( 1 - k \frac{ \sum_{K= k }^ {\infty}  p(K )/K  R_{\bN ,k}^{K, \alpha}}
  { \sum_{K= k }^ {\infty}  p(K )  R_{\bN ,k}^{K, \alpha}} \right).
 \end{eqnarray*}}

\begin{table}[t!]
  \centering
  \begin{tabular}{lccc}
    \hline
    $K-1 \sim p_t$ &  $p(K)$ & $\Ew{K-1}$ & $\Ew{K}$\\\hline\\[-4mm]
    $\Poiss{\lambda}$  &  $ \frac{\lambda^{K-1}}{ \Gamfun{K}} e ^ {- \lambda}$ &  $\lambda$ &  $\lambda+1$ \\[1mm]
    $ \NegBin{\alphaNB,\beta}$   &  $ \Bincoef {\alphaNB + K -2}{\alphaNB-1}
     \left( \frac{\beta }{\beta+1} \right)^{\alphaNB}  \left(  \frac{1}{\beta+1} \right) ^{ K -1} $ &                                                                                                                                                                               $  \frac{\alphaNB}{\beta}$ &   $ 1 + \frac{\alphaNB}{\beta}$\\[1mm]
    $\Geo{\pigeo}$     &   $   \pigeo (1 - \pigeo) ^{ K -1}$  &   $ \frac{1-\pigeo}{\pigeo}$  &   $ \frac{1}{\pigeo}$ \\[1mm]
    $\BNB{\alphaNB, \pia,\pib}$    & $ \frac{\Gamfun{\alphaNB + K -1}\Betafun{\alphaNB+\pia, K -1 + \pib}}{\Gamfun{\alphaNB} \Gamfun{K}\Betafun{\pia,\pib}}$    &  $ \alphaNB \frac{\pib}{\pia-1}$  & $  1+ \alphaNB \frac{\pib}{\pia-1} $ \\[1mm]
    \hline
  \end{tabular}
  \caption{Priors on the number of components based on various
    translated priors $K-1 \sim p_t$. The corresponding pmf
    $p(K)=p_t(K-1)$ as well as $\Ew{K-1}$ and $\Ew{K}$ \comment{(which exist
    for the $\BNB{\alphaNB, \pia,\pib}$-prior, iff $\pia>1$)} are
    given.}\label{tab:priorK}
\end{table}

\section{The beta-negative-binomial distribution}\label{app:beta-negat-binom}

The beta-negative-binomial (BNB) distribution is a hierarchical
generalization of the Poisson, the geometric and the negative-binomial
distribution. This can be derived in the following way: The starting
point is the translated Poisson distribution
$K-1 \sim \Poiss{\lambda}$ introduced by \citet{mil-har:mix} with a
fixed value of $\lambda$ which also determines the prior mean
$\Ew{K-1}=\lambda$.  A typical choice is $\lambda =1$, but this choice
might be influential and it appears promising to consider hierarchical
priors.

Assuming the Gamma prior $\lambda \sim \Gammad{\alphaNB,\beta}$ on
$\lambda$ leads to the translated negative-binomial distribution
$K-1 \sim \NegBin{\alphaNB,\beta}$.  For $\alphaNB=1$, this
distribution reduces to the translated geometric distribution
$K-1 \sim \Geo{\pigeo}$ with success probability
$\pigeo=\beta/(1+\beta)$, modeling the number of failures before the
first success.  The pmf of the negative-binomial distribution can be
combined with the hierarchical prior $\pigeo \sim \Betadis{\pia,\pib}$
on $\pigeo=\beta/(1+\beta)$. Marginally, this yields the translated
BNB distribution $K-1 \sim \BNB{\alphaNB, \pia,\pib}$.

\begin{figure}[t!]
  \centering
  \includegraphics[width=0.60\textwidth]{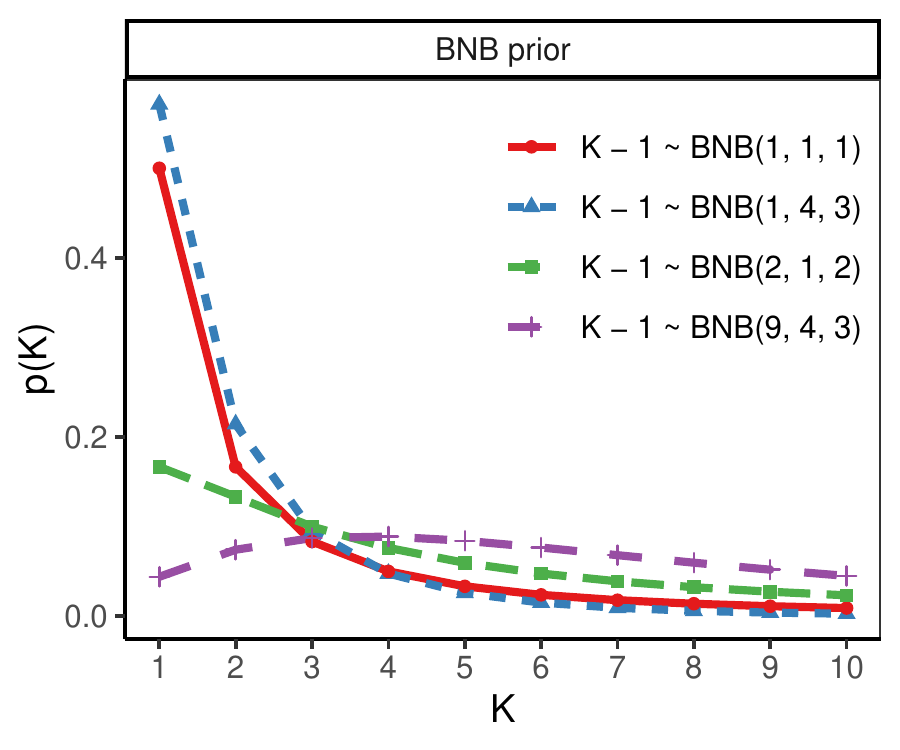}
  \caption{Priors \comment{$p(K)$ derived from the translated prior} $K-1 \sim \BNB{\alphaNB, \pia,\pib}$ for various parameter values \commentSF{$\alphaNB, \pia$, and $\pib$}.}\label{plot:BNB}
\end{figure}

Table~\ref{tab:priorK} gives an overview on the beta-negative-binomial
(BNB) distribution including its special cases given by the Poisson,
negative-binomial and the geometric distribution. The translated pmf
is provided as well as the prior mean values $\Ew{K-1}$ and
$\Ew{K}$. The different shapes of the BNB distribution possible for
various values of the parameters are illustrated in
Figure~\ref{plot:BNB}.

\section{Inference algorithm: Telescoping sampling} \label{sec:telescoping-sampler}

In the following we provide
more details about using telescoping sampling for MCMC estimation of
MFMs.
Algorithm~\ref{TELE} can be easily modified for static MFMs, as
outlined in Algorithm~\ref{TELEstat}.

\paragraph*{Starting values and burn-in.}

We define starting values in Algorithms~\ref{TELE} and \ref{TELEstat},
respectively, in the following way. \emph{k-means} \citepApp{mac:som}
or \emph{k-modes} \citepApp{hua:fas} clustering is used to split the
data into $K_0$ initial clusters, where $K_0$ is clearly overfitting
the number of clusters, e.g., $K_0=10$ or $15$, if about 5 clusters are
expected.  The cluster centers returned by \emph{k-means} or
\emph{k-modes} are the initial values for the component means. In case
the component distributions have a variance parameter independent of
the mean, e.g., \commentSF{for Gaussian distributions}, sufficiently large values
are specified to encourage merging of the components in the first
classification \commentSF{steps}. The component weights are initialized using
uniform weights.

We repeat Algorithms~\ref{TELE} and \ref{TELEstat}, respectively, for
$M_0 +M$ iterations and discard the draws from the first $M_0$
iteration as burn-in. In general only a rather small number of burn-in
iterations (e.g., 1,000) is required to reach a region of the parameter
space with high posterior values, while many iterations (e.g., 100,000)
need to be recorded in order to sufficiently well explore regions of
the parameter space with high posterior values. Convergence of the
MCMC \comment{sampler} \commentSF{is} assessed by exploring trace plots of the posterior of the
number of clusters $\Kn$ or the component weights.

\paragraph*{Details on Step~1(b).}

To reorder the components, determine the indices
$\{i_1, \ldots, i_{\Kn}\}$ $\subset \{1,\ldots,K\}$ of the $\Kn$
non-empty components and let $i_{\Kn+1}, \ldots, i_{K}$ be the
remaining sub-indices corresponding to the $K-K_+$ empty
components. Note that $\{i_1, \ldots, i_{K}\}$ is not unique, \comment{but the algorithm is invariant to the specific choice}. Given
$\{i_1, \ldots, i_{K}\}$, the cluster sizes, the component
parameters and the component weights are reordered using:
\begin{eqnarray} \label{reorder}
  N_k:=N_{i_k},\quad \thetav_{k}:= \thetav_{i_k},\quad \eta_{k}:= \eta_{i_k}, \qquad k=1,\ldots,K.
\end{eqnarray}
To reorder the allocations variables $S_i$, use the permutation
$\rho:\{1,\ldots,K\}\rightarrow \{1,\ldots,K\}$ underlying
\eqref{reorder}:
\begin{eqnarray*}
S_i:=\rho(S_i),  \qquad i=1,\ldots,N.
\end{eqnarray*}
Note that
$\{i_1, \ldots, i_{K}\}= \{\rho^{-1}(1),\ldots,\rho^{-1}(K)\}$, i.e.,
$i_k= \rho^{-1}(k)$. Therefore, $\rho$ can be recovered by ordering
the pairs $(i_k,k)$, $k=1,\ldots,K$ with respect to the first element.
Since $i_k=\rho^{-1}(k)$, the reordered pairs are equal to $(k,\rho(k))$.

\begin{algorithm}[t!] \caption{Telescoping sampling for a static MFM.}   \label{TELEstat}
  \footnotesize Perform Steps~1,~2, and~4(a)  as  in Algorithm~\ref{TELE} and substitute  Step~3 and~4(b) by the following  steps:
  \begin{enumerate} \itemsep 0mm
  \item[3(a*)]  Conditional on $\cP$ and $\eFM$, sample $K$  from
    \begin{eqnarray*}
      p(K|\cP,\eFM) % &\propto& p(K) p(\cP|\eFM,K) , ,\\
      \propto  p(K) 	\frac{K!}{(K-K_+)!} \frac{\Gamma(\eFM K)}{\Gamma(N+\eFM K)},   \quad K=K_+, K_+ +1,\ldots .
      % \prod_{k=1}^{K_+} \frac{\Gamma(N_k+\frac{\alpha}{K})}{\Gamma(\frac{\alpha}{K})}	,
    \end{eqnarray*}
  \item[3(b*)]   Use a  random walk Metropolis-Hastings with proposal
    $\log(\eFM^{\new}) \sim \Normal{\log(\eFM),s_{\eFM}^2}$ to   sample $\eFM|\cP,K$ from
    \begin{eqnarray*}
      p(\eFM |\cP,K) \propto % p(\eFM)p(\cP|K,\eFM)=
      p(\eFM)\frac{\Gamma(\eFM K)}{\Gamma(N+\eFM K)} \prod_{k=1}^{K_+} \frac{\Gamma(N_k+\eFM)}{\Gamma(\eFM)}.
    \end{eqnarray*}
    Numeric stability for small values of $\eFM$ is achieved through $\Gamma(\eFM)=\Gamma(1+\eFM)/\eFM$.
  \item[4(b*)] Sample $\etav_K|K,\eFM, \Siv \sim \Dir{e_1,\ldots,e_K}$, where $e_k=\eFM + N_k  $.
  \end{enumerate}
\end{algorithm}
	
\paragraph*{Details on Step~3(a).}

In Step~3(a),
$\Gamma(\frac{\alpha}{K})=\frac{K}{\alpha}\Gamma(1+\frac{\alpha}{K})$
is used to evaluate the posterior \eqref{postKK2} to increase the
numeric stability for large values of $K$ or small values of $\alpha$,
respectively.
\begin{eqnarray}  \label{postKK3}
  \frac{K!}{(K-K_+)!}
  \prod_{j=1}^{K_+} \frac{\Gamma(N_j+\frac{\alpha}{K})}{\Gamma(\frac{\alpha}{K})}
= 	\frac{ \alpha ^{\Kn}   K!}{K ^{\Kn} (K-K_+)!}
  \prod_{j=1}^{K_+} \frac{\Gamma(N_j+\frac{\alpha}{K})}{\Gamma(1+ \frac{\alpha}{K})}	.
\end{eqnarray}

\section{Empirical demonstrations \comment{-- Details and additional results}}

\setcounter{equation}{0}
\setcounter{figure}{0}
\setcounter{table}{0}

\subsection{Benchmarking the telescoping sampler}\label{app:benchm-telesc-sampl}

\begin{table}[t!]
  \centering
  {\small
    \begin{tabular}{lcccccccc}
      \toprule
      Sampler & 1 & 2 & 3 & 4 & 5 & 6 & 7 \\
      \midrule
      TS & 0.000 & 0.000 & 0.060 & 0.135 & 0.188 & 0.195 & 0.158 \\
              & (0.000) & (0.000) & (0.005) & (0.004) & (0.002) & (0.002) & (0.002) \\
      RJ & 0.000 & 0.000 & 0.061 & 0.134 & 0.187 & 0.194 & 0.157 \\
              & (0.000) & (0.000) & (0.004) & (0.008) & (0.011) & (0.011) & (0.009) \\
      JN & 0.000 & 0.000 & 0.061 & 0.135 & 0.188 & 0.195 & 0.158 \\
              & (0.000) & (0.000) & (0.001) & (0.002) & (0.001) & (0.001) & (0.001) \\
      \midrule
      8 & 9 & 10 & 11 & 12 & 13 & 14 & 15 \\
      \midrule
      0.109 & 0.068 & 0.039 & 0.022 & 0.012 & 0.006 & 0.003 & 0.002 \\
			(0.002) & (0.001) & (0.001) & (0.001) & (0.000) & (0.000) & (0.000) & (0.000) \\
      0.108 & 0.067 & 0.039 & 0.022 & 0.012 & 0.006 & 0.003 & 0.002 \\
      (0.006) & (0.004) & (0.002) & (0.001) & (0.001) & (0.000) & (0.000) & (0.000) \\
      0.109 & 0.068 & 0.039 & 0.022 & 0.012 & 0.006 & 0.003 & 0.002 \\
      (0.001) & (0.001) & (0.000) & (0.000) & (0.000) & (0.000) & (0.000) & (0.000) \\
      \bottomrule
    \end{tabular}}
  \caption{Galaxy data. Estimates of the \textbf{posterior of}
    $\mathbf{K}$ for the telescoping (TS), the RJMCMC (RJ) and the
    Jain-Neal (JN) sampler. Means (and standard deviations in parentheses)
    over 100 MCMC runs are reported.}\label{tab:K_gal}
\end{table}

\begin{table}[t!]
  \centering
  {\small
    \begin{tabular}{lcccccccc}
      \toprule
      Sampler & 1 & 2 & 3 & 4 & 5 & 6 & 7 \\
      \midrule
      TS & 0.000 & 0.000 & 0.070 & 0.161 & 0.228 & 0.228 & 0.159 \\
              & (0.000) & (0.000) & (0.005) & (0.004) & (0.003) & (0.003) & (0.003) \\
      RJ & 0.006 & 0.000 & 0.070 & 0.161 & 0.227 & 0.226 & 0.158 \\
              & (0.058) & (0.000) & (0.005) & (0.010) & (0.013) & (0.013) & (0.009) \\
      JN & 0.000 & 0.000 & 0.070 & 0.162 & 0.228 & 0.228 & 0.159 \\
              & (0.000) & (0.000) & (0.002) & (0.002) & (0.002) & (0.002) & (0.001) \\
      \midrule
      8 & 9 & 10 & 11 & 12 & 13 & 14 & 15 \\
      \midrule
      0.087 & 0.040 & 0.017 & 0.006 & 0.002 & 0.001 & 0.000 & 0.000 \\
      (0.002) & (0.001) & (0.001) & (0.000) & (0.000) & (0.000) & (0.000) & (0.000) \\
			0.086 & 0.040 & 0.017 & 0.006 & 0.002 & 0.001 & 0.000 & 0.000 \\
      (0.005) & (0.003) & (0.001) & (0.000) & (0.000) & (0.000) & (0.000) & (0.000) \\
      0.087 & 0.040 & 0.017 & 0.006 & 0.002 & 0.001 & 0.000 & 0.000 \\
      (0.001) & (0.001) & (0.000) & (0.000) & (0.000) & (0.000) & (0.000) & (0.000) \\
      \bottomrule
    \end{tabular}}
  \caption{Galaxy data. Estimates of the \textbf{posterior of}
    $\mathbf{K_+}$ for the telescoping (TS), the RJMCMC (RJ) and the
    Jain-Neal (JN) sampler. Means (and standard deviations in
    parentheses) over 100 MCMC runs are reported.}\label{tab:Kp_gal}
\end{table}

Tables~\ref{tab:K_gal} and \ref{tab:Kp_gal} provide additional details
on the results obtained \comment{in Section~\ref{sec:benchm-telesc-sampl}} when using the telescoping sampler (TS),
RJMCMC (RJ) and the Jain-Neal sampler (JN) to fit a static MFM to the
Galaxy data set using the priors as suggested in
\citetApp{ric-gre:bay}. Table~\ref{tab:K_gal} gives the \commentSF{average} posterior
probabilities over 100 different initializations for the posterior of
$K$ together with the standard deviations in parentheses. The mean
values differ at most at the third decimal place. The standard
deviations differ more strongly, in particular the RJ sampler has
higher standard deviations than the other two samplers. Similar
observations apply to Table~\ref{tab:Kp_gal} depicting the mean
posterior probabilities over 100 different initializations for the
posterior of $\Kn$ together with the standard deviations in
parentheses.

\begin{figure}[t!]
  \centering
  \includegraphics[width=0.85\textwidth]{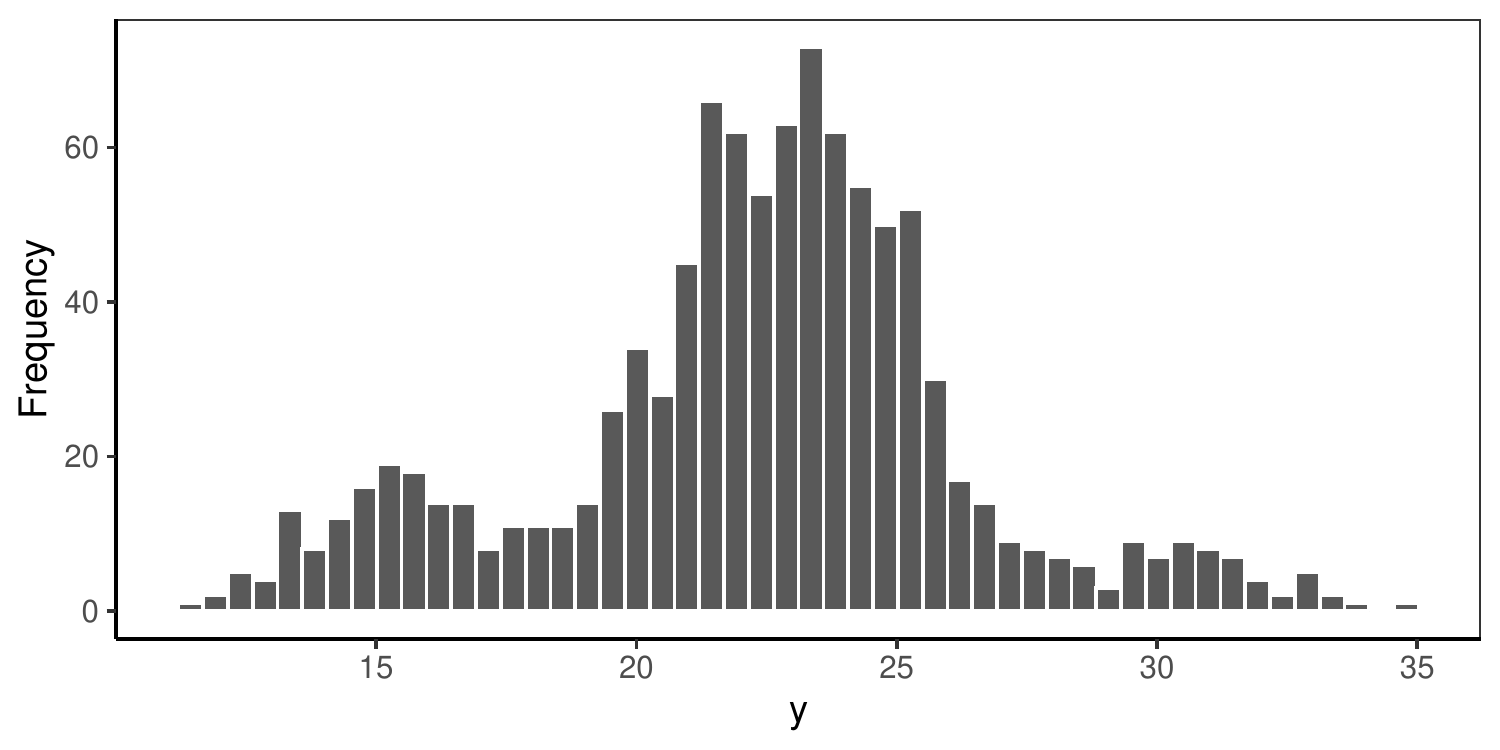}
  \caption{Histogram of the simulated data \comment{used in Section~\ref{sec:benchm-telesc-sampl}.  $N=1000$ data points} are
    simulated from a three-component univariate Gaussian mixture
    with parameters $(\mu_1,\mu_2,\mu_3)=(15,23,31)$,
    $(\sigma^2_1,\sigma^2_2,\sigma^2_3)=(3,5,3)$ and
    $\boldeta=(0.13,0.80,0.07)$.\label{fig:N1000}}
\end{figure}

Figure~\ref{fig:N1000} visualizes the empirical distribution of the
sample with $N=1000$ observations \commentSF{drawn} from a three-component mixture of
Gaussian distributions \commentSF{and} used \comment{in Section~\ref{sec:benchm-telesc-sampl}} for the performance comparison of the
three samplers. The three components differ in size with one large
component of size 0.8 \commentSF{and a small component of size 0.07}. While the three modes are guessable, there
is considerable overlap between the components.

\subsection{\commentBG{Sensitivity to the prior choice on the number of
  components}}
\label{app:revis-galaxy-data}

The Galaxy data set has been %previously
used \comment{numerous times} in the literature to
illustrate the use of Bayesian methods to fit a mixture model with
\comment{univariate} Gaussian components, in particular to address the issue of the number
of components and clusters. \citetApp{ait:lik} compares the results
obtained in \citetApp{esc-wes:bay}, \citetApp{car-chi:bay},
\citetApp{phi-smi:bay}, \citetApp{roe-was:pra} and \citetApp{ric-gre:bay} and
points out that the posterior probabilities for $K$ obtained in the
different analyses are rather diffuse over the range 4--9, except for
\citetApp{roe-was:pra} who conclude that the number of components is
almost certainly three. The five Bayesian analyses did not only differ
with respect to the prior specification on $K$ and $\edK$, but also
the priors specified for the component parameters.  \citetApp{ait:lik}
also compares the Bayesian results to those obtained using a maximum
likelihood analysis which shows strong evidence for 3 or 4 \comment{mixture} %data
components, depending on whether equal or unequal variances between
the components are considered.

\begin{figure}[t!]
  \centering
  \includegraphics[width=\textwidth]{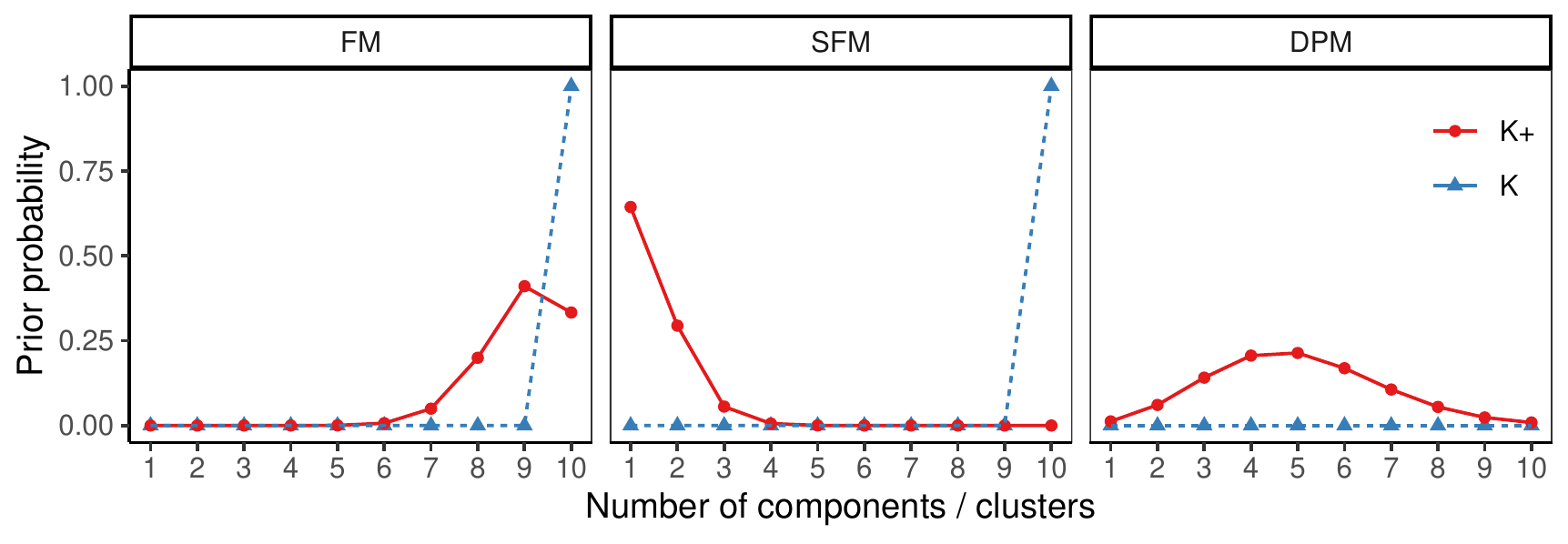}
  \caption{\comment{Galaxy data.}  Prior of $\Kn$ (solid \commentBG{red} lines, \commentBG{circles}) for the standard
    finite mixture model with $K=10$ fixed and $\gamma=1$ (left), the
    sparse finite (or overfitting) mixture model with $K=10$ fixed and
    $\gamma=0.01$ (middle), and a DPM with $\alpha=1$ (right),
    $N=82$. The prior on $K$ \commentBG{(dashed blue lines, triangles)} is a spike on $K=10$ (left and middle)
    and on $K=\infty$ (right, not shown).  \label{plot:DPM}}
  \includegraphics[width=\textwidth]{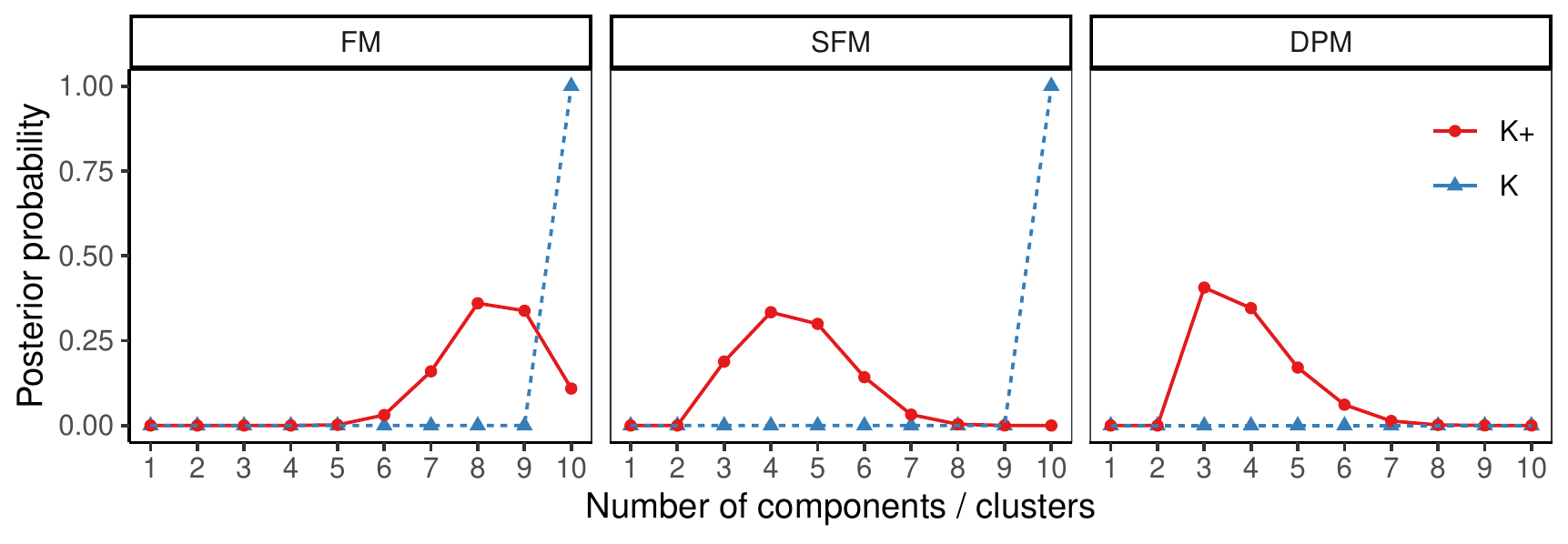}
  \caption{Galaxy data. Posterior of $\Kn$ \commentBG{(solid red lines, circles) and $K$ (dashed blue lines, triangles)}
    for a standard finite mixture with $K=10,\gamma=1$ (left), a sparse
    finite mixture with $K=10,\gamma=0.01$ (middle) and a DPM with
    $\alpha=1$ (right).
    \label{plot_postDPM}}
\end{figure}
\begin{figure}[t!]
  \includegraphics[width=\textwidth]{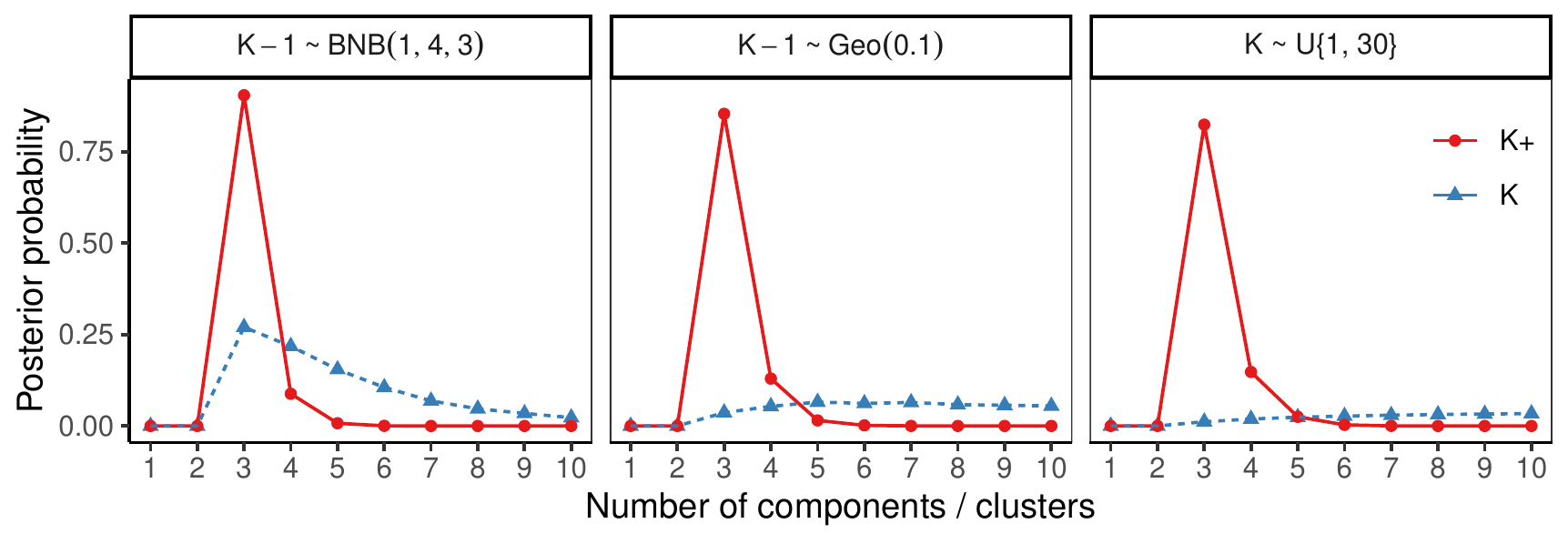}
  \caption{Galaxy data. Posteriors of $K$ (dashed blue lines,
    triangles) and $\Kn$ (solid red lines, \commentBG{circles}) under priors
    $K-1 \sim \BNB{1, 4, 3}$ (left), $K-1\sim \Geo{0.1}$ (middle)
    and $K \sim \cU\{1,30\}$ (right) for a dynamic MFM with
    $\alpha \sim \cG(1,20)$, for $N=82$. \label{plot_F_G}}
\end{figure}
In addition to static and dynamic MFMs \comment{considered in Section~\ref{revGalax}}
also a finite mixture model (FM), a sparse finite mixture model (SFM) and a Dirichlet process
mixture (DPM) model \comment{are fitted.} % considered for the Galaxy data set.
While all three modeling approaches might be seen as \comment{special cases of} %dynamic
 MFMs, they differ in the specification \comment{of % the prior on
 $K$ and the Dirichlet parameter.} %$\alpha$.
 The FM uses a fixed value of $K$ and $\gamma = 1$ inducing
a priori a uniform distribution on the component weights. The SFM \commentSF{combines} a fixed
value of $K$ with a fixed small value for $\gamma$ which induces that
a priori empty components occur. Figure~\ref{plot:DPM} visualizes the
priors for $K$ and $\Kn$ for the FM, SFM, and DPM. The fixed value for
$K$ is equal to $10$ and the number of observations is selected as
$N=82$. We use for the FM $\gamma = 1$ (corresponding to
$\alpha = 10$ for a dynamic MFM), for the SFM $\gamma = 0.01$
(corresponding to $\alpha = 0.1$ for a dynamic MFM) and for the DPM
$\alpha = 1$. For FM and SFM the prior on $K$ has a degenerate
distribution putting all \comment{prior} mass at 10, whereas the DPM puts all mass at
$K=\infty$. The FM with $\gamma = 1$  \comment{implies} %has
a mode at 9 for the prior on
$\Kn$ while also putting considerable mass on \comment{$\Kn = 8$ and
$\Kn = 10$}. Clearly a value of $\gamma = 1$ is not sufficiently large
to ensure that all components are filled. For the SFM, the prior on
$\Kn$ has its mode at 1 and is quickly decreasing putting also some
mass on $\Kn = 2$, but negligible mass \comment{on higher values of $\Kn$}. The DPM
prior for $\Kn$ is a unimodal distribution with mode at 4--5,
% and having some mass for the surrounding values, but
but essentially no mass
assigned to $\Kn = 1$ or $\Kn = 10$ and beyond.

The posterior distributions for the priors in Figure~\ref{plot:DPM}
are shown in Figure~\ref{plot_postDPM} when fitting the corresponding
 \comment{finite mixture} % MFM
and DPM specifications to the Galaxy data set. While
%the posteriors for
$K$ \comment{(being fixed) remains} unchanged, the differences in prior
distributions for $\Kn$ are also reflected in different posteriors for
$\Kn$. The FM  obtains a fine-grained approximation of the data
density with in general 8--9 components being filled in the mixture
model. The SFM obtains an approximation with only 4--5 components
being filled \comment{with} a high probability, with some probability also being
assigned to 3 or 6 components being filled. The approximation with the
DPM specification is the sparsest with a mode at $\Kn = 3$ and most
mass assigned to the values 3--5.

\commentBG{
If the
shrinkage prior $\alpha \sim \cG(1,20)$ is specified, the posterior of
$K_+$ becomes completely independent of both the prior and posterior
of $K$, as can be seen in Figure \ref{plot_F_G}.
In this case, regardless of $p(K)$, for each prior
specification three clusters are estimated, while the posteriors of
$K$ are very flat.}

\begin{table}[t!]
  \centering
  \begin{tabular}{@{}rrrrrrrrrrr@{}}
    \toprule
  &\multicolumn{3}{c}{F}&\multicolumn{3}{c}{C}&\multicolumn{4}{c}{M}\\
     \cmidrule(lr{.05em}){2-4}\cmidrule(lr{.05em}){5-7}\cmidrule(lr{.05em}){8-11}
    &1 &2 &3& 1 & 2 & 3& 1 & 2 & 3 & 4 \\
    \midrule
  1 & 0.62 & 0.28 & 0.09 & 0.68 & 0.11 & 0.21 & 0.22 & 0.57 & 0.13 & 0.08 \\
   & (0.10) & (0.09) & (0.06) & (0.08) & (0.05) & (0.07) & (0.06) & (0.08) & (0.06) & (0.04) \\
  2 & 0.07 & 0.29 & 0.64 & 0.26 & 0.31 & 0.43 & 0.15 & 0.17 & 0.41 & 0.28 \\
   & (0.06) & (0.09) & (0.10) & (0.09) & (0.08) & (0.09) & (0.06) & (0.08) & (0.09) & (0.08) \\
    \bottomrule
  \end{tabular}
  \caption{Fear data. Posterior means (and standard deviations in
    parenthesis) for the cluster-specific success probabilities after
    model identification for $K-1\sim \BNB{1,4,3}$ and
    \commentSF{a dynamic MFM with} $\alpha\sim \cF(6,3)$ and a uniform Dirichlet prior on the
    component parameters. \label{tab:fear_prob}}
\end{table}

\subsection{Changing the clustering kernel}\label{sec:furth-illustr-appl}

\subsubsection{Multivariate Gaussian mixtures: Thyroid data}\label{app:mult-gauss-mixt}

A simplified version of the priors proposed in \citeApp{mal-etal:mod}
are specified on the component parameters. That is,
$\bmu_k \sim \cN(\bb_0,\bB_0)$, $\bb_0=\text{median}(\by)$,
$\bB_0=\Diag{R_1^2,\ldots,R_r^2}$, where $R_j$ is the range of the
data in dimension $j$, and $r=5$. For the component covariance
matrices the hierarchical prior $\bSigma_k^{-1} \sim\cW(c_0,\bC_0)$,
$\bC_0 \sim \cW(g_0,\bG_0)$, where $c_0=2.5+(r-1)/2$,
$g_0=0.5+(r-1)/2$ and $\bG_0=100g_0/c_0$
$\Diag{1/R_1^2,\ldots,1/R_r^2}$, is assumed. Note that the same priors
on the component parameters are used in the simulation study with
artificial data \comment{in Section~\ref{TSartificial}} where also multivariate Gaussian mixtures are fitted.

% A simplified version of the priors proposed in \citeApp{mal-etal:mod} are
% specified on the component parameters. That is,
% $\bmu_k \sim \cN(\bb_0,\bB_0)$, $\bb_0=\text{median}(\by)$,
% $\bB_0=\Diag{R_1^2,\ldots,R_r^2}$, where $R_j$ is the range of the
% data in dimension $j$, and $r=5$. For the component covariance
% matrices the hierarchical prior $\bSigma_k^{-1} \sim\cW(c_0,\bC_0)$,
% $\bC_0 \sim \cW(g_0,\bG_0)$, where $c_0=2.5+(r-1)/2$,
% $g_0=0.5+(r-1)/2$ and
% $\bG_0=100g_0/c_0$ $\Diag{1/R_1^2,\ldots,1/R_r^2}$, is assumed.

\subsubsection{Latent class analysis: Fear data}\label{app:latent-class-analys}

Table~\ref{tab:fear_prob} summarizes the cluster-specific parameter
estimates obtained for a dynamic MFM model after model
identification. A dynamic MFM was fitted with the following prior
specifications: $K-1\sim \BNB{1,4,3}$, $\alpha\sim \cF(6,3)$ and
uniform Dirichlet priors on the component parameters. Model
identification is performed by first selecting the mode of the
posterior on $\Kn$ as suitable number of clusters. In the following
only the MCMC draws are considered where the number of filled
components equals the estimated number of clusters $\Kn$ and unique
labels are assigned by clustering the component parameters of
filled components in the point process representation and retaining
only MCMC draws where the cluster labels assigned to the
component parameters of filled components from the same MCMC
draw represent a permutation of the numbers 1 to the estimated number
of clusters.

The posterior distributions of the cluster-specific parameters
obtained in this way are summarized in Table~\ref{tab:fear_prob} by
the posterior mean and standard deviation.  \commentBG{Note that the
  categories can be interpreted as scores with higher scores
  indicating a stronger behavior.}  Whereas children belonging to
class 2 are more likely to have higher scores in all three variables,
children in class 1 show less motor activity, crying behavior and fear
at the same time. This clustering result coincides with both the
results reported in \cite{fru-mal:fro} and the psychological theory
behind the experiments, according to which all three behavioral
variables are regularized by the same physiological mechanism, see
\cite{ste-etal:sta} for more details.

\subsection{Investigating the telescoping sampler with artificial data}
\label{app:invest-telesc-sampl}

\begin{figure}[t!]
  \includegraphics[width=0.7\textwidth, trim = 0 20 0 10, clip]{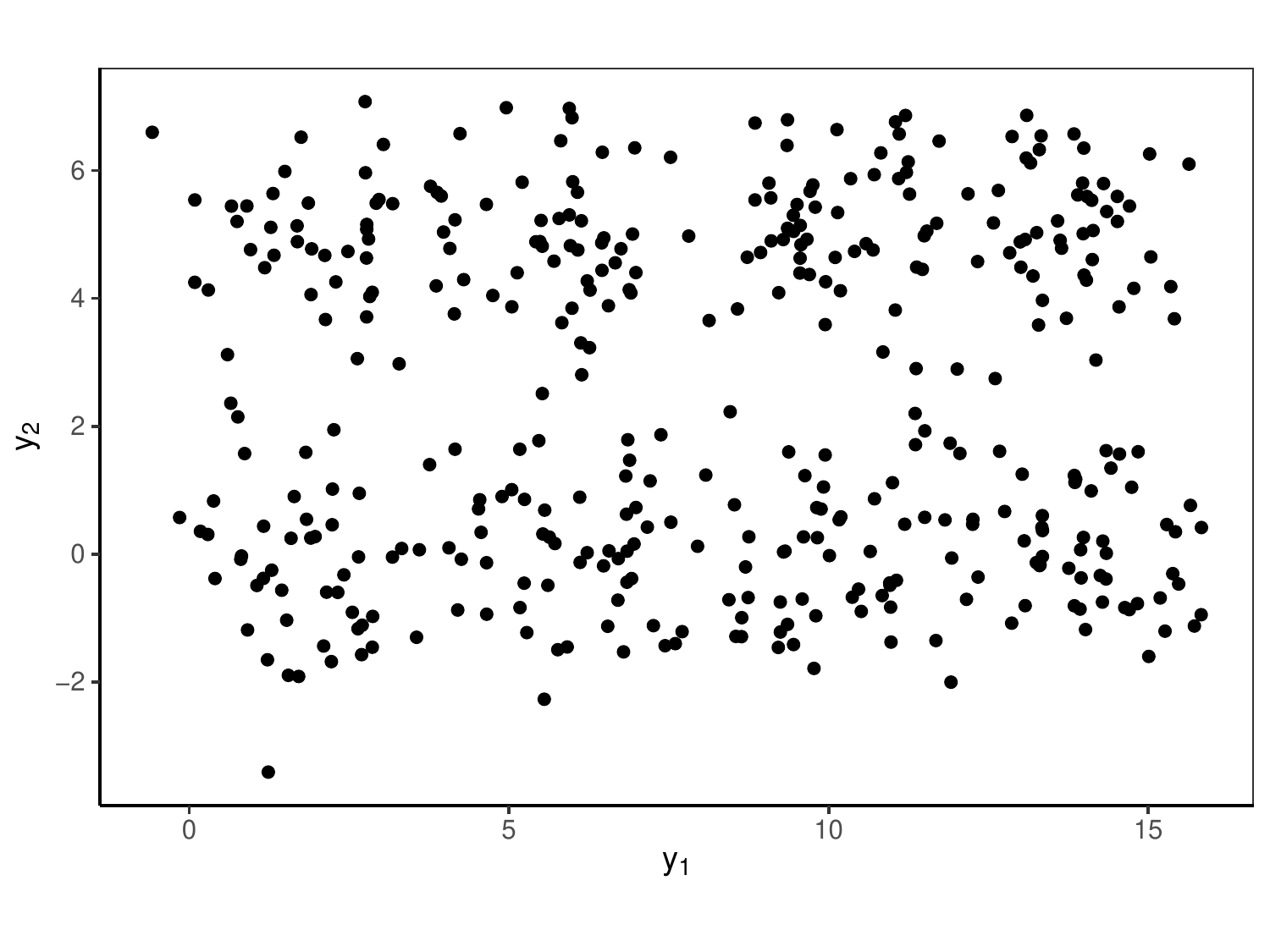}
  \caption{Simulation study, one artificial data
    set with eight components, $N=400$, $r = 2$. \label{fig:plot_K8}}
\end{figure}

For the simulation study \comment{in Section~\ref{TSartificial}}, we draw artificial data from finite mixtures
of \comment{multivariate} Gaussian distributions with eight components.
%For each finite mixture,
\comment{The} component weights are set to be equal, i.e., \comment{$\eta_k=1/8$}
% are equal to one eighth
 for all $k = 1,\ldots, \comment{8}$. The mean vectors for each of the components are determined in the following way.
%  by specifying four different values (2, 6, 10 and 14) in one dimension
% and two different values (0 and 5) in a second dimension which are crossed
\comment{The four values $\{2, 6, 10, 14\}$ are combined in  one dimension
with the two values $\{0, 5\} $ in a second dimension through a full factorial design
to define} eight different two-dimensional mean values. To
obtain the mean vectors for higher dimensional data (where $r$ is an
even number) the two dimensions are replicated but also multiplied with the
square root of the number of replicates to ensure that the Euclidean
distance between mean vectors remains the same. The
variance-covariance matrices of the component distributions are
assumed to be all equal to the identity matrix.
For drawing the artificial data, the number of dimensions $r$ and the
sample sizes $N$ are varied using the following settings:
$(N = 400, r = 2)$, $(N = 4000, r = 8)$ and $(N = 10000, r = 12)$. For
each setting 100 data sets are drawn. \comment{For illustration, an} example data set with
$N = 400$ and $r = 2$ is shown in Figure~\ref{fig:plot_K8}.

% The same priors are used for the component parameters as in the
% analysis of the Thyroid data where MFMs with multivariate Gaussian
% component distributions are fitted. The details are given in
% Section~\ref{app:mult-gauss-mixt}.

\comment{As} prior on $K$, we use the beta-negative-binomial
distribution $\BNB{1, 4, 3}$ for $K-1$ (see Section~\ref{sectionpK}),
the Poisson distribution with $\lambda = 1$ for $K-1$ \citep[similar
to][]{nob-fea:bay}, $\BNB{1, 1, 1}$  for $K-1$ \citep[as suggested
by][]{gra-etal:los}, $\Geo{0.1}$ for $K-1$ \citep[as suggested
by][]{mil-har:mix} and $\mathcal{U}\{1, 30\}$ for $K$ \citep[as
suggested by][]{ric-gre:bay}.
For the Dirichlet parameter $\gamma_K$ we consider different priors
for static as well as dynamic MFMs. For the static MFM where
$\gamma_K \equiv \gamma$ we use $\gamma \in \{1, 1 / \log(N), 0.01\}$.
$\gamma = 1$ corresponds to the value used in \citet{ric-gre:bay} and
\citet{mil-har:mix}; $\gamma = 0.01$ induces a sparse solution as
suggested by \citet{mal-etal:mod}. In addition we consider a specification for
$\gamma$ where $\gamma$ decreases in an indirectly proportional way to
the log of the sample size $N$. For the sample sizes considered, the
values of $\log(N)$ vary only moderately and take values between 6.0
and 9.2. For the dynamic MFM with $\gamma_K = \alpha / K$, we consider
a fixed value for $\alpha$ where $\alpha = 1$ and settings where a
prior on $\alpha$ is assumed. In addition to the prior
$\alpha \sim \cF(6, 3)$ \comment{(see Section~\ref{sectionpra})}, we consider $\alpha \sim \mathcal{G}(2, 4)$
\citep[see][]{esc-wes:bay} and $\alpha \sim \mathcal{G}(1, 20)$
\citep[see][]{fru-mal:fro}.
For the component parameters the same simplified version of the priors
proposed in \citeApp{mal-etal:mod} is used as for the Thyroid data set
(see Section~\ref{sec:mult-gauss-mixt} and
Appendix~\ref{app:mult-gauss-mixt}).

MCMC sampling is performed using the TS sampler. The sampler is
initialized using 15 filled components and then run for 10,000 burn-in
iterations and 100,000 iterations are recorded without any
thinning. The number of data clusters are estimated using the mode of
the posterior of the number of clusters. \comment{Applying the TS
  sampler to fit static and dynamic MFMs is straightforward, whereas
  \commentBG{the} RJMCMC and JN implementations, used in
  Section~\ref{sec:benchm-telesc-sampl} as benchmarks for the TS
  sampler, would require major changes to be applicable for this
  simulation setup where multivariate data and hierarchical priors are
  considered.}

\commentBG{In this simulation study the true data generating process
  is included in the fitted model. For larger sample sizes, we would
  thus expect to have the sampler concentrate on the part of the
  parameter space coinciding with the true data generating
  process. Results indicate that the TS sampler succeeds in converging
  during burn-in to the part of the parameter space where the
  estimated number of clusters $K_+$ corresponds to the true number of
  clusters. Note that the TS sampler is initialized with 15 filled
  components which implies that during burn-in filled components are
  merged and emptied. Overall the results indicate the feasibility of
  the TS sampler to be successfully applied in Bayesian cluster
  analysis for data with a clear clustering structure for sample sizes
  up to 10,000 and dimensions up to 12.}

\bibliographystyleApp{ba}
\bibliographyApp{sylvia_kyoto}